\documentclass{mimosis}

\usepackage{metalogo}

\usepackage{etoolbox}

\usepackage[%
  colorlinks = true,
  citecolor  = RoyalBlue,
  linkcolor  = RoyalBlue,
  urlcolor   = RoyalBlue,
  unicode,
  ]{hyperref}

\usepackage{bookmark}

\usepackage[%
  autocite     = plain,
  backend      = biber,
  doi          = true,
  url          = true,
  giveninits   = true,
  hyperref     = true,
  maxbibnames  = 99,
  maxcitenames = 99,
  sortcites    = true,
  style        = numeric,
  sorting      = none,
  ]{biblatex}

\AtEveryBibitem{\clearfield{month}}
\AtEveryCitekey{\clearfield{month}}

\AtBeginBibliography{%
  \renewcommand*{\finalnamedelim}{%
    \ifthenelse{\value{listcount} > 2}{%
      \addcomma
      \addspace
      \bibstring{and}%
    }{%
      \addspace
      \bibstring{and}%
    }
  }
}

\renewbibmacro{in:}{%
  \ifentrytype{article}
  {%
  }%
  {%
    \printtext{\bibstring{in}\intitlepunct}%
  }%
}

\renewbibmacro*{issue+date}{%
  \setunit{\addcomma\space}
    \iffieldundef{issue}
      {\usebibmacro{date}}
      {\printfield{issue}%
       \setunit*{\addspace}%
       \usebibmacro{date}}%
  \newunit}

\renewbibmacro*{volume+number+eid}{%
  \printfield{volume}%
  \setunit*{\addcolon}%
  \printfield{number}%
  \setunit{\addcomma\space}%
  \printfield{eid}%
}

\renewbibmacro*{publisher+location+date}{%
  \printlist{publisher}%
  \setunit*{\addcomma\space}%
  \printlist{location}%
  \setunit*{\addcomma\space}%
  \usebibmacro{date}%
  \newunit%
}

\renewbibmacro*{organization+location+date}{%
  \printlist{location}%
  \setunit*{\addcomma\space}%
  \printlist{organization}%
  \setunit*{\addcomma\space}%
  \usebibmacro{date}%
  \newunit%
}

\DeclareFieldFormat{labelnumberwidth}{#1\adddot}

\DeclareFieldFormat{doi}{%
  \mkbibacro{DOI}\addcolon\addnbspace
    \ifhyperref
      {\href{http://dx.doi.org/#1}{\nolinkurl{#1}}}
      {\nolinkurl{#1}}
}

\DeclareLanguageMapping{english}{english-mimosis}

\DeclareFieldFormat{citehyperref}{%
  \DeclareFieldAlias{bibhyperref}{noformat}%
  \bibhyperref{#1}%
}

\DeclareFieldFormat{textcitehyperref}{%
  \DeclareFieldAlias{bibhyperref}{noformat}%
  \bibhyperref{%
    #1%
    \ifbool{cbx:parens}
      {\bibcloseparen\global\boolfalse{cbx:parens}}
      {}%
    }%
}

\savebibmacro{cite}
\savebibmacro{textcite}

\renewbibmacro*{cite}{%
  \printtext[citehyperref]{%
    \restorebibmacro{cite}%
    \usebibmacro{cite}}%
}

\renewbibmacro*{textcite}{%
  \ifboolexpr{
    ( not test {\iffieldundef{prenote}} and
      test {\ifnumequal{\value{citecount}}{1}} )
    or
    ( not test {\iffieldundef{postnote}} and
      test {\ifnumequal{\value{citecount}}{\value{citetotal}}} )
  }%
  {\DeclareFieldAlias{textcitehyperref}{noformat}}
  {}%
  \printtext[textcitehyperref]{%
    \restorebibmacro{textcite}%
    \usebibmacro{textcite}}%
}

\newcommand{\vtg}{V_\mathrm{{TG}}}
\newcommand{\vbg}{V_\mathrm{{BG}}}
\newcommand{\vr}{V_\mathrm{{R}}}
\newcommand{\vl}{V_\mathrm{{L}}}
\newcommand{\bfA}{{\textbf{a}}}
\newcommand{\bfB}{{\textbf{b}}}
\newcommand{\bfC}{{\textbf{c}}}
\newcommand{\bfD}{{\textbf{d}}}
\newcommand{\bfE}{{\textbf{e}}}
\newcommand{\bfF}{{\textbf{f}}}
\newcommand{\xfree}{X_\mathrm{{2D}}}
\newcommand{\xqc}{X_\mathrm{{QC}}}

\usepackage[toc, title, titletoc]{appendix}

\usepackage[%
figurename=FIG.,
format=plain]{caption}
\ifxetexorluatex
  \usepackage{unicode-math}
\else
  \usepackage[lf]{ebgaramond}
  \usepackage[oldstyle,scale=0.7]{sourcecodepro}
\fi

\makeindex
\makeatletter
\@ifundefined{st}{%
  
}{}
\@ifundefined{rd}{%
  
}{}
\@ifundefined{nd}{%
  
}{}
\makeatother

\title{Electrically tunable quantum confinement of neutral excitons}
\author{DEEPANKUR THUREJA}

\begin{document}

\frontmatter
  \begin{titlepage}
  \begin{center}
    \begin{large}
      \textsc{Diss.\ ETH No.\ 29550}
    \end{large}\\[2.25cm]
    \begin{Huge}
      \makeatletter%
      \@title
      \makeatother%
    \end{Huge}\\[2.25cm]
    \begin{large}
      {A thesis submitted to attain the degree of}\\[0.4em]
      \textsc{Doctor of Sciences of ETH Zurich}\\
      (Dr.\,sc.\ ETH Zurich)
    \end{large}\\[1.75cm]
    \begin{large}
        {presented by}\\[0.2em]
        \textsc{DEEPANKUR THUREJA}\\[1.25em]
        {Master of Science ETH in Mechanical Engineering, ETH Zurich}\\[0.8em]
        {born on} 30.04.1992\\[0.8em]
        {citizen of Switzerland}    
    \end{large}\\[1.75cm]
    \begin{large}
        {accepted on the recommendation of}\\[0.8em]
        Prof.\,Dr.~David J. Norris, examiner\\
        Prof.\,Dr.~Ata\c{c} \.{I}mamo\v{g}lu, co-examiner\\
        Prof.\,Dr.~Lukas Novotny, co-examiner\\
        Prof.\,Dr.~Mete Atatüre, co-examiner
    \end{large}\\[1.75cm]
    {\large 2023}
  \end{center}
\end{titlepage}

\newpage
\null
\thispagestyle{empty}
\newpage

  \begin{center}
  \textsc{Abstract}
\end{center}
\noindent

\noindent Confining particles to distances below their de Broglie wavelength discretizes their motional state. This fundamental effect is observed in many physical systems, ranging from electrons confined in atoms or quantum dots to ultracold atoms trapped in optical tweezers. In solid-state photonics, a long-standing goal has been to achieve fully tunable quantum confinement of optically active electron--hole pairs known as excitons. To confine excitons, existing approaches mainly rely on material modulation, which suffers from poor control over the energy and position of trapping potentials. This has severely impeded the engineering of large-scale quantum photonic systems. \\

\noindent In this thesis, we demonstrate electrically controlled quantum confinement of neutral excitons in two-dimensional semiconductors. By combining gate-defined in-plane electric fields with inherent interactions between excitons and free charges in a lateral p-i-n junction, we achieve tunable exciton confinement lengths reaching values below $10\,$nm. Quantization of excitonic motion manifests in the measured optical response as a ladder of discrete voltage-dependent states below the continuum. Moreover, we observe that our confining potentials lead to a strong modification of the relative wave function of excitons. We further highlight the versatility of our approach by extending our confinement scheme to create quantum-dot-like structures with a fully tunable confinement length. \\

\noindent Our technique provides an experimental route towards achieving polariton blockade and creating scalable arrays of identical single-photon sources, which has wide-ranging implications for realizing strongly correlated photonic phases and on-chip optical quantum information processors.

  \begin{center}
  \textsc{Zusammenfassung}
\end{center}
\noindent

\noindent Durch den Einschluss von Teilchen auf einer L\"{a}nge unterhalb ihrer De-Broglie-Wellenl\"{a}nge wird ihr Bewegungszustand diskretisiert. Dieser grundlegende Effekt wird in vielen physikalischen Systemen beobachtet, von Elektronen, die in Atomen oder Quantenpunkten eingeschlossen sind, bis hin zu ultrakalten Atomen, die in optischen Pinzetten gefangen sind. In der Festk\"{o}rperphotonik besteht seit langem das Ziel darin, einen vollst\"{a}ndig einstellbaren Quanteneinschluss optisch aktiver Elektron-Loch-Paare, bekannt als Exzitonen, zu erreichen. Um Exzitonen einzufangen, st\"{u}tzen sich bestehende Ans\"{a}tze haupts\"{a}chlich auf Materialmodulation, die jedoch unter einer schlechten Kontrolle \"{u}ber die Energie und Position der Einfangpotentiale leidet. Dies hat die Entwicklung von gross angelegten Quantenphotonik-Systemen stark behindert. \\

\noindent In dieser Dissertation demonstrieren wir einen elektrisch gesteuerten Quanteneinschluss neutraler Exzitonen in zweidimensionalen Halbleitern. Durch die Kombination von Gate-definierten elektrischen Feldern in der Ebene mit inh\"{a}renten Wechselwirkungen zwischen Exzitonen und freien Ladungstr\"{a}gern in einem lateralen p-i-n \"{U}bergang erreichen wir einstellbare Einschlussl\"{a}ngen f\"{u}r Exzitonen, die Werte unter $10\,$nm erreichen. Die Quantisierung der exzitonischen Bewegung manifestiert sich in den optischen Messungen als eine Leiter mit diskreten, spannungsabh\"{a}ngigen Zust\"{a}nden unterhalb des Kontinuums. Dar\"{u}ber hinaus beobachten wir, dass unsere Einschlusspotentiale zu einer starken Modifikation der relativen Wellenfunktion von Exzitonen f\"{u}hren. Wir unterstreichen die Vielseitigkeit unseres Ansatzes, indem wir unser Einschlussschema erweitern, um quantenpunktartige Strukturen mit einer vollst\"{a}ndig einstellbaren Einschlussl\"{a}nge zu erzeugen. \\

\noindent Unsere Methode bietet einen experimentellen Weg zur Erreichung der Polariton-Blockade und zur Erzeugung skalierbarer Arrays von identischen Einzelphotonenquellen, was weitreichende Auswirkungen auf die Realisierung stark korrelierter photonischer Phasen und chip-basierter optischer Quanteninformationsprozessoren hat.
  \begin{center}
  \textsc{Acknowledgements}
\end{center}
\noindent I would not have been able to complete this thesis without the collective efforts of many dedicated individuals, to whom I am immensely grateful. \\

\noindent First and foremost, I would like to express my deepest gratitude to my advisors, Prof. David J. Norris and Prof. Atac Imamoglu for their constant support and encouragement. It has been a privilege to work with you. Prof. Norris, your meticulous feedback and the freedom you gave me to explore various research directions has significantly contributed to my academic growth. Prof. Imamoglu, your profound expertise has always been a source of inspiration. I am deeply grateful for our stimulating discussions and your patience in answering all my questions, no matter how numerous. \\

\noindent I am grateful to Prof. Mete Atat{\"u}re and Prof. Lukas Novotny for their time and effort in reviewing this thesis and serving on the thesis committee. \\

\noindent My heartfelt thanks go out to my fellow researchers and friends. Puneet Murthy, for your role in my development as a scientist and for teaching me many lessons along the way which I may not have learnt otherwise. Martin Kroner, your support in aspects of all kind, be it the most intricate of issues, conceptual physics or career planning, has helped me navigate many roadblocks. Tomasz Smole\'{n}ski, your endless passion and dedication to your work were infectious and inspiring, always propelling me to reach further in my own endeavors. Your exceptional resourcefulness has fundamentally shaped my ability to work effectively in the lab. Ivan Amelio, our illuminating discussions and your patience in explaining basic concepts have broadened my understanding. Yuya Shimazaki and Alexander Popert, the meticulous approach to device fabrication that you taught me was invaluable. Prof. Thomas Ihn, for your help in implementing the electrostatic simulations.\\

\noindent Nolan Lassaline, for constantly challenging me to aim higher. Your insights into the potential of thermal scanning probe lithography have opened my eyes to new possibilities. Thibault Chervy, for your help in the optics lab. Li Bing Tan, for sharing her expertise on open microcavity measurements. Having you as an office mate was delightful. The stimulating discussions and valuable advice from Andrea Bergschneider, Francesco Colangelo, Ido Schwartz, and Ajit Srivastava along the way greatly contributed to my learning experience. I extend my gratitude also to Livio Ciorciaro, Bertrand Evrard, Olivier Huber, Natasha Kiper, Patrick Knüppel, Felix Helmrich, Sarah Hiestand, Stefan Fält, Xiaobo Lu for the help and support they have offered me throughout these years. \\

\noindent My heartful thanks to Tobia Nova and Alperen Tugen for their friendship and our stimulating discussions about science, life, and philosophy, and to Gian-Marco Schnüriger for teaching me relentless optimism in the face of adversity. Furthermore, I want to thank my students Emre Yazici, Sarah Hiestand, Si Wang, Klemens Fl{\"o}ge, and Miray Ko\c{s}ucu for their fresh ideas and contributions that brought new perspectives to my projects. I also appreciate the enthusiasm and support of Daniel Petter and Juri Crimmann on the h-BN-lens project, which sparked intriguing research directions. I would like to thank Felipe Antolinez and Jan Winkler, for your guidance during the early days of my PhD and for encouraging me to venture beyond my comfort zone, and Ann-Katrin Michel, for being a constant source of moral support. Special thanks to Manuela Weber-Semler and Pascale Bachmann, for working tirelessly behind the scenes to handle all the organizational matters. I would like to express my gratitude to all present and past members of QPG and OMEL for creating an environment filled with inspiration, support, and unforgettable memories. \\

\noindent To my lifelong friends Varun Kumaran, Jonathan Theodore, Gaurav Parthasarathy, thank you for standing by my side since our early school days, supporting all my endeavors. Most importantly, I am deeply grateful to my family -- my parents and sister, who deserve more gratitude than words can convey. Your unwavering belief, constant encouragement, and infinite love have been my pillars of strength.

  \tableofcontents

\mainmatter

  %%%%%%%%%%%%%%%%%%%%%%%%%%%%%%%%%%%%%%%%%%%%%%%%%%%%%%%%%%%%%%%%%%%%%%%%
\chapter{Introduction}
%%%%%%%%%%%%%%%%%%%%%%%%%%%%%%%%%%%%%%%%%%%%%%%%%%%%%%%%%%%%%%%%%%%%%%%%

Experiments with light have served as one of the foundational pillars in deciphering the captivating consequences of the laws of quantum mechanics \cite{Freedman1972,Aspect1982,Pan1998}. In these explorations, the focus typically revolved around the generation, manipulation, and detection of a small set of photons -- the fundamental quanta of the electromagnetic field. More recently, technological advances in creating hybrid excitations, which are part light and part matter, have challenged the conventional perspective of photons as entities that always interact weakly \cite{Chang2014}. Under appropriate conditions, it has become possible to consider light as a macroscopic collection of bosonic particles, interacting in a manner akin to a fluid \cite{Carusotto2013}. This analogy with the field of condensed matter naturally compels us to ponder the extent of this parallelism. \emph{Is it possible to engineer synthetic materials composed of photons? Could the scale of interactions be magnified to the point of generating strongly correlated photonic many-body systems?} \cite{Chang2014,Carusotto2013,Noh2016,Angelakis2017,Carusotto2020}

Researching these avenues is expected to result in both fundamental and technological advancements. Strongly correlated phenomena are typically investigated in a solid-state setting with electrons, or more recently, with ultracold atomic gases. In comparison, utilizing photons for this purpose may allow for a more direct access to microscopic observables and faster repetition rates, thereby greatly facilitating the evaluation of quantum many-body correlations. In addition, their low-loss propagation serves as a crucial enabler for generating large-scale entanglement across distributed systems \cite{Carusotto2013,Carusotto2020,Noh2016,Angelakis2017,Kimble2008}. As such, these efforts will additionally contribute to the development of optical quantum information processors \cite{OBrien2009,Guzik2012,Rudolph2017} and quantum communication networks \cite{Kimble2008}, in which programmable arrays of single-photon sources serve as a key building block \cite{Lodahl2015,Uppu2021}.

Realizing such single-photon emission is a telltale signature for the presence of interactions among photons -- where the introduction of one photon in a system prevents simultaneous addition of another, a phenomenon referred to as photon blockade \cite{Kim1999,Michler2000}. Such a scenario naturally arises in media where the energy level spectrum is anharmonic and thus effectively approximates a two-level system. Consequently, it comes as no surprise that pioneering experimental demonstrations of such single-photon emission were based on atoms \cite{Kimble1977}, ions \cite{Diedrich1987} and molecules \cite{Basche1992}.
In an attempt to replicate such behavior in the solid-state, and in particular semiconductors, material modulation techniques \cite{Davies1997} are commonly employed to create nanoscale objects, such as quantum dots \cite{Harrison2016}. The quantum confinement of excitons, which are bound electron--hole pairs, within such structures leads to discrete energy levels, effectively allowing them to function as ``artificial atoms''. These approaches, however, suffer from poor control over the energy and position of excitonic trapping potentials, which has severely hindered the engineering of large-scale quantum photonic systems. \emph{Consequently, a long-standing challenge in the field of solid-state quantum photonics has been the realization of scalable arrays of quantum emitters which are identical, individually addressable, and strongly couple to light} \cite{Lodahl2015,Tuerschmann2019,Montblanch2023}.

It is the central aim of this thesis to present a viable path towards achieving this goal. Beginning with \textbf{chapter 2}, we set the theoretical foundations upon which our research findings are based. In \textbf{chapter 3}, we demonstrate our approach for realizing electrically controlled quantum confinement of neutral excitons by exploiting key attributes of transition metal dichalcogenide (TMD) monolayer semiconductors. By utilizing inhomogeneous in-plane electric fields which naturally arise in the neutral region of a lateral p-n junction, we realize one-dimensional (1D) excitonic trapping potentials with a lateral confinement below $10\,$nm. In \textbf{chapter 4}, we explore the consequences of such 1D confinement on the polarization properties of emission from the confined states, and in this manner, infer modifications in their center-of-mass (COM) wave function as the confinement potential is dynamically altered. Conversely, in \textbf{chapter 5}, we investigate the nature of the relative wave function of such quantum confined excitons by applying an out-of-plane magnetic field. In \textbf{chapter 6}, we elaborate on how our electric-field-induced confinement mechanism can be extended to realize quantum-dot-like structures with a fully tunable COM confinement length. In addition, we present preliminary optical signatures of quantum confined excitons in such structures. Finally, we discuss potential limitations of our confinement method and discuss strategies to overcome these. We conclude by highlighting the potential of our approach in realizing polariton blockade and discussing viable experimental strategies for the scale-up of our device architecture towards implementing a scalable, solid-state experimental platform for quantum many-body photonics exploration.

  %%%%%%%%%%%%%%%%%%%%%%%%%%%%%%%%%%%%%%%%%%%%%%%%%%%%%%%%%%%%%%%%%%%%%%%%
\chapter{Excitons in 2D semiconductors}
%%%%%%%%%%%%%%%%%%%%%%%%%%%%%%%%%%%%%%%%%%%%%%%%%%%%%%%%%%%%%%%%%%%%%%%%

%%%%%%%%%%%%%%%%%%%%%%%%%%%%%%%%%%%%%%%%%%%%%%%%%%%%%%%%%%%%%%%%%%%%%%%%
\section{Band structure}
%%%%%%%%%%%%%%%%%%%%%%%%%%%%%%%%%%%%%%%%%%%%%%%%%%%%%%%%%%%%%%%%%%%%%%%%

\begin{figure}[htb]
    \centering
	\includegraphics[width=14cm]{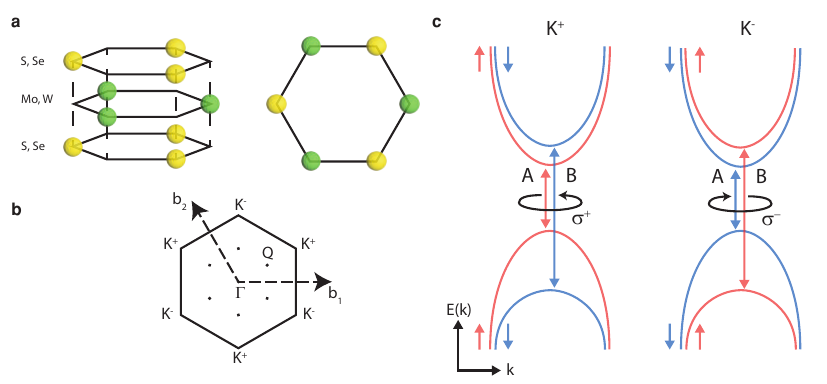}
	\caption{\textbf{Band structure and optical selection rules in monolayer TMDs.}
    (\bfA) Trigonal prismatic crystal structure of a monolayer TMD (left) and the resulting honeycomb lattice in top view (right). Transition metal atoms are shown in green, chalcogen atoms in yellow. (\bfB) Hexagonal Brillouin zone along with high-symmetry points and the reciprocal lattice vectors $\mathbf{b_1}$ and $\mathbf{b_2}$. (\bfC) Schematic illustration of the direct bandgap optical transition at $K^\pm$ points for Mo$X_2$-type compounds. The spin state of the electronic bands is indicated in red (spin up) and blue (spin down). The spin splitting in the VB is much greater than for the CB. This leads to valley-dependent optical selection rules, in which spin-allowed A and B interband transitions at $K^+$ and $K^-$, respectively, couple to $\sigma^+$ and $\sigma^-$ circularly polarized light.
    }  
	\label{fig:2_TMD}
\end{figure}

Atomically thin transition metal dichalcogenides (TMDs) form a key element of our experimental investigations, in particular the material MoSe$_2$. This compound belongs to a broader class of the semiconducting TMD family denoted by the general formula $M\!X_2$. Here, $M$ is transition metal, such as molybdenum (Mo) or tungsten (W), whereas $X$ signifies a chalcogen atom, typically sulfur (S) or selenium (Se). Each monolayer of this material is composed of a sheet of metal atoms sandwiched between two planes of chalcogen atoms arranged in a trigonal prismatic configuration \cite{Mak2016} (Fig.\,\ref{fig:2_TMD} \bfA). Evidently, the monolayer possesses an out-of-plane mirror symmetry, while lacking an in-plane inversion symmetry. Viewed from the top, the TMD monolayer exhibits hexagonal symmetry akin to graphene, albeit with an alternating arrangement of $M$ and $X$ atoms. Consequently, a hexagonal Brillouin Zone (BZ) is formed, with the high-symmetry point $\Gamma$ ($\mathbf{k} = 0$) located at the center , and $K^+$ and $K^-$ points situated at the corners, as illustrated in Fig.\,\ref{fig:2_TMD} \bfB. Here, the parameter $\mathbf{k}$ represents the wavevector in momentum space. These $K^\pm$ points are inherently distinct and cannot be linked to one another through the two elementary reciprocal lattice vectors $\mathbf{b_1}$ and $\mathbf{b_2}$ \cite{Liu2015,Wang2018a}.

In adjacent $M\!X_2$ layers, the spatial overlap between orbitals situated at the $\Gamma$ point of the valence band (VB) and those at the midpoint along $\Gamma$-$K$, also referred to as $Q$ point, of the conduction band (CB) is significant. In contrast, the CB and VB states at the $K^\pm$ points are strongly localized in the plane of metal atoms \cite{Wang2018a}. Therefore, as the bulk crystals are progressively thinned down, the magnitude of the indirect bandgap between $\Gamma$ and $Q$ points increases significantly \cite{Yun2012,Zhao2013}, while the $K^\pm$ point CB and VB energies remain largely unaltered. Upon reaching the monolayer limit, the semiconductor transitions from an indirect to direct bandgap, with the band extrema situated at the $K^\pm$ points \cite{Mak2010,Splendiani2010}.

In addition, owing to their heavy constituent elements, the spin-orbit interaction in TMDs is substantial \cite{Zhu2011}, greatly exceeding that found in graphene. In combination with the broken inversion symmetry in monolayer TMDs, this effect lifts the spin degeneracy of both CB and VB in each $K^\pm$ valley in opposite directions \cite{Cheiwchanchamnangij2012,Xiao2012}. Consequently, an effective coupling between spin and valley degrees of freedom emerges, a phenomenon often referred to as spin-valley locking \cite{Wang2018a}. Conversely, in bilayer and bulk TMDs inversion symmetry is re-established, and thus the spin degeneracy in each valley recovered \cite{Mak2016,Cheiwchanchamnangij2012}. These features further result in valley-dependent optical selection rules \cite{Yao2008,Xiao2012,Cao2012,Mak2012,Zeng2012,Sallen2012}: Spin-allowed A and B interband transitions at $K^+$ and $K^-$ respectively couple to $\sigma^+$ and $\sigma^-$ circularly polarized light (Fig.\,\ref{fig:2_TMD} \bfC).

%%%%%%%%%%%%%%%%%%%%%%%%%%%%%%%%%%%%%%%%%%%%%%%%%%%%%%%%%%%%%%%%%%%%%%%%
\section{Neutral excitons}
\label{chap:theory:sec:excitons}
%%%%%%%%%%%%%%%%%%%%%%%%%%%%%%%%%%%%%%%%%%%%%%%%%%%%%%%%%%%%%%%%%%%%%%%%

\begin{figure}[htb]
    \centering
	\includegraphics[width=13.5cm]{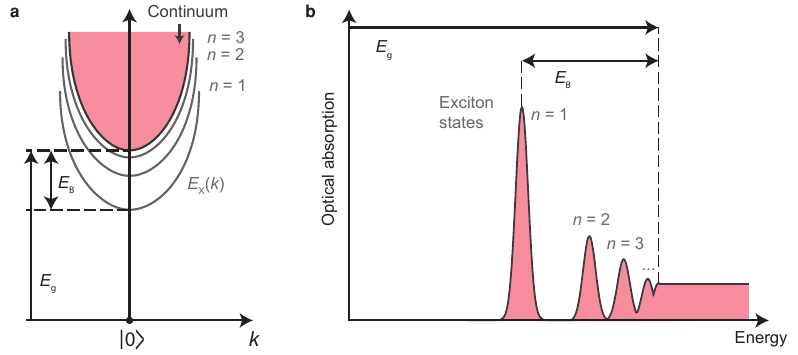}
	\caption{\textbf{Exciton states in semiconductors.}
    (\bfA) Dispersion relation of hydrogenic exciton states with principal quantum number $n=1,2,3,\dots$ as a function of their COM momentum $k$, depicted in gray. The energy of the lowest exciton state ($n=1$) is separated by the binding energy $E_\mathrm{B}$ from the continuum (shaded area), in which excitons are ionized into free electrons and free holes. The bandgap $E_\mathrm{g}$ denotes the energy separation from the vacuum state $\ket{0}$ to the continuum. (\bfB) Schematic depiction of the optical absorption spectrum in a semiconductor, which features a series of sharp transitions associated with excitonic states below the absorption into the continuum. Adapted from Ref.\,\cite{Wang2018a}.
    }  
	\label{fig:2_Excitons}
\end{figure}

Fundamentally, in such an interband transition the absorption of an incident photon in a semiconductor will excite an electron from the VB into the CB. The resulting empty electronic state in the VB can be equivalently described as a positively charged hole. Owing to their opposite charge, these photo-generated electrons and holes experience Coulombic attraction, which correlates their motion and leads to formation of bound electron--hole pairs, termed \emph{excitons} \cite{Yu2010,Haug2009,Klingshirn2012,Combescot2015}. Their energy can be determined by solving a two-particle Schr{\"o}dinger equation for the exciton envelope wave function $\Psi_\mathrm{X}(\mathbf{r}_n,\mathbf{r}_p)$:
\begin{equation}
    \left(
    -\frac{\hbar^2}{2m_\mathrm{n}^*}\nabla_{\mathbf{r}_n}^2
    -\frac{\hbar^2}{2m_\mathrm{p}^*}\nabla_{\mathbf{r}_p}^2
    -\frac{e^2}{4\pi\epsilon_0\epsilon_r\left|\mathbf{r}_n-\mathbf{r}_p \right|}\right) \Psi_\mathrm{X} =
    E_\mathrm{X}\Psi_\mathrm{X}
    \label{eqn:X-2-particle}
\end{equation}
where $\hbar$ is the reduced Planck constant, $e$ is the elementary charge, $\epsilon_0$ is the vacuum permittivity, $\epsilon_r$ is the relative permittivity of the excitonic host medium, and $\mathbf{r}_{n,p}$ are the electron and hole position coordinates, respectively. In Eqn.\,\ref{eqn:X-2-particle} we have implicitly assumed the effective mass approximation \cite{Yu2010}, where the impact of the crystal potential on both electron and hole motion is captured by an effective mass $m_\mathrm{n,p}^*$ for each. The first two terms describe the single-particle kinetic energies of the electron and hole, respectively. The third term denotes the Coulomb interaction potential\footnote{Exchange contributions to the Coulomb interaction have been neglected at this point for simplicity and will be addressed in further detail in section \ref{chap:theory:sec:fine_structure}.}$^{\text{,}}$\footnote{The lack of spherical symmetry in a 2D system introduces a deviation from the $1/r$ scaling in the electron--hole Coulomb interaction, which is discussed at the end of this section.} between them, which, notably, depends \emph{only} on the separation between the two particles. This allows the exciton motion to be separated into two parts: a \emph{center-of-mass} (COM) motion and a \emph{relative} motion of the two particles around the COM \cite{Yu2010,Griffiths2018}. This is accomplished by introducing a COM coordinate $\mathbf{R}$ and a relative coordinate $\mathbf{r}$ as
\begin{equation}
    \mathbf{R} = \frac{m_\mathrm{n}^*\mathbf{r}_n+m_\mathrm{p}^*\mathbf{r}_p}{m_\mathrm{n}^*+m_\mathrm{p}^*}\mathrm{\enspace and \enspace}\mathbf{r}=\mathbf{r}_n-\mathbf{r}_p
\end{equation}
The two respective decoupled equations for COM and relative motion are
\begin{align}
    \left(-\frac{\hbar^2}{2M}\nabla_{\mathbf{R}}^2\right)
    \psi(\mathbf{R})
    &=E_R\psi(\mathbf{R}) \label{eqn:free_particle}\\
    \left(-\frac{\hbar^2}{2\mu}\nabla_{\mathbf{r}}^2 - \frac{e^2}{4\pi\epsilon_0\epsilon_r |\mathbf{r}|}\right)
    \varphi(\mathbf{\mathbf{r}}) &= E_r\varphi(\mathbf{r}) \label{eqn:wannier}
\end{align}
where $M=m_\mathrm{n}^*+m_\mathrm{p}^*$ is the total mass of the exciton, and $\mu=\frac{m_\mathrm{n}^* m_\mathrm{p}^*}{m_\mathrm{n}^* + m_\mathrm{p}^*}$ is the reduced exciton mass. Eqn.\,\ref{eqn:wannier} is referred to as the \emph{Wannier equation}. Evidently, based on Eqn.\,\ref{eqn:free_particle}, the exciton COM behaves as a free particle with total momentum $\mathbf{k}=\mathbf{k}_\mathrm{n}+\mathbf{k}_\mathrm{p}$, and $\mathbf{k}_{\mathrm{n,p}}$ being the electron and hole momentum, respectively. Its total kinetic energy can be denoted as
\begin{equation}
    E_R(k) = \frac{\hbar^2k^2}{2M}
\end{equation}
In bulk semiconductors the eigenfunctions of the electron--hole relative motion $\varphi_{n,l,m}$ are labelled by the principal, orbital and magnetic quantum numbers $n$, $l$ and $m$, respectively, in analogy to the hydrogen atom. As such, the corresponding eigenenergies will only depend on the quantum number $n=1,2,3,\dots$, as follows
\begin{equation}
    E_r^{(n)} = E_\mathrm{g} - E_\mathrm{B}^{(n)}
\end{equation}
Here we have introduced the semiconductor bandgap energy $E_\mathrm{g}$ representing the minimum energy required for exciton ionization into free electrons and free holes. In a three-dimensional (3D) system the binding energy $E_\mathrm{B}^{(n)}$ of the $n^\mathrm{th}$ excitonic Rydberg state will be given by
\begin{equation}
\begin{aligned}
    E_\mathrm{B,3D}^{(n)} &=
    \mathrm{Ry}\frac{\mu/m_\mathrm{e}}{\epsilon_r^2}\cdot\frac{1}{n^2} \\
    &=\frac{e^2}{4\pi\epsilon_0\epsilon_r\left(2 a_\mathrm{B,3D}\right)}\cdot\frac{1}{n^2} \\
    &=\frac{\hbar^2}{2\mu}\frac{1}{a_\mathrm{B,3D}^2}\cdot\frac{1}{n^2}
\end{aligned}
\end{equation}
with $\mathrm{Ry}=13.6\,$eV being the Rydberg energy, $m_\mathrm{e}$ being the free-electron mass and $a_\mathrm{B,3D}$ being the Bohr radius defined below. In the more relevant case for us where the excitonic motion is restricted to a two-dimensional (2D) plane the exciton binding energy is given by
\begin{equation}
    E_\mathrm{B,2D}^{(n)} =\mathrm{Ry}\frac{\mu/m_\mathrm{e}}{\epsilon_r^2}\cdot\frac{1}{\left(n-\frac{1}{2}\right)^2}
    \label{eqn:EB_2D}
\end{equation}
The corresponding excitonic Bohr radius for the $1s$ exciton in the 3D and 2D case, respectively, can be denoted as 
\begin{align}
    a_\mathrm{B,3D} &= \frac{4\pi\epsilon_0\epsilon_r\hbar^2}{e^2\mu} =a_\mathrm{0}\cdot\frac{\epsilon_r}{\mu/m_\mathrm{e}} \\
    a_\mathrm{B,2D} &= \frac{2\pi\epsilon_0\epsilon_r\hbar^2}{e^2\mu} =a_\mathrm{0}\cdot\frac{\epsilon_r}{2\mu/m_\mathrm{e}} = \frac{a_\mathrm{B,3D}}{2}
    \label{eqn:a_B_2D}
\end{align}
with $a_\mathrm{0}$ defined as the atomic Bohr radius. Having determined the excitonic binding energies, the complete exciton dispersion relation as a function of the COM momentum $k$ can simply be derived as the sum of $E_R(k)$ and $E_r^{(n)}$, which is given by
\begin{equation}
    E_\mathrm{X}(k) = E_R(k) + E_r^{(n)} =  E_\mathrm{g} + \frac{\hbar^2k^2}{2M} - E_\mathrm{B}^{(n)}
\end{equation}
and illustrated in (Fig.\,\ref{fig:2_Excitons} \bfA). For momenta within the light cone, a photon absorption process will lead to the creation of an exciton with energy $E_\mathrm{X}$ from the vacuum state $\ket{0}$. Subsequently, this exciton can be annihilated through re-emission of a photon. The existence of such excitonic states will manifest in the optical absorption spectra of semiconductors as sharp optical transitions below the absorption into the continuum, where excitons are ionized into free electrons and free holes (Fig.\,\ref{fig:2_Excitons} \bfB).

Based on these relations it is straightforward to see that Coulomb effects in 2D TMDs are particularly pronounced \cite{Wang2018a}. Changing the dimensionality from a 3D to a 2D semiconductor leads to a four-fold increase in the binding energy for the exciton ground state, i.e.\,$E_\mathrm{B,2D}^{(1s)}=4\cdot E_\mathrm{B,3D}^{(1s)}$. This phenomenon can be rationalized as follows: As the thickness of a slab of semiconductor material is progressively reduced, the wave function tries to maintain spherical symmetry, since the admixture with higher-lying $p$-type wave functions is energetically unfavorable \cite{Haug2009}. Thus, the electron--hole wave function overlap increases, giving rise to a smaller Bohr radius. Additionally, in comparison to III-V semiconductors, electrons and holes in the $K^\pm$ valleys of TMDs have larger effective masses of around $0.5\,m_\mathrm{e}$ \cite{Larentis2018,Zhang2014,Goryca2019}, leading to a reduced mass which is greater than that of gallium arsenide (GaAs, $\approx 0.06\,m_\mathrm{e}$ \cite{Nakwaski1995}). Furthermore, in TMD devices the monolayer semiconductors are typically encapsulated by dielectrics with a comparatively smaller permittivity ($\epsilon_r\approx5$ \cite{Wang2018a}). Therefore, as compared to III-V heterostructures, which feature a substantially greater relative permittivity ($\epsilon_r\approx13$ \cite{Aspnes1986}), screening of Coulomb interactions in TMDs is greatly suppressed. Using these material parameters, typical binding energies for excitons in TMDs can be estimated to be around $500\,$meV, as opposed to approximately $5\,$meV for bulk GaAs. Equivalently, the TMD exciton Bohr radius is substantially reduced to around $0.5\,$nm, as compared to roughly $10\,$nm in bulk GaAs.

While the relations developed above provide a good order-of-magnitude estimate of the relevant energy and length scales of excitons in 2D semiconductors, their agreement with experimental data is limited, especially in the case of TMD monolayers \cite{Chernikov2014}. The main issue is the inability of a static relative permittivity $\epsilon_r$ to accurately capture the screening of electron--hole Coulomb interactions in a 2D system owing to its lack of spherical symmetry. A more appropriate description can be achieved by introducing non-local screening, i.e.\,a permittivity $\epsilon_r$ which depends on the in-plane wavevector \cite{Cudazzo2011}. This will introduce a deviation from a pure $1/r_\parallel$ scaling in the Coulomb interaction potential $V_\mathrm{2D}(r_\parallel)$, where $r_\parallel$ is the coordinate of in-plane relative motion. This is accomplished using the Rytova-Keldysh potential given by
\begin{equation}
    V_\mathrm{2D}(r_\parallel) = -\frac{\pi}{2}\frac{e^2}{4\pi\epsilon_0\epsilon_r}[H_0(r/r_0) - Y_0(r/r_0)]
\end{equation}
with $r_0$ being the ``screening length'' (approximately $3-6\,$nm for TMDs), and $H_0$ and $Y_0$ being Struve and Bessel functions, respectively. At large electron--hole separations, $V_\mathrm{2D}(r_\parallel)$ still adheres to the $1/r_\parallel$ scaling of the bare Coulomb interaction, while at shorter separations it exhibits a weaker $\log(r_\parallel)$ dependence. The crossover length scale between the two regimes is set by the screening length $r_0$. As a result, excitons in monolayer TMDs will exhibit deviations from pure hydrogenic character. These will be most pronounced for lower principal quantum number $n$, for which the excitonic wave functions have a reduced spatial extent in relative coordinates.

%%%%%%%%%%%%%%%%%%%%%%%%%%%%%%%%%%%%%%%%%%%%%%%%%%%%%%%%%%%%%%%%%%%%%%%%
\section{Excitons in external fields}
%%%%%%%%%%%%%%%%%%%%%%%%%%%%%%%%%%%%%%%%%%%%%%%%%%%%%%%%%%%%%%%%%%%%%%%%

The application of external fields constitutes an invaluable tool in semiconductor research. By systematically reducing the symmetry of the overall system, such experiments offer rich insights into the nature of electronic states and their transitions. In this section, we will focus on developing qualitative principles concerning excitons interacting with static electric and magnetic fields. We will thereby limit our discussion to the case of 3D semiconductors, owing to the simplicity of expressions resulting from the spherical symmetry of the problem. Similar results can also be derived for 2D semiconductors. We provide the corresponding references in the respective sections.

%%%%%%%%%%%%%%%%%%%%%%%%%%%%%%%%%%%%%%%%%%%%%%%%%%%%%%%%%%%%%%%%%%%%%%%%
\subsection{Interaction with electric fields}
%%%%%%%%%%%%%%%%%%%%%%%%%%%%%%%%%%%%%%%%%%%%%%%%%%%%%%%%%%%%%%%%%%%%%%%%

\begin{figure}[htb]
    \centering
	\includegraphics[width=7cm]{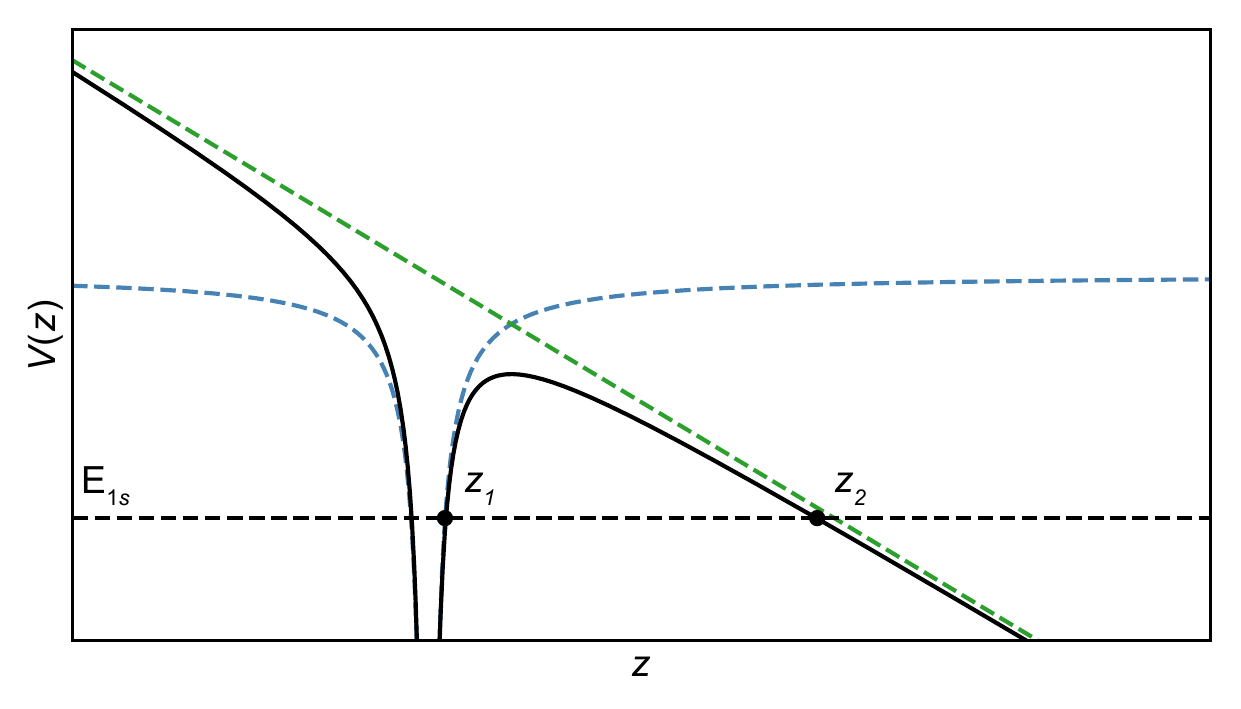}
	\caption{\textbf{Potential for electron--hole relative motion under combined influence of Coulomb interaction and static electric field.}
    The total potential is indicated in black, which is a sum of the Coulomb interaction (blue dashed line) and the electrostatic potential (green dashed line). The exciton state at energy $E_{1s}$ (black dashed line) can dissociate by tunnelling of a charge carrier through the potential barrier from position $z_1$ to $z_2$.
    }  
	\label{fig:2_electric-field}
\end{figure}

We consider an exciton subjected to a uniform electric field $F_\mathrm{z}$ oriented along the $z$-axis. The effect of this field can be incorporated in the excitonic Hamiltonian, given in Eqn.\,\ref{eqn:X-2-particle}, by adding the electrostatic potential energy $e F_\mathrm{z} z_\mathrm{n} - e F_\mathrm{z} z_\mathrm{p}$ of the electron and hole, respectively, where $z_\mathrm{n}$ and $z_\mathrm{p}$ denote the particle positions along the $z$-axis. Since this electric field is constant in magnitude a separation into COM and relative coordinates is still possible, where the modified Wannier equation can be written as
\begin{equation}
    \left(-\frac{\hbar^2}{2\mu}\nabla_{\mathbf{r}}^2 - \frac{e^2}{4\pi\epsilon_0\epsilon_r |\mathbf{r}|} + e F_\mathrm{z} z\right)
    \varphi(\mathbf{r}) = E_r\varphi(\mathbf{r})
\end{equation}
with $z = z_\mathrm{n}-z_\mathrm{p}$ being the $z$-component of the electron--hole relative motion. The total excitonic potential under the combined influence of Coulomb interaction and a static electric field is illustrated in Fig.\,\ref{fig:2_electric-field}.

Owing to the additional linear term, the electron and hole will be pulled in opposite directions. As depicted in Fig.\,\ref{fig:2_electric-field}, if the electric field is large enough, one of the carriers can tunnel from the Coulomb well (position $z_1$) through the tunnel barrier into the continuum of states (position $z_2$). This can eventually lead to exciton dissociation if the energy gain provided by the electrostatic potential, assessed on a length scale given by the exciton Bohr radius, is sufficient to overcome the exciton binding energy. With this condition we can obtain a rough estimate for the exciton dissociation field $F_\mathrm{d}$:
\begin{equation}
    eF_\mathrm{d}a_\mathrm{B}\stackrel{!}{\approx}E_\mathrm{B} \rightarrow F_\mathrm{d} \approx \frac{E_\mathrm{B}}{ea_\mathrm{B}}
    \label{eqn:dissociation_field}
\end{equation}
In view of this relation, it is apparent that excitons in TMD monolayers will be much more resilient to electric fields than in III-V semiconductors, owing to the large binding energies and small excitonic Bohr radii in the former. Conversely, Rydberg excitons feature a much greater spatial extent, and correspondingly, lower binding energies, causing a substantial reduction in their dissociation field. We emphasize that in practice, the actual field necessary for dissociation will be much lower than $F_\mathrm{d}$, since the probability that charge carriers tunnel through the potential barrier increases exponentially as the barrier width decreases.

For conditions in which the applied electric field $F_\mathrm{z}$ is sufficiently small, i.e.\,$|F_\mathrm{z}|\ll F_\mathrm{d}$, the influence of the electrostatic potential can be treated perturbatively. Notably, the first-order energy correction to the $1s$ exciton state $\bra{\varphi_{1,0,0}}e F_\mathrm{z} z\ket{\varphi_{1,0,0}}$ vanishes. This is because the wave function $\varphi_{1,0,0} = \frac{1}{\sqrt{\pi a_\mathrm{B}^3}}e^{-r/a_\mathrm{B}}$ is spherically symmetric and thus has even parity, while the position operator $z$ has odd parity. We can generalize the evaluation of such matrix elements using parity selection rules \cite{Griffiths2018}:
\begin{equation}
    \bra{\varphi_{n,l,m}}e F_\mathrm{z} z\ket{\varphi_{n',l',m'}} \neq 0\;\Leftrightarrow\;m'=m,\;l'=l\pm1
\end{equation}
This expression allows us to efficiently evaluate the second-order corrections to the energy of the $1s$ exciton state:
\begin{equation}
    \Delta E_{1s} = \sum_{n>1,l,m}\frac{|\bra{\varphi_{1,0,0}}e F_\mathrm{z} z\ket{\varphi_{n,l,m}}|^2}{E_1-E_n} = 
    e^2 F_\mathrm{z}^2\sum_{n>1}\frac{|\bra{\varphi_{1,0,0}} z\ket{\varphi_{n,1,0}}|^2}{E_1-E_n}
    \label{eqn:deltaE_1s_a}
\end{equation}
The resulting shift in energy is proportional to the square of the electric field, and is often referred to as the \emph{quadratic Stark effect}. It has been observed in various material systems \cite{Miller1985,Roch2018}. Intuitively, it can be understood in terms of the electric field inducing a finite dipole moment, which is proportional to $F_\mathrm{z}$ and is acquired through admixture with higher-lying $p$-type wave functions. The interaction of this dipole moment in turn with the electric field leads to the aforementioned quadratic field dependence. The energy shift $\Delta E_{1s}$ is typically expressed in terms of a polarizability $\alpha_{1s}$ as
\begin{equation}
    \Delta E_{1s} = -\frac{1}{2}\alpha_{1s} F_\mathrm{z}^2.
    \label{eqn:deltaE_1s_b}
\end{equation}
Equivalently, the magnitude of the induced dipole moment is given by $\alpha_{1s}F_\mathrm{z}$. Comparing Eqns. \ref{eqn:deltaE_1s_a} and \ref{eqn:deltaE_1s_b}, we find
\begin{equation}
    \alpha_{1s} = 2e^2 \sum_{n>1}\frac{|\bra{\varphi_{1,0,0}} z\ket{\varphi_{n,1,0}}|^2}{E_n-E_1}
\end{equation}
which is a positive quantity, since $E_n>E_1$ for $n>1$. The precise value of the sum can be found using algebraic manipulations \cite{Griffiths2018} and is given by
\begin{equation}
    \alpha_{1s} = \frac{9}{2}\cdot4\pi\epsilon_0\epsilon_r \cdot a_\mathrm{B}^3
    \label{eqn:polarizability}
\end{equation}
which indicates a pronounced sensitivity of the polarizability to the exciton Bohr radius, given its dependence on $a_\mathrm{B}^3$. Similar calculations can also be performed for 2D excitons and reveal the same $a_\mathrm{B}^3$ dependency, albeit with a different prefactor \cite{Yang1991,Lederman1976,Tanaka1987,Fernandez1992}.

While the results above underscore that a quadratic field dependence is relevant for the $1s$ exciton state, we emphasise that a \emph{linear Stark effect} may arise for Rydberg excitons. This is due to the degeneracy of states for $n>1$, which necessitates that matrix elements amongst all degenerate states at a given eigenenergy need to be calculated to determine the correction term. As an example, already for the $n=2$ states non-zero matrix elements can be found:
\begin{equation}
    \bra{\varphi_{2,0,0}}e F_\mathrm{z} z\ket{\varphi_{2,1,0}} = 
    3eF_\mathrm{z}a_\mathrm{B}
\end{equation}
This results in lifting of the 4-fold degeneracy of the $n=2$ state into three levels: a 2-fold degenerate state at the original energy $E_2$ with eigenfunctions $\varphi_{2,1,\pm1}$, and two distinct states at energy $E_2 \pm 3eF_\mathrm{z}a_\mathrm{B}$ with eigenfunctions $\frac{1}{\sqrt{2}}\left(\varphi_{2,0,0}\pm\varphi_{2,1,0}  \right)$ \cite{Griffiths2018}. Hence, while none of the individual states $\varphi_{n,l,m}$ possess the requisite permanent dipole moment, which could give rise to a linear field dependence, this need not be the case for their linear combinations at a particular energy level. 

%%%%%%%%%%%%%%%%%%%%%%%%%%%%%%%%%%%%%%%%%%%%%%%%%%%%%%%%%%%%%%%%%%%%%%%%
\subsection{Interaction with magnetic fields}
\label{chap:theory:sec:B-field}
%%%%%%%%%%%%%%%%%%%%%%%%%%%%%%%%%%%%%%%%%%%%%%%%%%%%%%%%%%%%%%%%%%%%%%%%

By applying an external magnetic field, the energies and angular momenta of charged particles get quantized, which further can result in modifications of exciton energies and their wave functions. To gain a qualitative understanding of such effects, we assume a uniform magnetic field $\mathbf{B} = \left( 0,0,B_\mathrm{z}\right)^\intercal $ oriented along the $z$-axis. The effect of this field can be included in the Wannier equation \ref{eqn:wannier} by performing the minimal coupling substitution, in which we replace the canonical momentum $\mathbf{p}$ by $\mathbf{p}+e\mathbf{A}$, with $-e$ being the electron charge and $\mathbf{A}$ being the magnetic vector potential related to the magnetic field according to $\nabla \times \mathbf{A} = \mathbf{B}$. This gives rise to the Hamiltonian $\hat{H}$
\begin{equation}
    \hat{H} = \frac{(\mathbf{p}+e\mathbf{A}^2)}{2\mu}-\frac{e^2}{4\pi\epsilon_0\epsilon_r |\mathbf{r}|} = 
    \underbrace{\frac{\mathbf{p}^2}{2\mu} - 
    \frac{e^2}{4\pi\epsilon_0\epsilon_r |\mathbf{r}|}}_{\hat{H}_0} +
    \underbrace{\frac{e\mathbf{p}\cdot\mathbf{A}}{\mu} +
    \frac{e^2\mathbf{A}^2}{2\mu}}_{\hat{H}_\mathrm{B}}
    \label{eqn:Hamiltonian_B-field}
\end{equation}
where we have used $[\mathbf{p},\mathbf{A}]=0$ \cite{Ciftja2020}. $\hat{H}_0$ denotes the original Hamiltonian from Eqn.\,\ref{eqn:wannier}, $\hat{H}_\mathrm{B}$ contains the additional terms due to inclusion of a magnetic field. We adopt the symmetric gauge $\mathbf{A}=\frac{1}{2}\mathbf{B} \times \mathbf{r}$, where $\mathbf{r}$ is the relative coordinate for electron--hole motion, and obtain for $\hat{H}_\mathrm{B}$
\begin{equation}
    \hat{H}_\mathrm{B} = 
    \frac{e}{\mu}\frac{1}{2}\mathbf{p}\cdot(\mathbf{B}\times\mathbf{r})+
    \frac{e^2}{2\mu}\frac{1}{4}(\mathbf{B}\times\mathbf{r})^2 
    = \frac{e}{2\mu}\mathbf{B}\cdot\mathbf{L} +
    \frac{e^2}{8\mu}(\mathbf{B}\times\mathbf{r})^2
\end{equation}
where in the last step we have used the definition for the angular momentum operator $\mathbf{L}=\mathbf{r}\times\mathbf{p}$. In addition, we also need to consider the coupling of the magnetic field with the spin $\frac{e}{\mu}\mathbf{B}\cdot\mathbf{S}$, leading us to the final form of $\hat{H}_\mathrm{B}$
\begin{equation}
    \hat{H}_\mathrm{B} =
    \frac{e}{2\mu}B_\mathrm{z}\cdot\left(L_\mathrm{z}+2S_\mathrm{z}\right) +
    \frac{e^2}{8\mu} B_\mathrm{z}^2 \left(x^2+y^2 \right)
\end{equation}
where we also used the fact that $\mathbf{B}$ is oriented along $z$. We emphasize that the derivation above is only valid in the case where one of the charge carriers comprising the exciton is much heavier than the other. For instance, this is the case in GaAs where $m_\mathrm{n}^* \ll m_\mathrm{p}^*$, such that $\mu \approx m_\mathrm{n}^*$. In material systems where this is not the case, e.g.\,TMD monolayers, the effect of the magnetic field needs to be considered on each charge carrier separately. A rigorous procedure for determining the Hamiltonian for electron--hole relative motion in this case is discussed in Refs.\,\cite{Gorkov1968,Lozovik2002,Cong2018}. In such a scenario, additional terms may arise in the Hamiltonian, potentially leading to new phenomena \cite{Chestnov2021}. However, these effects are typically smaller in magnitude compared to the more dominant effects, which our approximate model also allows to adequately capture.

Furthermore, to verify whether magnetic field effects can be treated perturbatively, it is convenient to introduce the relevant energy scales and length scales of the problem: the exciton binding energy $E_\mathrm{B,3D}$ and the corresponding Bohr radius $a_\mathrm{B,3D}$ to characterize the Coulomb interaction; the magnetic length $l_B=\sqrt{\frac{\hbar}{eB}}$ and the cyclotron energy $\hbar\omega_c=\frac{\hbar eB}{\mu}$ to describe the effect of the magnetic field. By defining a dimensionless parameter $\gamma$ we can quantify the relative importance of the two contributions in the Hamiltonian:
\begin{equation}
    \gamma = \frac{\hbar\omega_c}{2} / E_\mathrm{B,3D} = \left(\frac{a_\mathrm{B,3D}}{l_B} \right)^2
\end{equation}
Our experimental capabilities allow accessing magnetic fields of up to $16\,$T. Assuming binding energies on the order of a few $100\,$meV for TMD monolayers, we obtain $\gamma \ll 1$, which puts us in the weak $B$-field limit and justifies the treatment of $\hat{H}_\mathrm{B}$ as a perturbation. Using perturbation theory we can determine the first-order energy correction
\begin{equation}
\begin{aligned}
    \Delta E_{n,l,m} &= \bra{\varphi_{n,l,m}} \hat{H}_\mathrm{B} \ket{\varphi_{n,l,m}} \\
    &= \left\langle \varphi_{n,l,m}\left| \frac{e^2}{8\mu} B_\mathrm{z}^2 \left(x^2+y^2 \right)\right|\varphi_{n,l,m} \right\rangle + 
    \left\langle \varphi_{n,l,m}\left| \frac{e}{2\mu}B_\mathrm{z}\left(L_\mathrm{z}+2S_\mathrm{z}\right) \right|\varphi_{n,l,m} \right\rangle \\
    &= \frac{e^2}{8\mu}B_\mathrm{z}^2\left\langle x^2+y^2 \right\rangle + \frac{e}{2\mu}B_\mathrm{z}\left( \left\langle L_\mathrm{z} \right\rangle + 2\left\langle S_\mathrm{z} \right\rangle \right)
\end{aligned}
\label{eqn:B-field_correction}
\end{equation}
The first term in Eqn.\,\ref{eqn:B-field_correction} is the \emph{diamagnetic-shift} term. It features a quadratic dependence with the magnetic field and is directly proportional to the spatial extent of the respective wave function in the plane perpendicular to the applied $B$-field \cite{Stier2016,Cong2018}. The proportionality constant $\sigma$ is referred to as the diamagnetic-shift coefficient.
\begin{equation}
    \Delta E_\mathrm{dia}=\frac{e^2}{8\mu} \left\langle r_\perp^2 \right\rangle B_\mathrm{z}^2 = \sigma B_\mathrm{z}^2
\end{equation}
Provided the reduced mass $\mu$ is a known quantity, by experimentally determining $\sigma$ a value for the root mean squared (r.m.s.) radius $r_\mathrm{rms}$ of the exciton can be estimated
\begin{equation}
    r_\mathrm{rms} = \sqrt{\left\langle r_\perp^2 \right\rangle} = \sqrt{8\mu\sigma}/e
    \label{eqn:diamag}
\end{equation}
Notably, Eqn.\,\ref{eqn:diamag} denotes a general definition, which is independent of $V(\mathbf{r})$ \cite{Stier2016}. In the particular case of an unscreened Coulomb interaction in 2D (Eqn.\,\ref{eqn:Hamiltonian_B-field}), the following correspondence with the Bohr radius can be established \cite{Stier2016}
\begin{equation}
    r_\mathrm{rms} = \sqrt{\frac{3}{2}}a_\mathrm{B,2D}
\end{equation}

The second term in Eqn.\,\ref{eqn:B-field_correction} is the \emph{Zeeman shift} term. It has a linear dependence with the magnetic field and lifts the degeneracy of different spin and angular momentum states of a particular energy level. In real semiconductor materials the actual magnitude of the resulting energy splittings might vary from those predicted by this term due to existence of additional contributions. These effects are typically bundled together and captured by introducing phenomenological ``$g$-factors'' in the Zeeman energy $E_\mathrm{Z} = -\boldsymbol{\mu}\cdot\mathbf{B} = -g \cdot \mu_\mathrm{B} B_\mathrm{z}$ where $\boldsymbol{\mu}$ is the total magnetic moment and $\mu_\mathrm{B} = \frac{e\hbar}{2m_\mathrm{e}}$ is the Bohr magneton.

\sloppy In particular, the Zeeman shift for exciton states in TMD monolayers is quantified as ${-(\boldsymbol{\mu}^\mathrm{c}-\boldsymbol{\mu}^\mathrm{v})\cdot \mathbf{B}}$, where $\boldsymbol{\mu}^\mathrm{c,v}$ is the total magnetic moment of the CB and VB state, respectively. Since the $K^\pm$ valleys are linked by time-reversal symmetry, the CB and VB states are degenerate in energy, but possess a total magnetic moment that is equal in magnitude and opposite in sign, i.e.\,$\boldsymbol{\mu}_{K^+}^{\mathrm{c,v}}=-\boldsymbol{\mu}_{K^-}^{\mathrm{c,v}}$. Therefore, observing a discernible energy splitting between $K^+$ and $K^-$ excitons evidently requires symmetry breaking between electron and hole. As such, the spin magnetic moment will not contribute, since optically allowed excitonic transitions only couple CB and VB states with the same spin. The main contribution to this splitting arises from the atomic orbital magnetic moment. Being composed of $d_{z^2}$ orbitals, the CB states have an azimuthal orbital angular momentum $l_z=0$. In contrast, the VB states consist of hybridized $d_{x^2-y^2}\pm id_{xy}$ orbitals with $l_z=\pm2 \hbar$. Thus, a total Zeeman splitting of $\Delta E_\mathrm{Z}^{K^\pm} = 4\mu_\mathrm{B}B_\mathrm{z}$ results. In addition, a weaker contribution from the valley orbital angular momentum is also expected, which has its origin in the unequal phase winding of CB and VB Bloch functions at $K^\pm$ points \cite{Srivastava2015,Stier2016,Schaibley2016}.

%%%%%%%%%%%%%%%%%%%%%%%%%%%%%%%%%%%%%%%%%%%%%%%%%%%%%%%%%%%%%%%%%%%%%%%%
\section{Light--matter coupling via excitons}
\label{chap:theory:sec:LM_coupling}
%%%%%%%%%%%%%%%%%%%%%%%%%%%%%%%%%%%%%%%%%%%%%%%%%%%%%%%%%%%%%%%%%%%%%%%%

In section \ref{chap:theory:sec:excitons} we established that excitons are the elementary optical excitations in a semiconductor. This insight allowed us to comprehend the discrete nature of the peaks observed in the optical absorption spectrum of a semiconductor (Fig.\,\ref{fig:2_Excitons}). However, to fully understand the specific line shape of these transitions, we must delve deeper into how excitons interact with light.

Since an exciton constitutes a bound electron--hole pair in the relative coordinate frame, while being delocalized in its COM, the total wave function of an exciton with COM momentum $\mathbf{k}$ can be expressed as
\begin{equation}
    \Psi_\mathrm{X,\nu,\mathbf{k}}(\mathbf{r}_n,\mathbf{r}_p)=\frac{1}{\sqrt{V_\mathrm{exc}}}e^{i\mathbf{k}\cdot \mathbf{R}}\varphi_\nu(\mathbf{r})
\end{equation}
Here, we introduced the index $\nu=(n,l,m)$, which denotes the quantum numbers associated with a particular Rydberg state, and $V_\mathrm{exc}$, which represents the crystal volume over which the exciton is delocalized. Specifically, $V_\mathrm{exc}$ designates the coherence volume of the exciton, indicating the spatial region where its momentum $\mathbf{k}$ can be precisely defined. With this relation as a starting point, a quantitative description of the light--matter coupling strength can be obtained by adopting the second-quantization formalism. In such a setting, the aforementioned exciton wave function can be conveniently expressed as
\begin{equation}
    \ket{\Psi_\mathrm{X,\nu,\mathbf{k}}} = x^\dagger_{\nu,\mathbf{k}} \ket{\underline{0}} = \frac{1}{\sqrt{V_\mathrm{exc}}}
    \sum_{\boldsymbol{\rho}} \tilde{\varphi}_\nu^*(\boldsymbol{\rho})\hat{e}^\dagger_{\boldsymbol{\rho}+\gamma_e\mathbf{k}}
    \hat{h}^\dagger_{-\boldsymbol{\rho}+\gamma_h\mathbf{k}}\ket{\underline{0}} 
    \label{eqn:exciton_2nd_quantization}
\end{equation}
where $x^\dagger_{\nu,\mathbf{k}}$ represents the exciton creation operator, $\hat{e}^\dagger_{\boldsymbol{\rho}+\gamma_e\mathbf{k}}$ ($\hat{h}^\dagger_{-\boldsymbol{\rho}+\gamma_h\mathbf{k}}$) denotes the creation of an electron (hole) with momentum $\boldsymbol{\rho}+\gamma_e\mathbf{k}$ ($-\boldsymbol{\rho}+\gamma_h\mathbf{k}$), with mass ratios $\gamma_{e,h}$ defined as $\frac{m_\mathrm{n,p}^*}{m_\mathrm{n}^* + m_\mathrm{p}^*}$. The state $\ket{\underline{0}}$ is the semiconductor many-body ground state with a filled VB and an empty CB. The weighting factor $\tilde{\varphi}_\nu(\boldsymbol{\rho})$ is the Fourier transform of the real-space exciton relative wave function $\varphi_\nu(\mathbf{r})$, which is obtained as a solution to the Wannier equation (Eqn.\,\ref{eqn:wannier}). As such, the sum in its entirety represents a superposition state of correlated electron--hole excitations $\hat{e}^\dagger_{\boldsymbol{\rho}+\gamma_e\mathbf{k}}\hat{h}^\dagger_{-\boldsymbol{\rho}+\gamma_h\mathbf{k}}$ with total COM momentum $\left(\boldsymbol{\rho}+\gamma_e\mathbf{k}\right) + \left(-\boldsymbol{\rho}+\gamma_h\mathbf{k} \right) = \mathbf{k}$. It can be demonstrated that the exciton wave function shown in Eqn.\,\ref{eqn:exciton_2nd_quantization} forms an eigenstate of the following \textit{matter} Hamiltonian \cite{Imamoglu2022}, which describes electrons and holes in a semiconductor and their Coulomb interaction:
\begin{equation}
    \hat{H}_\mathrm{matter} = \sum_\mathbf{k}
    \left(E_\mathrm{g}+\frac{\hbar^2 k^2}{2m_\mathrm{n}^*} \right)
    \hat{e}^\dagger_\mathbf{k} \hat{e}_\mathbf{k} + 
    \sum_\mathbf{k} \frac{\hbar^2 k^2}{2m_\mathrm{p}^*}
    \hat{h}^\dagger_\mathbf{k} \hat{h}_\mathbf{k} - 
    \sum_{\mathbf{k}, \mathbf{k'},\boldsymbol{\rho}\neq0} V(\boldsymbol{\rho})
    \hat{e}^\dagger_{\mathbf{k}+\boldsymbol{\rho}}
    \hat{h}^\dagger_{\mathbf{k'}-\boldsymbol{\rho}}
    \hat{h}_\mathbf{k'}\hat{e}_\mathbf{k}
\end{equation}
Here, $V(\boldsymbol{\rho})$ is the Fourier transform of the Coulomb interaction potential. In our initial examination of the light--matter coupling strength, we will momentarily set aside the final term in the matter Hamiltonian and concentrate solely on interband transitions. This requires extending the overall Hamiltonian to also consider terms describing the quantized electromagnetic field and the light--matter interaction:
\begin{equation}
    \hat{H} = \hat{H}_\mathrm{matter} + \hat{H}_\mathrm{light} + \hat{H}_\mathrm{LM}
\end{equation}
$\hat{H}_\mathrm{light}$ can be defined by introducing the operators $\hat{a}^\dagger_\mathbf{q}$ ($\hat{a}_\mathbf{q}$), which create (annihilate) a single photon with momentum $\mathbf{q}$
\begin{equation}
    \hat{H}_\mathrm{light} = \sum_\mathbf{q} \hbar \omega_\mathbf{q}
    \hat{a}^\dagger_\mathbf{q} \hat{a}_\mathbf{q}
\end{equation}
The term $\hbar \omega_\mathbf{q}$ represents the energy of a single photon with momentum $\mathbf{q}$, the dispersion relation of which is given by $\omega_\mathbf{q} = q\frac{c}{n_r}$, with $n_r=\sqrt{\epsilon_r}$ being the material refractive index.

To obtain a description of the light--matter interaction Hamiltonian $\hat{H}_\mathrm{LM}$ we need to incorporate the electromagnetic field via the appropriate scalar and vector potentials of the electric and magnetic field, respectively. By adopting the Coulomb gauge [$\nabla\cdot\mathbf{A}(\mathbf{r})=0$] it can be shown that the only non-vanishing component of these potentials, and thus the one entering the Hamiltonian, is the vector potential $\mathbf{A}_\perp$, which is oriented transverse to the photon momentum $\mathbf{q}$ \cite{Imamoglu2022}. As elaborated in the previous section, the effect of such a vector potential is captured by performing the minimal-coupling substitution and replacing the canonical momentum $\mathbf{p}$ by $\mathbf{p}+e\mathbf{A}_\perp$:
\begin{equation}
    \frac{\mathbf{p}^2}{2m} \rightarrow \frac{\left(\mathbf{p}+e\mathbf{A}_\perp\right)^2}{2m} = \frac{\mathbf{p}^2}{2m} + \frac{e}{m} \mathbf{p} \cdot \mathbf{A}_\perp + \frac{e^2}{2m} |\mathbf{A}_\perp|^2
\end{equation}
The last term, which is proportional to $|\mathbf{A}_\perp|^2$, is thereby typically neglected for small light intensities. This leaves the $\mathbf{p} \cdot \mathbf{A}_\perp$ term, which defines the interaction Hamiltonian as follows:
\begin{equation}
    \hat{H}_\mathrm{LM} = \int \mathrm{d}^3\mathbf{r} \, \hat{\Psi}^\dagger(\mathbf{r})\frac{e}{m} \mathbf{p} \cdot \mathbf{A}_\perp\hat{\Psi}(\mathbf{r})
    \label{eqn:H_LM}
\end{equation}
where we introduced the Schr\"{o}dinger field operator describing the creation of an electron--hole pair with net zero momentum
\begin{equation}
    \hat{\Psi}^\dagger(\mathbf{r}) = \sum_\mathbf{k} \left(
    \phi_{c,\mathbf{k}}^*(\mathbf{r})\hat{e}^\dagger_\mathbf{k} + 
    \phi_{v,\mathbf{k}}^*(\mathbf{r})\hat{h}^\dagger_\mathbf{-k}
    \right)
    \label{eqn:field_operator}
\end{equation}
in which $\phi_{c,\mathbf{k}}(\mathbf{r})$ and $\phi_{v,\mathbf{k}}(\mathbf{r})$ are the Bloch wave functions of the CB electron and VB hole, respectively. These can be denoted as
\begin{equation}
    \phi_{a,\mathbf{k}}(\mathbf{r})=\frac{1}{\sqrt{V}}e^{i\mathbf{k}\cdot \mathbf{r}} u_{a,\mathbf{k}}(\mathbf{r})\text{\enspace for \enspace} a=c,v
    \label{eqn:Bloch_wf}
\end{equation}
where $V$ is a quantization volume and $u_{a,\mathbf{k}}(\mathbf{r})$ is the periodic part of the electronic wave function. In addition, the vector potential $\mathbf{A}_\perp$ can be expressed in terms of photon creation and annihilation operators:
\begin{equation}
    \mathbf{A}_\perp(\mathbf{r}) = \sum_\mathbf{q}
    \sqrt{\frac{\hbar}{2\epsilon_0\epsilon_r V_\mathrm{ph} \omega_\mathbf{q}}} \left(
    \hat{a}_\mathbf{q}e^{i\mathbf{q}\cdot \mathbf{r}} + 
    \hat{a}^\dagger_\mathbf{q} e^{-i\mathbf{q}\cdot \mathbf{r}} 
    \right)
    \label{eqn:vector_potential}
\end{equation}
where $V_\mathrm{ph}$ represents the photon quantization volume. By plugging in the expressions\,\ref{eqn:field_operator}$-$\ref{eqn:vector_potential} in Eqn.\,\ref{eqn:H_LM} and focusing only on resonant interband transitions, in which photon annihilation coincides with electron--hole pair creation or vice versa, the following form of the light--matter interaction Hamiltonian can be established \cite{Imamoglu2022}:
\begin{equation}
    \hat{H}_\mathrm{LM} = \hbar \sum_{\mathbf{k},\mathbf{q}} g_\mathbf{q} \hat{e}^\dagger_{\mathbf{k}+\mathbf{q}}
    \hat{h}^\dagger_{-\mathbf{k}}
    \hat{a}_{\mathbf{q}} + \mathrm{H.c.}
    \label{eqn:H_LM_final}
\end{equation}
with the light--matter coupling strength $g_\mathbf{q}$ given by
\begin{equation}
    g_\mathbf{q} = \frac{e}{m}\langle \mathbf{p} \rangle_{cv} \sqrt{\frac{1}{2\epsilon_0\epsilon_r\hbar V_\mathrm{ph} \omega_\mathbf{q}}}
\end{equation}
in which we have used the matrix element $\langle \mathbf{p} \rangle_{cv} = \bra{u_c}\mathbf{p}\ket{u_v} = \int \mathrm{d}^3\mathbf{r}\,u^*_{c,\mathbf{k}}(\mathbf{r}) \mathbf{p} u_{v,\mathbf{k}}(\mathbf{r})$. We remark that in using the Coulomb gauge we arrived at a description of the light--matter interaction in terms of the parameter $\frac{e}{m} \mathbf{p} \cdot \mathbf{A}$. An equivalent description of this interaction can be obtained by working in the electric-dipole gauge and making a long-wavelength approximation, according to which the wavelength of light ($\frac{2\pi}{q}$) can be taken to be much greater than the size of dipole (which in our case is the exciton) with which the light interacts. In such a setting, the interaction Hamiltonian is given by the term $e\mathbf{E} \cdot \mathbf{r}$, where $\mathbf{E}$ represents the electric field and $\mathbf{r}$ the position coordinate. The two descriptions are linked through the Power-Zienau-Woolley unitary transformation, which allows to switch from the Coulomb to the dipole gauge. In the latter setting, the interaction Hamiltonian $\hat{H}_\mathrm{LM}$ will have the same general expression as in Eqn.\,\ref{eqn:H_LM_final}, however the coupling strength will be given by
\begin{equation}
    g_\mathbf{q} = \langle \mathbf{\boldsymbol{\mu}} \rangle_{cv}
    \sqrt{\frac{\omega_\mathbf{q}}{2\epsilon_0\epsilon_r\hbar V_\mathrm{ph}}}
\end{equation}
where the transition dipole moment is $\langle \boldsymbol{\mu} \rangle_{cv} = e\bra{u_c}\mathbf{r}\ket{u_v} = e\int \mathrm{d}^3\mathbf{r}\,u^*_{c,\mathbf{k}}(\mathbf{r}) \mathbf{r} u_{v,\mathbf{k}}(\mathbf{r})$, and $\hbar \cdot \sqrt{\frac{\omega_\mathbf{q}}{2\epsilon_0\epsilon_r\hbar V_\mathrm{ph}}}$ represents the vacuum electric field amplitude $E_0$. With this expression, we can readily estimate the light--matter coupling strength in a typical GaAs semiconductor cavity structure, assuming a photon volume $V_\mathrm{ph}$ of around $10\,\mu\mathrm{m}\times10\,\mu\mathrm{m}\times1\,\mu\mathrm{m}$, with $1\,\mu\mathrm{m}$ being the cavity length. The transition dipole  $\langle \mathbf{\mu} \rangle_{cv}$ moment can be taken to be approximately $3\text{\AA}\cdot e$ and the photon energy $\hbar \omega_\mathbf{q}$ is estimated to be $1.5\,$eV. With these parameters, a coupling strength $g_\mathbf{q}$ of about $1.5\times10^9\,$Hz is obtained \cite{Imamoglu2022}, which corresponds to a rate substantially smaller than phonon scattering even at cryogenic temperatures. Such a diminished coupling strength would negate the possibility of discerning distinct excitonic transitions, as schematically illustrated in Fig.\,\ref{fig:2_Excitons} \bfB. Evidently, the incorporation of electron--hole Coulomb interaction is indispensable to our analysis. This can be accomplished by first rewriting the light--matter Hamiltonian (Eqn.\,\ref{eqn:H_LM_final}) in a more compact form:
\begin{equation}
    \hat{H}_\mathrm{LM} = \sum_\mathbf{q}\left(\hat{U}^\dagger_\mathbf{q} + \hat{U}_\mathbf{q} \right)\text{\enspace with \enspace} \hat{U}^\dagger_\mathbf{q} = \hbar g_\mathbf{q} \hat{a}_\mathbf{q}
    \sum_\mathbf{k}\hat{e}^\dagger_{\mathbf{k}+\mathbf{q}}
    \hat{h}^\dagger_{-\mathbf{k}}
\end{equation}
Moreover, having already established that excitonic excitations of the form $\hat{x}^\dagger_{\nu} \ket{\underline{0}}$ (Eqn.\,\ref{eqn:exciton_2nd_quantization}) are eigenstates of the matter Hamiltonian, with Coulomb interaction included, we do not need to retrace the comprehensive steps previously taken to deduce the coupling strength $g_\mathbf{q}$. It suffices to recast $\hat{U}^\dagger_\mathbf{q}$ in terms of excitonic creation operators, as opposed to electron--hole excitations \cite{Imamoglu2022}, which effectively corresponds to a change of basis. Without loss of generality, we focus on a scenario with zero photon momentum $\mathbf{q}$. In this case, Eqn.\,\ref{eqn:exciton_2nd_quantization} can be restated as
\begin{equation}
    \hat{x}^\dagger_{\nu} = \frac{1}{\sqrt{V_\mathrm{exc}}}
    \sum_{\boldsymbol{\rho}} \tilde{\varphi}_\nu^*(\boldsymbol{\rho})\hat{e}^\dagger_{\boldsymbol{\rho}}
    \hat{h}^\dagger_{-\boldsymbol{\rho}}
\end{equation}
We multiply both sides with $\sqrt{V_\mathrm{exc}}\cdot \varphi_\nu(\mathbf{r}=0)$ and sum over all $\nu$:
\begin{equation}
\begin{aligned}
    \sqrt{V_\mathrm{exc}}\sum_\nu\varphi_\nu(\mathbf{r}=0)\hat{x}^\dagger_{\nu} &= \sum_\nu\varphi_\nu(\mathbf{r}=0)\sum_{\boldsymbol{\rho}} \tilde{\varphi}_\nu^*(\boldsymbol{\rho})\hat{e}^\dagger_{\boldsymbol{\rho}}
    \hat{h}^\dagger_{-\boldsymbol{\rho}} \\
    &= \sum_{\boldsymbol{\rho}} \hat{e}^\dagger_{\boldsymbol{\rho}}
    \hat{h}^\dagger_{-\boldsymbol{\rho}}\sum_\nu\varphi_\nu(\mathbf{r}=0)
    \tilde{\varphi}_\nu^*(\boldsymbol{\rho})
\end{aligned}
\end{equation}
Furthermore, since $\varphi_\nu$ and $\tilde{\varphi}_\nu$ are linked through a Fourier transform, we can express $\varphi_\nu(\mathbf{r}=0)$ as $\sum\limits_{\mathbf{k}}\tilde{\varphi}_\nu(\mathbf{k})$. In this manner, we can state the expression above as
\begin{equation}
\begin{aligned}
    \sqrt{V_\mathrm{exc}}\sum_\nu\varphi_\nu(\mathbf{r}=0)\hat{x}^\dagger_{\nu} &= \sum_{\boldsymbol{\rho}} \hat{e}^\dagger_{\boldsymbol{\rho}}
    \hat{h}^\dagger_{-\boldsymbol{\rho}} \sum_{\nu,\mathbf{k}} \tilde{\varphi}_\nu(\mathbf{k}) \tilde{\varphi}_\nu^*(\boldsymbol{\rho}) \\
    &= \sum_{\boldsymbol{\rho}} \hat{e}^\dagger_{\boldsymbol{\rho}}
    \hat{h}^\dagger_{-\boldsymbol{\rho}} \sum_{\nu,\mathbf{k}} \braket{\mathbf{k}|\nu}\braket{\nu|\mathbf{\rho}} \\
    &= \sum_{\boldsymbol{\rho}} \hat{e}^\dagger_{\boldsymbol{\rho}}
    \hat{h}^\dagger_{-\boldsymbol{\rho}} \sum_{\mathbf{k}} \delta_{\mathbf{k},\boldsymbol{\rho}} \\
    &= \sum_\mathbf{k} \hat{e}^\dagger_{\mathbf{k}}
    \hat{h}^\dagger_{-\mathbf{k}}
\end{aligned}
\end{equation}
which is exactly the expression we desired to recast. This result can be generalized to finite photon momentum $\mathbf{q}$, however the final outcome remains unchanged. The revised light--matter interaction Hamiltonian reads
\begin{equation}
    \hat{H}_\mathrm{LM} = \sum_\mathbf{q}\left(\hat{U}^\dagger_\mathbf{q} + \hat{U}_\mathbf{q} \right) = \hbar 
    \sum_\mathbf{\nu,\mathbf{q}} g_{\mathbf{q},\nu}^{\mathrm{exc}}
    \hat{a}_\mathbf{q} \hat{x}^\dagger_{\nu} + \mathrm{H.c.}
\end{equation}
in which the modified light--matter coupling strength is given by
\begin{equation}
    g_{\mathbf{q},\nu}^{\mathrm{exc}} = g_{\mathbf{q}}\cdot \sqrt{V_\mathrm{exc}} \cdot \varphi_\nu(\mathbf{r}=0) = 
    \langle \mathbf{\boldsymbol{\mu}} \rangle_{cv}
    \sqrt{\frac{\omega_\mathbf{q}}{2\epsilon_0\epsilon_r\hbar }} \cdot \sqrt{\frac{V_\mathrm{exc}}{V_\mathrm{ph}}} \cdot \varphi_\nu(\mathbf{r}=0)
    \label{eqn:exc_phot_mode_overlap}
\end{equation}
which forms the core result of this section. The inclusion of Coulomb interactions greatly amplifies the light--matter coupling strength, by a factor $\sqrt{V_\mathrm{exc}} \cdot \varphi_\nu(\mathbf{r}=0)$. Here, $\varphi_\nu(\mathbf{r})$ signifies the exciton relative wave function, derived from solving the Wannier equation (Eqn.\,\ref{eqn:wannier}). Consequently, $\varphi_\nu(\mathbf{r}=0)$ indicates the probability amplitude of the electron and hole, comprising the exciton, to coexist at the same location. Additionally, $V_\mathrm{exc}$ denotes the exciton coherence volume, which reveals the spatial range over which the exciton COM momentum is well-defined. The validity of the expression above becomes apparent by considering that a diminishing Coulomb interaction strength would progressively increase the distance between the electron and hole. Therefore, in the absence of any electron--hole interaction, the relative wave function $\varphi_\nu$ would be uniformly delocalized over the entire excitonic coherence volume, leading to $\varphi_\nu(\mathbf{r}=0) = \frac{1}{\sqrt{V_\mathrm{exc}}}$. Consequently, this would restore the light--matter coupling $g_{\mathbf{q}}$ given solely by interband transitions.

Overall, we underscore that attributes of \emph{both the COM and relative wave function of the exciton} jointly dictate the strength of light--matter coupling. The significance of this result becomes evident in various settings. Firstly, reevaluating the coupling strength in GaAs for the $1s$ exciton with a Bohr radius $a_\mathrm{B}$ of around $10\,$nm and $\varphi_{1s}(\mathbf{r}=0) = \frac{1}{\sqrt{\pi a_\mathrm{B}^3}}$, we find $g_{\mathbf{q},\nu}^{\mathrm{exc}} \approx 7.5 \times 10^{12}\,$Hz, which corresponds to an enhancement of three orders of magnitude in comparison to the case of bare non-interacting electrons and holes, making radiative decay the dominant decay channel, which is in line with experimental observations at liquid helium temperatures. Moreover, as excitonic Rydberg states inherently possess an extended spatial extent in relative coordinates, their wave function content at $\mathbf{r} = 0$ is significantly diminished, leading to weaker light--matter coupling. Given that the oscillator strength of optical transitions proportionally scales with $|g_{\mathbf{q},ns}^{\mathrm{exc}}|^2$, Rydberg excitons exhibit a diminishing oscillator strength with an increasing principal quantum number $n$, as illustrated previously in Fig.\,\ref{fig:2_Excitons} \bfB. A similar phenomenon occurs for indirect excitons, where the electron and hole reside in spatially separated quantum wells. This spatial separation considerably reduces their light--matter coupling, thus making their resonant excitation challenging. Ultimately, situations where $\varphi_{\nu}(\mathbf{r}=0)$ vanishes entirely, as exemplified by $p$-type wave functions, result in optical transitions that are completely dark.

Lastly, the dependence of light--matter coupling on the ratio $\sqrt{\frac{V_\mathrm{exc}}{V_\mathrm{ph}}}$ gains relevance when the dimensionality of the semiconductor is altered. For bulk 3D semiconductors, it is reasonable to assume that excitons and photons share a similar coherence volume, i.e.\,$V_\mathrm{exc} \approx V_\mathrm{ph}$. However, in 2D semiconductors $V_\mathrm{exc}$ is substantially smaller than $V_\mathrm{ph}$, which can potentially reduce the light--matter coupling significantly. To overcome this limitation it is crucial to enhance the mode overlap between excitons and photons, achievable, for instance, through photonic confinement in optical cavities.

%%%%%%%%%%%%%%%%%%%%%%%%%%%%%%%%%%%%%%%%%%%%%%%%%%%%%%%%%%%%%%%%%%%%%%%%
\section{Exciton fine structure}
\label{chap:theory:sec:fine_structure}
%%%%%%%%%%%%%%%%%%%%%%%%%%%%%%%%%%%%%%%%%%%%%%%%%%%%%%%%%%%%%%%%%%%%%%%%

In all the considerations above we have only taken into account the direct part of the Coulomb interaction potential. This contribution is of \emph{electrostatic} nature and thus, in a semiclassical picture, describes the interplay between positive and negative charge distributions and the resulting mutual attractive forces. As such, it leads to formation of an electron--hole bound state and defines an energy scale for the exciton binding energy $E_\mathrm{B}$ \cite{Wang2018a}. In addition to the direct part, the Coulomb potential also gives rise to an exchange contribution in which the Coulomb interaction is evaluated jointly with the Pauli exclusion principle. In essence, our considered case of an interacting electron and a hole is actually a many-body system comprised of numerous indistinguishable fermions, where one charge carrier from a filled VB is promoted to the CB. Therefore, to properly satisfy Fermi statistics the wave function of this system needs to be antisymmetrized with respect to particle exchange. As a result, akin to atomic systems, the exciton energy will depend on the specific configuration of electron and hole spins, and as a unique characteristic of monolayer TMDs, also on the respective valley index \cite{Wang2018a}.

\begin{figure}[htb]
    \centering
	\includegraphics[width=14cm]{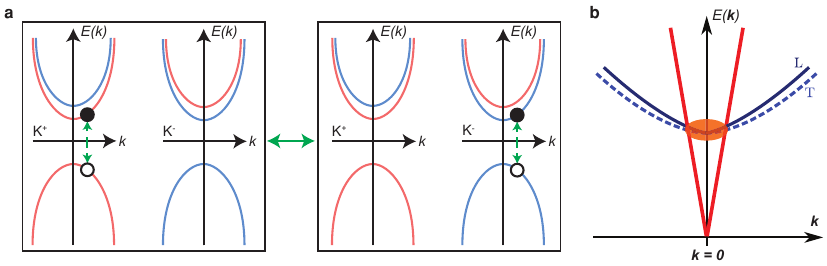}
	\caption{\textbf{Exciton fine structure due to electron--hole exchange.}
    (\bfA) Electron--hole exchange processes in Mo$X_2$-type compounds. The dashed green arrow denotes intra-valley exchange, the solid green arrow represents inter-valley exchange. (\bfB) Exciton dispersion relation with consideration of intra-valley and inter-valley exchange processes. These interactions split the exciton dispersion into two branches, with the splitting scaling linearly with COM momentum $\mathbf{k}$. The upper and lower branch exhibit linear polarized emission which is oriented longitudinal (L) and transverse (T) to the in-plane wavevector. The red solid line indicates the light cone. The exciton dispersion is depicted not to scale. Reproduced with permission from Ref.\,\cite{Glazov2015}, \textcopyright\,$2015$ WILEY-VCH Verlag GmbH \& Co. KGaA, Weinheim.
    }  
	\label{fig:2_exchange}
\end{figure}

The exchange contribution itself is typically separated into two parts, a long-range and a short-range interaction term. Here, we will only focus on the former\footnote{The short-range exchange term influences the splitting between bright and dark excitons. In this work, we exclusively focus on bright excitons, and thus limit our discussion to the long-range exchange interaction.}. Contrary to the direct term, the long-range exchange interaction is of \emph{electrodynamic} origin. It can be conceptualized by considering an electron--hole pair, which generates an electromagnetic field through a virtual recombination process. Subsequently, this field in turn leads to creation of another electron--hole pair either in the same valley, a process denoted as intra-valley exchange, or in the opposite valley, a process labelled as inter-valley exchange (Fig.\,\ref{fig:2_exchange} \bfA) \cite{Wang2018a,Yu2014,Yu2015,Glazov2014,Glazov2015}.

As a consequence, for zero COM momentum $k$, excitons in $K^\pm$ valleys will remain $\sigma^\pm$ polarized and degenerate in energy. However, for excitons with momenta outside the light cone, these additional coupling mechanisms will result in new excitonic eigenstates. These can be determined by expressing the exchange Hamiltonian in the valley pseudospin basis \cite{Xu2014}, in which spin up/down signifies an exciton in the $K^+$/$K^-$ valley, respectively \cite{Shimazaki2021}:
\begin{equation}
    \hat{H}_\mathrm{ex}=\sum_\mathbf{k}
    \begin{pmatrix}
        x_{\mathbf{k},K^+}^\dagger \\
        x_{\mathbf{k},K^-}^\dagger
    \end{pmatrix}^\intercal
    \left[
    \frac{\hbar^2\mathbf{k}^2}{2M} +
    \frac{|\mathbf{k}|}{|K|}J + 
    \frac{|\mathbf{k}|}{|K|}J
    \begin{pmatrix}
        0 & e^{-2i\theta} \\
        e^{+2i\theta} & 0
    \end{pmatrix}
    \right]
    \begin{pmatrix}
        x_{\mathbf{k},K^+} \\
        x_{\mathbf{k},K^-}
    \end{pmatrix}
    \label{eqn:exchange_hamiltonian}
\end{equation}
where $x_{\mathbf{k},K^+}^\dagger$ ($x_{\mathbf{k},K^-}^\dagger$) leads to creation of an exciton in $K^+$ ($K^-$) valley with COM momentum $\mathbf{k}$, $\theta=\arctan(\mathrm{k}_y/\mathrm{k_x})$, and $|K|=\frac{4\pi}{3a}$ being the distance from the $K^\pm$ to the $\Gamma$ point in the Brillouin zone with a lattice constant $a$, and $J$ being the exchange coupling strength. The first term in the exchange Hamiltonian describes the exciton kinetic energy. The second term denotes intravalley exchange, which leads to an overall shift in energy irrespective of the valley. The last term represents intervalley exchange and thus describes the coupling between exciton COM and valley pseudospin, typically referred to as valley--orbit coupling. The resulting exciton dispersion relation can be obtained by diagonalizing $\hat{H}_\mathrm{ex}$ and is given by
\begin{equation}
    E(\mathbf{k})_\pm=\frac{\hbar^2\mathbf{k}^2}{2M} +
    \frac{|\mathbf{k}|}{|K|}J \pm
    \frac{|\mathbf{k}|}{|K|}J
\end{equation}
Therefore, for finite $\mathbf{k}$ two branches emerge with an energy splitting
\begin{equation}
    \Delta E = E_+ - E_- \propto |\mathbf{k}|J
\end{equation}

First-principle calculations show that the valley--orbit coupling strength $J$ itself depends on the exciton radiative decay, and thus $|\varphi(\mathbf{r}=0)|^2$, where $\varphi(\mathbf{r})$ denotes the wave function for the electron--hole relative motion \cite{Glazov2014,Yu2014}. Therefore, owing to the strong light--matter coupling in TMDs, the valley--orbit coupling is expected to be quite large and estimated to be around $1\,$eV. However, in experimentally relevant conditions TMD monolayers are typically encapsulated with dielectrics which can lead to a screening of the Coulomb interaction and thus an effective reduction in $J$ to around $300\,$meV \cite{Shimazaki2021,Qiu2015}. A schematic illustration of the resulting dispersion relation for the two branches is shown in Fig.\,\ref{fig:2_exchange} \bfB. For momenta inside the light cone the upper branch exhibits a linear dependence, while the lower branch shows a quadratic one. Furthermore, since the new eigenstates are equal superpositions of excitons in both valleys, the upper and lower branch will couple to photons with linear polarization oriented longitudinal (L) and transverse (T) to the $\mathbf{k}$ direction, respectively. The intrinsic 3-fold rotational (C3) symmetry of the lattice ensures that this LT-splitting vanishes for $\mathbf{k}=0$. A finite splitting can be achieved by breaking this rotational symmetry, for instance through application of an in-plane uniaxial strain \cite{Yu2014,Yu2015}.

  %%%%%%%%%%%%%%%%%%%%%%%%%%%%%%%%%%%%%%%%%%%%%%%%%%%%%%%%%%%%%%%%%%%%%%%%
\renewcommand{\thefootnote}{\fnsymbol{footnote}}
\chapter[Tunable exciton quantum confinement in 1D]{Tunable exciton quantum confinement in 1D\footnote{This chapter is adapted from the following publication \cite{Thureja2022}:\\ \\
\textbf{Electrically tunable quantum confinement of neutral excitons}\\
Thureja, D.; Imamoglu, A.; Smole\'{n}ski, T.; Amelio, I.; Popert, A.; Chervy, T.; Lu, X.; Liu, S.; Barmak, K.; \mbox{Watanabe}, K.; Taniguchi, T.; Norris, D. J.; Kroner, M.; Murthy, P. A.\\
\textit{Nature}\textbf{ 606}, 298 (2022)
}}
\label{chap:1D}
%%%%%%%%%%%%%%%%%%%%%%%%%%%%%%%%%%%%%%%%%%%%%%%%%%%%%%%%%%%%%%%%%%%%%%%%

%%%%%%%%%%%%%%%%%%%%%%%%%%%%%%%%%%%%%%%%%%%%%%%%%%%%%%%%%%%%%%%%%%%%%%%%
\section{Introduction}
%%%%%%%%%%%%%%%%%%%%%%%%%%%%%%%%%%%%%%%%%%%%%%%%%%%%%%%%%%%%%%%%%%%%%%%%

Owing to their zero net charge, neutral excitons are intrinsically more challenging to electrically confine than charged particles. One possible route for exciton confinement involves the dc Stark effect which ensures that excitons experience an attractive potential around the absolute maximum of an inhomogeneous electric field distribution. However, achieving quantum confinement in such a potential requires that the energy splitting between discrete motional excitonic states ($\hbar\omega$) exceeds the exciton line broadening $\Gamma$ as well as the characteristic energy of thermal fluctuations $k_{\mathrm{B}}T$. For excitons in semiconductor heterostructures at $T=4$~K, this implies $\hbar\omega \gtrsim 1\,$meV. This -- in turn -- requires a confinement length scale of  $\ell = \sqrt{\hbar/m_\mathrm{X}\omega} \lesssim 10\,$nm for exciton mass $m_\mathrm{X}$ comparable with free electron mass. Engineering electrically tunable confinement potentials at such small length scales is a technical challenge. In addition, unless the exciton binding energy is much larger than $\hbar\omega$, the requisite applied fields will lead to fast ionization of excitons, drastically reducing their radiative efficiency.

Previous experiments have mainly approached the problem of electrical confinement by relying on indirect excitons, in which the electron and hole comprising an exciton are spatially separated in different quantum wells. This gives rise to a permanent electric dipole moment which makes them easier to electrically manipulate, but at the expense of weaker coupling to light (see section \ref{chap:theory:sec:LM_coupling}). In III-V semiconductor heterostructures with coupled quantum wells, different potential landscapes have been demonstrated for indirect excitons, such as ramps, lattices, and harmonic traps \cite{Hagn1995,Rapaport2005,Gartner2007,Vogele2009,Schinner2013,Andreev2016,Butov2017}. However, the requirement to suppress exciton ionization in these systems \cite{Hammack2006} prevented the observation of quantum confinement. Moreover, the quantum wells typically need to be buried deep within the heterostructure, which limits the electric field gradients that can be applied using lithographically patterned gates outside the structure. Electrical manipulation as well as confinement of spatially indirect excitons at large length scales has also been reported in transition metal dichalcogenide (TMD) heterostructure devices \cite{Unuchek2018,Wang2018a,Liu2020,Jauregui2019,Shanks2021,Shanks2022}. Nevertheless, tunable confinement and quantization of motional states of direct excitons, which couple strongly to light, has not been demonstrated previously.

Here, we overcome these challenges and demonstrate an electrically tunable quantum confining potential for spatially direct excitons in a monolayer semiconductor. In achieving this goal, we not only solve a long-standing experimental problem in quantum photonics, but reveal intriguing aspects of quantum confined excitons unique to our system. In addition to being highly relevant for technological applications, our work highlights that the electrically controlled quantum confined system is a promising experimental platform for exploring strongly correlated physics of excitons and photons.

%%%%%%%%%%%%%%%%%%%%%%%%%%%%%%%%%%%%%%%%%%%%%%%%%%%%%%%%%%%%%%%%%%%%%%%%
\section{Confinement scheme}
%%%%%%%%%%%%%%%%%%%%%%%%%%%%%%%%%%%%%%%%%%%%%%%%%%%%%%%%%%%%%%%%%%%%%%%%

\begin{figure}[htb]
    \centering
	\includegraphics[width=8.5cm]{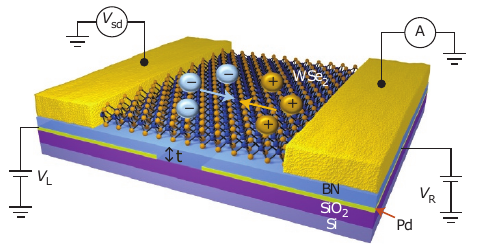}
	\caption{\textbf{2D lateral p-n junction.} Schematic of a monolayer semiconductor (WSe$_2$) p-n junction device featuring split bottom gate (BG) electrodes, and source and drain contacts. The semiconductor layer is separated from the BG by an insulating dielectric spacer (h-BN) of thickness $t$. The entire device is placed on a silicon dioxide layer on a silicon substrate. Reproduced with permission from Springer Nature \cite{Ross2014}.
    }  
	\label{fig:3_1_Splitgate_Device}
\end{figure}

Our proposed scheme to confine excitons relies on the creation of large electric fields and, in particular, large electric field gradients. To realize an electrostatic landscape with these features one can leverage in-plane electric fields that naturally arise in the insulating region of a 2D lateral p-i-n diode. A possible method for realizing such a structure is through electrostatic doping of a monolayer semiconductor using split bottom gate (BG) electrodes separated by a narrow gap (Fig.\,\ref{fig:3_1_Splitgate_Device}). This approach has already been demonstrated to be an effective means for controlling the respective carrier densities in the p- and n-doped regions individually. Various works have confirmed the feasibility of such devices for use in optoelectronics \cite{Pospischil2014,Baugher2014,Ross2014,Buscema2014,Memaran2015,Bie2017}, electronics \cite{Pang2019} and neuromorphic applications \cite{Pan2020}.

To verify whether such a device architecture will allow for creating the requisite electric field strengths and gradients for exciton quantum confinement, we derive an analytical model for estimating the depletion width $d$ and the in-plane electric field strength $F_\mathrm{x}$ associated with a gate-tunable lateral p-n junction (further details elaborated in section 7.1). We make the simplifying assumption that the gap between the split BG is infinitesimally small, such that the BG potential changes in a step-like fashion. This should allow us to establish a conservative estimate for the minimum width of the depletion zone. Furthermore, we constrain our analysis to the case of symmetric biasing conditions, i.e.\,$V_\mathrm{L} = -V_\mathrm{R}$, where $V_\mathrm{L/R}$ is the potential applied to the left and right BG, respectively (Fig.\,\ref{fig:3_1_Splitgate_Device}). Under these settings, one can approximate the depletion width as
\begin{equation}
    d\approx\frac{\pi}{\mathrm{cosh}^{-1}\left( \frac{8}{\pi} \frac{|V_\mathrm{L}|}{E_\mathrm{g}} \right)} \cdot t
    \label{eqn:depletion_width}
\end{equation}
with $E_\mathrm{g}$ being the bandgap of the semiconductor under consideration, and $t$ being the thickness of the BG dielectric. We emphasize that the depletion width strongly depends on the thickness of the BG dielectric and only exhibits a very weak dependence on its dielectric constant. The peak in-plane electric field strength will then be given by
\begin{equation}
    F_\mathrm{x,max} \approx \frac{\pi}{2} \frac{E_\mathrm{g}}{d}
    \label{eqn:field_max}
\end{equation}
Using these relations as a starting point we aim to realize tunable exciton quantum confinement using monolayer TMD heterostructures as our material platform of choice, as it offers several key advantages over conventional III-V semiconductor heterostructures. The most important one is the strong exciton binding energy ($E_\mathrm{B}$, around $200\,$meV \cite{Goryca2019}), which renders the excitons more resilient to in-plane electric fields. In addition, owing to the layered nature of Van der Waals (VdW) dielectrics, such as hexagonal boron nitride (h-BN), the thickness of the dielectric spacer layer can be tuned with a sub-nanometer precision, granting precise control of in-plane electric field distribution in the device. Assuming typical material and operational parameters for TMD heterostructures ($E_\mathrm{g} = 2$\,eV, $t = 30$\,nm and $V_\mathrm{L} = 5$\,V) the resulting depletion width would be approximately $d\approx40\,$nm. This would lead to an in-plane electric field maximum $F_\mathrm{x,max}$ of approximately $85\,$V$/\mu$m, creating highly favorable conditions for the experimental observation of exciton quantum confinement.

\begin{figure}[htb]
    \centering
	\includegraphics[width=\textwidth]{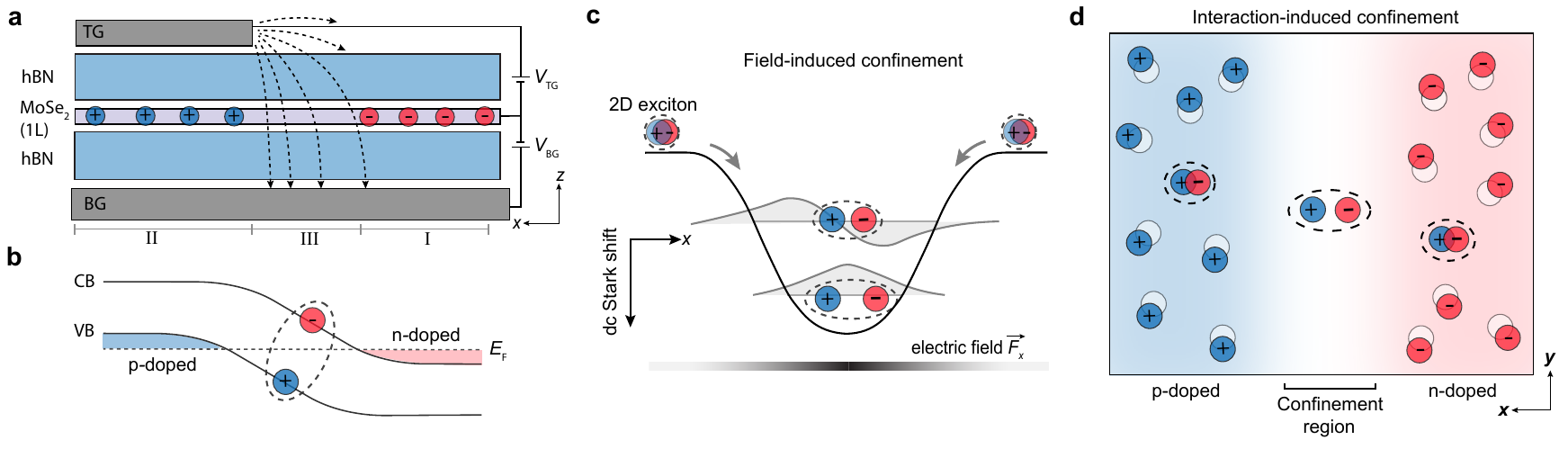}
	\caption{\textbf{Tunable quantum confinement of excitons.}
    (\bfA) Schematic side view of proposed device architecture for achieving tunable exciton quantum confinement. It consists of a monolayer TMD semiconductor, such as MoSe$_2$, encapsulated by h-BN, and partially overlapping gates (TG and BG). By applying voltage between TG and BG, we generate inhomogeneous in-plane electric fields along the TG edge (black dashed lines) and obtain a p-i-n charging configuration, in which holes (region II) and electrons (region I) are represented by blue and red circles, respectively. (\bfB) Schematic diagram of CB and VB edges in the p-i-n regime. (\bfC) Field-induced confinement originates from the local in-plane electric field $F_\mathrm{x}$, which induces a permanent dipole moment for 2D excitons and causes a dc Stark shift $\Delta E \propto |F_\mathrm{x}|^2$. (\bfD) Interaction-induced confinement arises from the repulsive interaction between excitons and itinerant charges, which imposes a density-dependent energy cost, $\Delta E \propto \sigma(x)$, for excitons entering the doped regions.
    }  
	\label{fig:3_2_Confinement_scheme}
\end{figure}

However, it should be noted that achieving these specific values represents an experimental challenge, since achieving the best results requires the gap between the split BG to be minimized. To avoid the need for nanoscale lithographic patterning, we propose an alternative device structure, shown in Fig.\,\ref{fig:3_2_Confinement_scheme}\,\bfA. It consists of a monolayer TMD semiconductor, such as MoSe$_2$, encapsulated by insulating dielectric spacer layers and two gate electrodes (top gate, TG; bottom gate, BG). Notably the TG and BG have partial spatial overlap as shown in Fig.\,\ref{fig:3_2_Confinement_scheme}\,\bfA, which allows to separately define adjacent n-doped (region I) and p-doped regions (region II), analogous to the aforementioned device architecture featuring a split BG. Fringing in-plane electric fields are generated along the TG edge (region III) giving rise to the requisite spatially inhomogeneous electric field distribution. The energies of conduction (CB) and valence band (VB) edges relative to the Fermi level $E_\mathrm{F}$ in this doping configuration are illustrated in Fig.\,\ref{fig:3_2_Confinement_scheme}\,\bfB.

Two fundamentally distinct effects contribute to quantum confinement of excitons in our structure. The first arises from the strong in-plane electric field $F_\mathrm{x}$, which polarizes the excitons along $x$ and lowers their energy due to the dc Stark shift, $\Delta E_\mathrm{S} = -\frac{1}{2} \alpha |F_\mathrm{x}|^2$, in which $\alpha$ is the exciton polarizability \cite{Cavalcante2018}. Because $F_\mathrm{x}$ vanishes in the doped regions on either side of the insulating i region, excitons experience an attractive confining potential towards the local maximum in $|F_\mathrm{x}(x)|$. Hence, we expect the dipole size to be the largest for excitons confined in the lowest eigenstates, as illustrated in Fig.\,\ref{fig:3_2_Confinement_scheme}\,\bfC. In addition to the Stark shift, we describe a new confinement mechanism that stems from the interaction between neutral excitons and itinerant charges present in the neighbouring p-doped and n-doped regions (Fig.\,\ref{fig:3_2_Confinement_scheme}\,\bfD). As neutral excitons generated in the i region enter the n-doped or p-doped regions, they experience a repulsive interaction which increases their energy proportional to the charge density, $\Delta E_\mathrm{P} \propto \sigma(x)$. This many-body dressed state can be described as a repulsive polaron (RP) \cite{Efimkin2017}, and has been previously observed in charge-tunable semiconductor heterostructure devices \cite{Sidler2017}. In this scenario, a gradient in charge density exerts a force on excitons that pushes them towards the local minimum in charge density \cite{Chervy2020}. A steep charge density gradient on both sides of the i region therefore acts as a repulsive potential barrier for excitons, which confines them spatially.

The total potential experienced by excitons in the center-of-mass (COM) frame is a sum of dc Stark shift and repulsive interaction shift contributions:
\begin{equation}
    V(x) = \underbrace{-\frac{1}{2}\alpha |F_\mathrm{x}(x)|^2}_\text{dc Stark shift} \,\,\,\, + \underbrace{\beta |\sigma(x)|}_\text{Interaction shift}
    \label{eqn:potential}
\end{equation}
Here, the proportionality constant $\beta$ is an effective exciton-charge coupling constant. We emphasize that the potential in Eqn.\,\ref{eqn:potential} provides confinement only for excitons whereas unbound electrons or holes experience a repulsive potential that accelerates them towards the n-doped and p-doped regions, respectively.

%%%%%%%%%%%%%%%%%%%%%%%%%%%%%%%%%%%%%%%%%%%%%%%%%%%%%%%%%%%%%%%%%%%%%%%%
\section{Electrostatic simulation of TMD heterostructure devices}
\label{sec:1DX_sims}
%%%%%%%%%%%%%%%%%%%%%%%%%%%%%%%%%%%%%%%%%%%%%%%%%%%%%%%%%%%%%%%%%%%%%%%%

To obtain quantitative information about the in-plane electric fields, charge densities and the corresponding confinement potentials achievable in our devices, we perform finite-element calculations using the \emph{Electrostatics} package in COMSOL (implementation details in appendix section \ref{appendix:comsol}). For all our simulations, we assume a temperature $T=0\,$K. We employ the Thomas--Fermi approximation and model the TMD monolayer as a single sheet of charge with density
\begin{align}
    \sigma(x) = \sigma_\mathrm{n}(x) + \sigma_\mathrm{p}(x)
    \label{eqn:sigma}
\end{align}
where $\sigma_\mathrm{n}$ and $\sigma_\mathrm{p}$ are the electron and hole charge densities, which are in turn given by
\begin{align}
    \sigma_\mathrm{n}(x) &= -e \int_{E_\mathrm{C}(V(x))}^{E_\mathrm{F}} \mathcal{D}(E) dE, &E_\mathrm{F} > E_\mathrm{C}(V(x)) \nonumber \\
    &= -e \mathcal{D}(E)(E_\mathrm{F}-E_\mathrm{C}), \label{eqn:sigma_n} \\
    \sigma_\mathrm{p}(x) &= e \int_{E_\mathrm{F}}^{E_\mathrm{V}(V(x))} \mathcal{D}(E) dE, &E_\mathrm{F} < E_\mathrm{V}(V(x)) \nonumber \\
    &= e \mathcal{D}(E)(E_\mathrm{V}-E_\mathrm{F}) \label{eqn:sigma_p}
\end{align}
Here, $\mathcal{D}(E) = g_\mathrm{S} g_\mathrm{V} m^*/2\pi\hbar^2$ is the 2D density of states for electrons and holes in the semiconductor, where $g_\mathrm{S} = 1$ is the spin degeneracy and $g_\mathrm{V} = 2$ is the valley degeneracy. $E_\mathrm{C}$ and $E_\mathrm{V}$ are the conduction and valence band edge energies, which depend on the local electrostatic potential. $E_\mathrm{F}$ is the  Fermi level determined by the alignment of the contact work function with respect to the band edges.

The simulated device geometry is depicted in Fig.\,\ref{fig:3_2_Confinement_scheme}\,\bfA. The sheet of charge is encapsulated by $30$-nm-thick h-BN slabs and contacted by ohmic electrodes. Furthermore, we include two gates with partial overlap. The BG is kept at a fixed bias of $4$\,V and ensures a global electron doping throughout the semiconductor. The voltage on the TG is varied from $0$\,V to $-10$\,V in our simulations. We take MoSe$_2$ to be our monolayer semiconductor and assume the following material parameters for this calculation: bandgap $E_\mathrm{g} = 1.85$\,eV, Fermi level offset relative to the valence band edge at zero potential $E_\mathrm{F} - E_\mathrm{V}(V=0)=0.99$\,eV \cite{Wilson2017}, electron effective mass $m_\mathrm{n}^* = 0.7\,m_\mathrm{e}$ \cite{Larentis2018}, hole effective mass $m_\mathrm{p}^* = 0.6\,m_\mathrm{e}$ \cite{Zhang2014,Goryca2019}, out-of-plane dielectric constant  $\varepsilon_{\perp} = 3.76$, and in-plane dielectric constant $\varepsilon_{\parallel} = 6.93$ for h-BN \cite{Laturia2018}.

\begin{figure}[htbp]
    \centering
	\includegraphics[width=\textwidth]{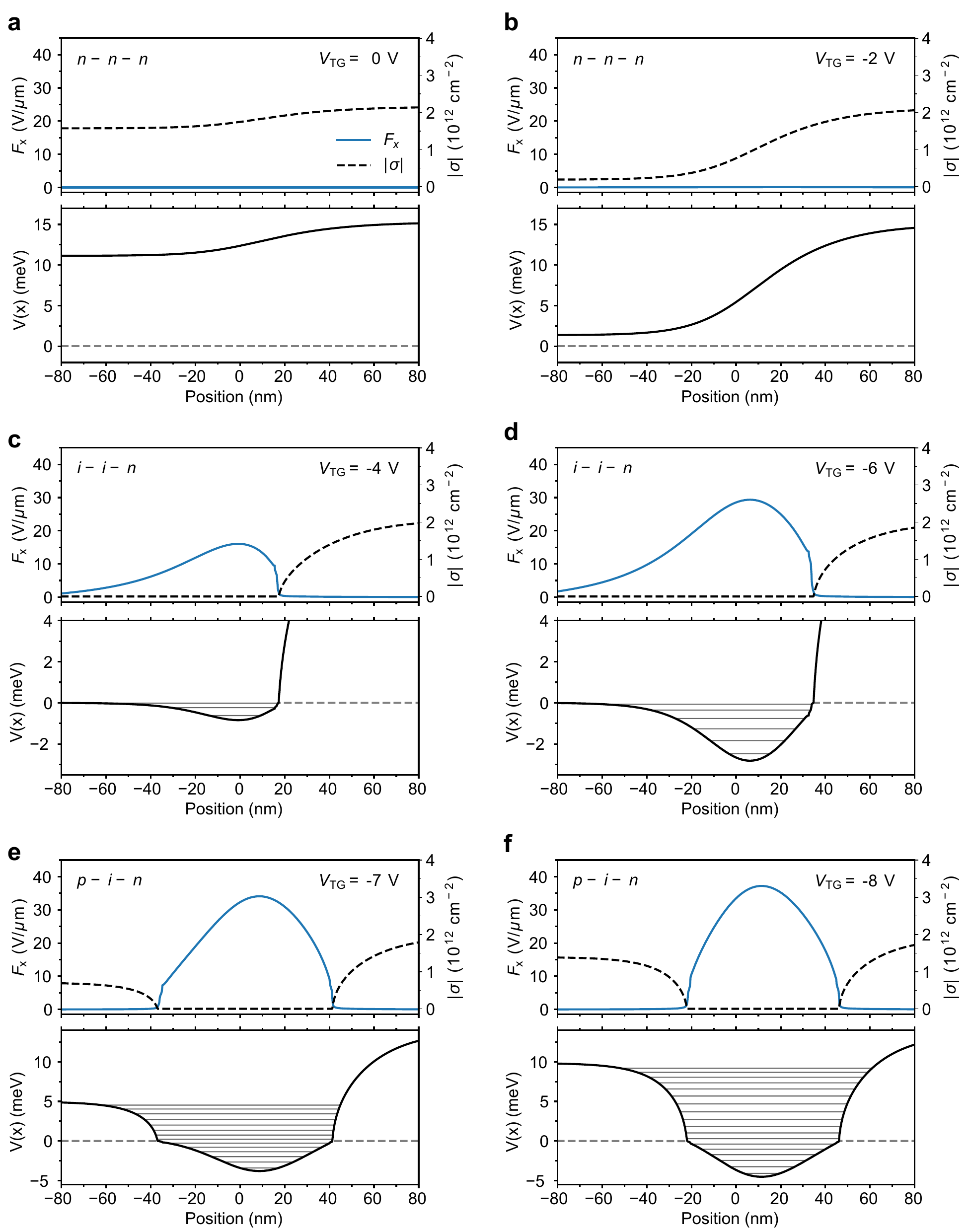}
	\caption{\textbf{Electrostatic simulations of the device.}
    Magnitude of charge density $|\sigma(x)|$, in-plane electric field $|F_x|$ and exciton confining potential $V(x)$ as a function of position for varying TG voltage $\vtg$ and fixed BG voltage $\vbg=4$\,V. The BG extends over the entire plotted range of the position from $-80$\,nm to $80$\,nm. The TG extends from $-80$\,nm to $0$\,nm, with the edge at $x = 0$. The different charging configurations, namely n-n-n (\bfA, \bfB), i-i-n (\bfC, \bfD) and p-i-n (\bfE, \bfF) show the evolution of the confinement potential as a function of $\vtg$.
    }  
	\label{fig:3_3_Electrostatic_sims}
\end{figure}

In this manner, we obtain the spatial charge density and in-plane electric field distribution for varying TG voltages $\vtg$, which are depicted in the top panel of each subplot in Fig.\,\ref{fig:3_3_Electrostatic_sims}. The BG extends over the entire plotted range of the position from $-80$\,nm to $80$\,nm, whereas the TG only ranges from $-80$\,nm to $0$\,nm, with the edge located at $x=0$. We can identify three distinct regimes as we vary $\vtg$, which we identify according to the doping state in the single-gated (region I), dual-gated (region II) and the intermediate region III. For $\vtg > -4\,$V, we are in the n-n-n regime, where all three regions are n-doped, but the electron density varies spatially, as shown in Fig.\,\ref{fig:3_3_Electrostatic_sims}\,{\bfA} and {\bfB}. As we decrease $\vtg$ further, we deplete regions II and III completely and hence we are in the i-i-n regime, which is accompanied by a large increase in magnitude of the in-plane electric field $|F_\mathrm{x}|$ (Fig.\,\ref{fig:3_3_Electrostatic_sims}\,{\bfC} and {\bfD}). While the maximum of the field distribution is located close to the TG edge, owing to the large lateral extent of the neutral region, the in-plane field persists even under the TG and exhibits a spatial asymmetry. The i-i-n regime persists until the onset of hole doping in region II, which occurs in our simulations at $\vtg < -6\,$V (Fig.\,\ref{fig:3_3_Electrostatic_sims}\,{\bfE} and {\bfF}). As hole doping starts, only a narrow approximately $60$\,nm wide neutral region remains, located at the edge of the TG and flanked by a steep increase in charge density. This is the p-i-n regime. As a consequence, the in-plane field distribution also becomes concentrated in this narrow region due to screening in the neighboring charged areas. Lowering $\vtg$ further pushes the neutral junction region further away from the TG, thereby making the electric field distribution increasingly symmetric. Ultimately, at $\vtg = -8$\,V we obtain a sizeable in-plane electric field, with a maximum of $|F_\mathrm{x}| \sim 40$\,V/$\mu$m.

\sloppy From these quantities we determine the total excitonic confining potential as outlined in Eqn.\,\ref{eqn:potential}. The dc Stark shift contribution is computed by assuming an exciton polarizability $\alpha = 6.5\,\mathrm{eV\,nm}^2/\mathrm{V}^2$ \cite{Cavalcante2018}. The repulsive polaron shift is determined by empirically extracting an effective exciton-electron coupling strength $\beta \simeq 0.7\,\mu\mathrm{eV}\mu\mathrm{m^2}$ from our experimental data. It corresponds to the slope of a linear function fitted to the density-dependent blueshift of the repulsive polaron in the reflectance data, shown in section \ref{chap:1D:sec:signatures}, Fig.\,\ref{fig:3_5_WL}.

The resulting potential experienced by the exciton in its COM frame for various $\vtg$ is depicted in the lower panel of each subplot in Fig.\,\ref{fig:3_3_Electrostatic_sims}. Additionally, we also show the calculated discrete eigenstates associated with the confining potential obtained by numerically solving the Schr\"{o}dinger equation in the COM frame \cite{Harrison2016}. A potential well starts to form only at $\vtg = -4$\,V, but is not strong enough to lead to discernible quantization (Fig.\,\ref{fig:3_3_Electrostatic_sims}\,{\bfC}). However, the 2D excitonic state may exhibit a small redshift. At $\vtg < -4$\,V the Stark shift contribution starts to become significant and leads to the formation of a narrower potential well localized in close proximity to the TG edge.  Here, in the i-i-n regime, the continuum is given by the 2D free exciton energy $E_\mathrm{X,2D}$ and the confinement is solely driven by the dc Stark shift. As the gate voltage is lowered further to $\vtg < -6$\,V, we enter the p-i-n regime (Fig.\,\ref{fig:3_3_Electrostatic_sims}\,{\bfE} and {\bfF}), where the continuum is solely determined by the hole or electron repulsive polaron (RP$^\pm$) energy, depending on whether the electron side or hole side has lower density. This leads to the striking observation of quantized states above the free exciton energy $E_\mathrm{X,2D}$.

%%%%%%%%%%%%%%%%%%%%%%%%%%%%%%%%%%%%%%%%%%%%%%%%%%%%%%%%%%%%%%%%%%%%%%%%
\section{Device fabrication and experimental setup}
\label{chap:1D:sec:devices}
%%%%%%%%%%%%%%%%%%%%%%%%%%%%%%%%%%%%%%%%%%%%%%%%%%%%%%%%%%%%%%%%%%%%%%%%

\begin{figure}[htbp]
    \centering
	\includegraphics[width=\textwidth]{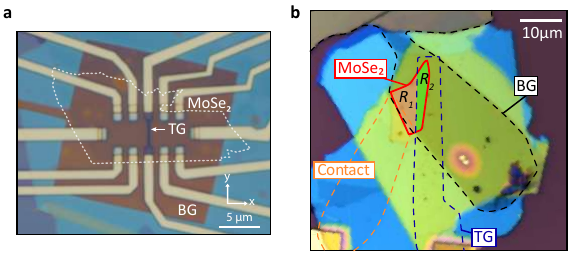}
	\caption{\textbf{Optical micrograph of devices.}
    (\bfA) Device 1, in which the white dashed line indicates the monolayer MoSe$_2$ flake. The split top gate electrode TG appears purple in the micrograph. The pre-patterned bottom gate BG is square-shaped. The remaining leads are contact electrodes to the monolayer MoSe$_2$. The scale bar denotes $5\,\mu$m. (\bfB) Device 2, in which the outline of the MoSe$_2$ monolayer is indicated by the solid red line. The TG and BG, made of few-layer graphene, are indicated by the dashed blue and black lines, respectively. R$_1$ indicates the single-gated, and R$_2$ the dual-gated region of the device. The contact to the monolayer MoSe$_2$ is formed by few-layer graphene as well, and indicated with an orange dashed line. The scale bar denotes $10\,\mu$m.
    }  
	\label{fig:3_4_Devices}
\end{figure}

We fabricate two van der Waals heterostructure devices, which we refer to as Device 1 and Device 2 (Fig.\,\ref{fig:3_4_Devices}). All the necessary flakes are obtained through mechanical exfoliation of bulk crystals and stacked into a heterostructure in an inert Ar atmosphere inside a glovebox using a standard dry polymer transfer technique \cite{Zomer2014,Pizzocchero2016}. Both devices consist of a MoSe$_2$ monolayer sandwiched between h-BN slabs and gated by two electrodes. For Device 1, the MoSe$_2$ monolayer is encapsulated using approximately $30$-nm-thick h-BN and deposited on a prepatterned Ti/Au ($3$\,nm/$10$\,nm) BG. Pd/Au ($20$\,nm/$30$\,nm) contacts to the MoSe$_2$ layer are embedded in the top h-BN layer and prepared using the via--contacting method \cite{Telford2018,Jung2019,Liu2022}. Subsequently, a $200$-nm-wide split TG electrode featuring a $100$-nm gap is formed by evaporating Ti/Au ($3$\,nm/$10$\,nm). This renders the TG optically transparent and therefore allows to investigate the optical properties of the monolayer underneath, while the local charge density is being altered. Metal electrodes to all contacts and both gates are formed with Ti/Au ($5$\,nm/$85$\,nm). Device 2 has a similar architecture as Device 1. However, all gate electrodes and the contact to the MoSe$_2$ layer were made using few-layer graphene. Furthermore, the top/bottom h-BN spacer layers were chosen to be thicker ($40$\,nm and $54$\,nm, respectively). Additional fabrication details can be found in Appendix section \ref{appendix:fabrication}.

We perform our optical experiments in a confocal microscope setup. The sample is mounted on $x$-$y$-$z$ piezo-electric stages located inside a stainless steel tube, which is immersed in a liquid helium bath cryostat. The steel tube is filled with $20$\,mbar helium exchange gas to maintain a sample temperature of about $4.2$\,K. Free-space optical access to the sample is enabled through a glass window on top of the tube. To investigate the excitonic properties in our devices, we measure white-light reflectance and photoluminescence (PL) using a broadband light-emitting diode (LED) centered at $760$\,nm and a single-mode Ti:sapphire laser tuned to $720$\,nm as the excitation source, respectively. The light from the source is focused to a diffraction-limited spot through a high-numerical-aperture ($0.68$) lens. The light reflected/emitted from the sample is then collected using the same lens, separated from the incident light by a beam splitter, coupled into a single-mode fibre and imaged on a spectrometer equipped with a liquid-nitrogen-cooled charge-coupled device. For white-light measurements an excitation power of a few tens of nW and for PL a few $\mu$W is maintained. The polarization-resolved PL measurements (discussed in the next chapter) are carried out with an angle-scanning polarizer placed in the emission path.

%%%%%%%%%%%%%%%%%%%%%%%%%%%%%%%%%%%%%%%%%%%%%%%%%%%%%%%%%%%%%%%%%%%%%%%%
\section{Signatures of quantum confinement}
\label{chap:1D:sec:signatures}
%%%%%%%%%%%%%%%%%%%%%%%%%%%%%%%%%%%%%%%%%%%%%%%%%%%%%%%%%%%%%%%%%%%%%%%%

The modification of excitonic states owing to confinement can be observed in the optical response of the narrow depleted region close to the TG edge. Owing to the diffraction-limited spot size of our optical setup (about $700\,$nm), our measurements include the combined optical response of three distinct spatial regions (see Fig.\,\ref{fig:3_2_Confinement_scheme}\,\bfA): (I) the electron-doped region away from the TG that is affected only by the BG, (II) the region directly underneath the TG, and (III) the narrow region between I and II. The contribution of region I to the total optical response remains unchanged as $\vtg$ is varied. Therefore, to discern the influence of the TG alone, we measure $\vtg$-dependent spectra for fixed values of $\vbg$, and obtain the normalized differential reflectance according to: 

\begin{equation}
\frac{\Delta R}{R} = \frac{R(V_{\mathrm{TG}}) - R(V_{\mathrm{TG}} = 0)}{R(V_{\mathrm{TG}} = 0)}.
\label{eqn:drr}
\end{equation}

\begin{figure}[htbp]
    \centering
	\includegraphics[width=12.7cm]{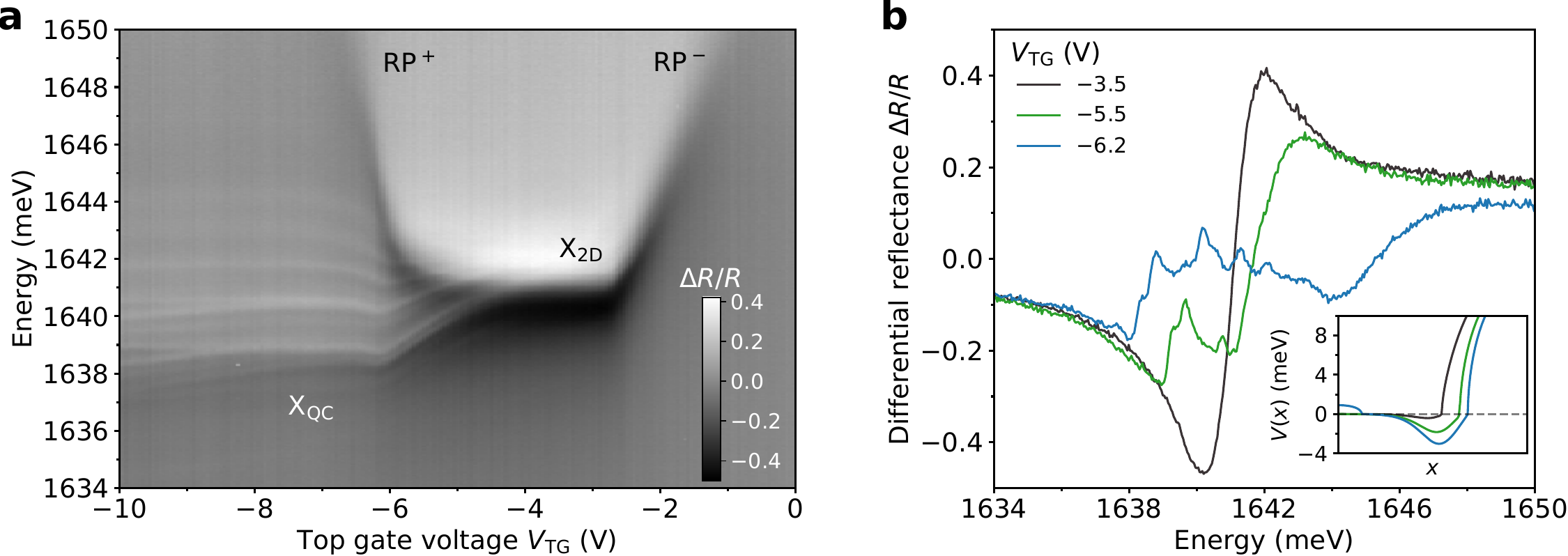}
	\caption{\textbf{Signatures of quantum confined excitons in differential reflectance.}
    (\bfA) The normalized differential reflectance $\Delta R/R$ as a function of $\vtg$, measured on the TG edge at $\vbg = 4\,$V. In addition to the typical features associated with the neutral exciton (X$_\mathrm{2D}$) and repulsive polaron branches (RP$^+$ and RP$^-$) from underneath the TG, we observe narrow discrete spectral lines that redshift as a function of $\vtg$ compared with the exciton or RP$^+$ energy at that voltage. (\bfB) Spectra at $\vtg = -3.5\,$V (black), $-5.5\,$V (green), and $-6.2\,$V (blue). The inset shows the estimated shape of the total potential $V(x)$ for excitons at the corresponding voltages (see equation (\ref{eqn:potential})).
    }  
	\label{fig:3_5_WL}
\end{figure}

In Fig.\,\ref{fig:3_5_WL}\,\bfA, we present ${\Delta R}/{R}$ as a function of $\vtg$ at fixed $\vbg = 4\,$V, which corresponds to an electron density $\sigma_\mathrm{n} = 2 \times 10^{12}\,\mathrm{cm}^{-2}$ in region I (as determined from our electrostatic simulations in section \ref{sec:1DX_sims}). First, we identify the typical doping-dependent optical response from region II directly under the TG. This includes a neutral regime ($X_\mathrm{2D}$: $-6\,\mathrm{V} \lesssim \vtg \lesssim -3\,\mathrm{V}$). As we dope the system with holes or electrons, we observe the typical density-dependent blueshift of the exciton resonance, which we refer to as the repulsive polaron branch (RP$^-$, RP$^+$). The charging behavior of the device is discussed in detail in section \ref{chap:1D:sec:alternative_configs}.

In addition to the expected optical response, we observe several narrow and discrete spectral lines for $\vtg \lesssim -4$\,V, which emerge below the 2D exciton and repulsive polaron continuum, and redshift with decreasing $\vtg$. In Fig.\,\ref{fig:3_5_WL}\,\bfB, we show representative reflectance spectra taken at $\vtg = -3.5\,$V, $-5.5\,$V and $-6.2\,$V, which highlight the appearance of new resonances with varying $\vtg$. For these voltages, estimates of the corresponding potential $V(x)$ (Eqn.\,(\ref{eqn:potential})) for excitons are shown in the inset.

We attribute the emergence of discrete lines from the 2D continuum as an unambiguous signature of  quantization of the COM motion of excitons resulting from strong confinement. Henceforth, we refer to this series of confined exciton states as $\xqc$. To perform a more comprehensive characterization of these states, we examine their reflectance line shape. While a rigorous approach typically would entail using the transfer matrix method \cite{Byrnes2016,Back2018}, a simpler alternative involves describing the measured reflectance signal as $\mathrm{Im}\left[ e^{i\zeta(E)} \chi(E)\right]$, where $\chi(E)$ is the MoSe$_2$ monolayer optical susceptibility and $\zeta(E)$ a wavelength-dependent effective phase shift \cite{Smolenski2018}. The parameter $\zeta(E)$  captures the effect of light interfering at different material interfaces in our device heterostructure (e.g.\,h-BN/Au). To first order we assume $\zeta$ to be wavelength-independent in our spectral range of interest. The reflectance spectral profile $S(E)$ associated with an optical resonance can then be modelled in the following manner:
\begin{align}
L_0(E) &= \frac{\Gamma/2}{(E-E_0)^2 + \Gamma^2/4} \\ 
L_D(E) &= \frac{E_0-E}{(E-E_0)^2 + \Gamma^2/4} \\
S(E) &= A \left( \mathrm{cos}(\zeta) L_0(E) + \mathrm{sin}(\zeta) L_D(E) \right) + C \label{eqn:spec_func}
\end{align}
where $L_0(E)$ and $L_\mathrm{D}(E)$ constitute a pure Lorentzian and a dispersive Lorentzian line shape, respectively, with $E_0$ being the center frequency and $\Gamma$ the linewidth. The parameter $A$ characterizes the overall amplitude of the resonance, while $C$ takes into account any broad background signal.

\begin{figure}[htbp]
    \centering
	\includegraphics[width=13cm]{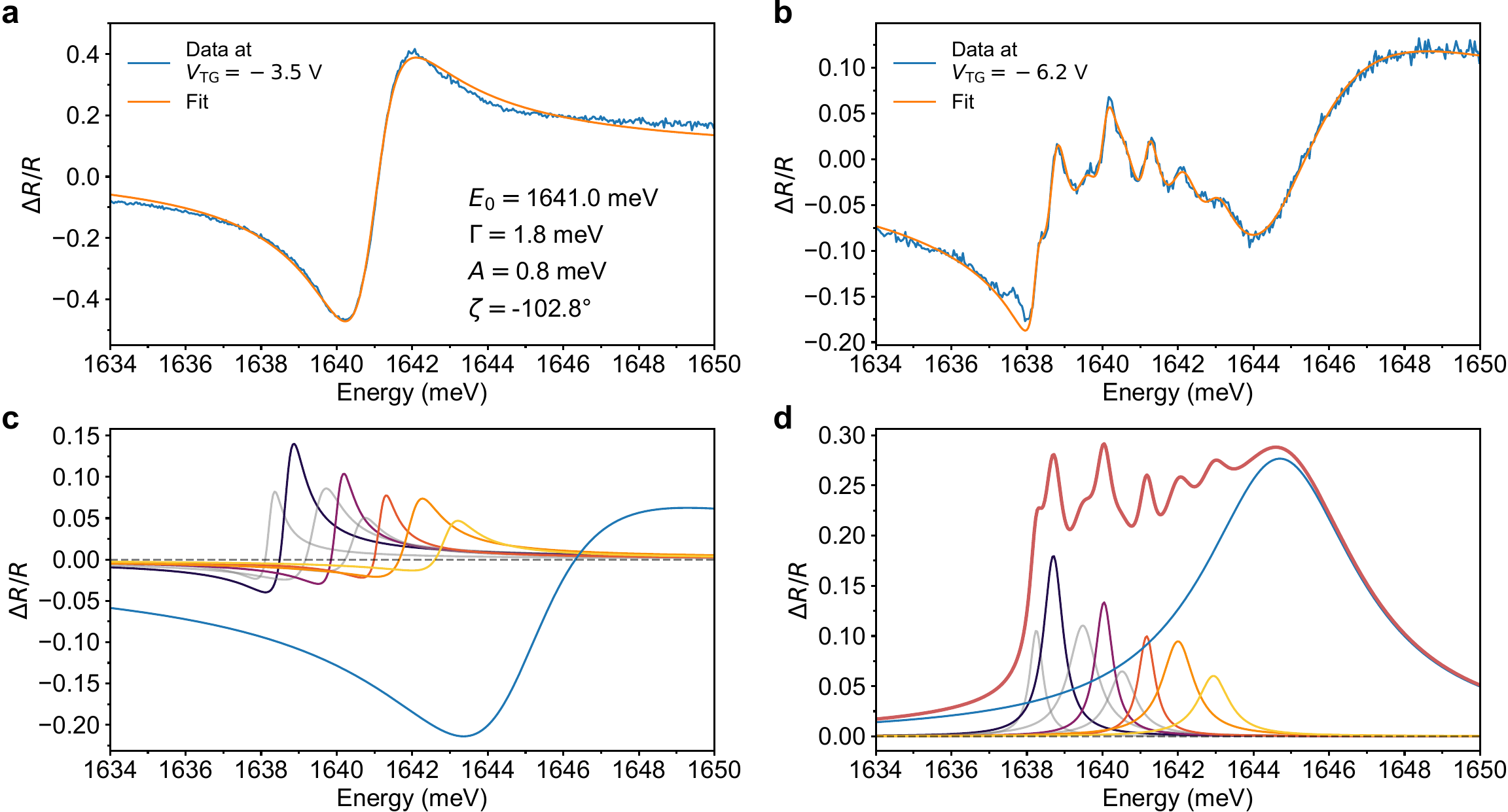}
	\caption{\textbf{Line shape analysis of reflectance data.}
    (\bfA) Spectral profile described by Eqn.\,(\ref{eqn:spec_func}) fit to the bare 2D exciton transition at $\vtg = -3.5$\,V. (\bfB) Reflectance linecut at $\vtg = -6.2$\,V fit by a superposition of multiple narrow spectral profiles	$\sum_i S(E;E_{0,i},\Gamma_i,A_i,\zeta_\mathrm{QC}) + S(E;E_\mathrm{RP},\Gamma_\mathrm{RP},A_\mathrm{RP},\zeta_\mathrm{RP})$. (\bfC) Individual components of the fit. (\bfD) Individual components after removing line asymmetry by setting $\zeta_\mathrm{QC}$ and $\zeta_\mathrm{RP}$ to $0^\circ$. The resulting overall line shape is shown in red.
    }  
	\label{fig:3_6_fits}
\end{figure}

The result of fitting this spectral profile firstly to the bare 2D exciton transition (corresponding to the line cut taken at $\vtg = -3.5$\,V in Fig.\,\ref{fig:3_5_WL}\,\bfB) is depicted in Fig.\,\ref{fig:3_6_fits}\,\bfA. When the energy splitting $\hbar\omega$ between quantized states becomes comparable to the linewidth of this resonance, we expect to observe a coherent superposition of overlapping discrete lines which are associated with individual states splitting off from a broad continuum resonance. This can be seen most clearly by considering the spectrum at $\vtg = -6.2$\,V (Fig.\,\ref{fig:3_5_WL}\,\bfB) and fitting it with a superposition of multiple narrow spectral profiles $\sum_i S(E;E_{0,i},\Gamma_i,A_i,\zeta_\mathrm{QC})$, which characterize the lines associated with quantized motional states, and a broad resonance $S(E;E_\mathrm{RP},\Gamma_\mathrm{RP},A_\mathrm{RP},\zeta_\mathrm{RP})$ that accounts for the repulsive polaron continuum. As shown in Fig.\,\ref{fig:3_6_fits}\,\bfB, a good fit of the measured data can be achieved over the whole spectral range of interest. We emphasize that during this procedure the same phase factor $\zeta_\mathrm{QC}$ is assumed for all lines associated with confined states. This can also be seen in Fig.\,\ref{fig:3_6_fits}\,\bfC, which depicts the individual components of the overall spectrum. Furthermore, the phase factor of the hole repulsive polaron resonance $\zeta_\mathrm{RP}$ is not the same as $\zeta_\mathrm{QC}$. This is justified considering that the origin of this resonance is rooted in a separate spatial region of the device, thus causing a different interference pattern. For the purpose of illustration, we also show in Fig.\,\ref{fig:3_6_fits}\,{\bfD} the resonances when the asymmetry in their line shape is removed by setting $\zeta_\mathrm{QC} = 0^\circ$ while retaining the other fit parameters. The full list of parameters obtained from the fit of this spectrum at $\vtg = -6.2\,$V are shown in table \ref{table:fit_params}.

\begin{table}[h!]
\centering
\begin{tabular}{||c|c|c|c|c||} 
 \hline
 i & $E_{0}$ (meV) & $\Gamma$ (meV) & $A$ (meV) & $\zeta$ ($^\circ$) \\ [0.5ex] 
 \hline\hline
 1  & 1638.2 & 0.42 & 0.022 & -55.5 \\ 
 2  & 1638.7 & 0.62 & 0.056 & -55.5 \\
 3  & 1639.5 & 0.91 & 0.050 & -55.5 \\
 4  & 1640.1 & 0.56 & 0.038 & -55.5 \\
 5  & 1640.5 & 0.88 & 0.028 & -55.5 \\
 6  & 1641.2 & 0.53 & 0.027 & -55.5 \\
 7  & 1642.0 & 1.02 & 0.048 & -55.5 \\
 8  & 1642.9 & 1.00 & 0.030 & -55.5 \\
 RP & 1644.7 & 4.95 & 0.685 & 236.8 \\ [1ex] 
 \hline
\end{tabular}
\caption{Fit parameters obtained by fitting $\Delta R/R$ at $\vtg = -6.2$\,V (see Fig.\,\ref{fig:3_6_fits}\,\bfB). }
\label{table:fit_params}
\end{table}

The marked decrease in the linewidth of the $\xqc$ states, in contrast to the unconfined 1s exciton $\xfree$, is further compelling evidence for the strong confinement of excitons. Whereas the $\xfree$ linewidth is typically $\Gamma_\mathrm{2D} \approx 2\,$meV (Fig.\,\ref{fig:3_6_fits}\,{\bfA}), the lowest energy $\xqc$ resonance exhibits a linewidth $\Gamma_\mathrm{QC} \gtrsim 500\,\mu$eV (table \ref{table:fit_params}). This value further decreases to approximately $300\,\mu$eV at lower $\vtg$, in which a stronger excitonic confinement is expected. Such linewidth narrowing is qualitatively expected to stem from three factors:  (i) lower inhomogeneous broadening, as exciton COM motion is restricted to a smaller area due to confinement; (ii) reduction of radiative decay of $\xqc$ as compared with their free 2D counterparts according to the ratio $\ell_\mathrm{x}/\lambda_\mathrm{photon}$, in which $\ell_x$ is the harmonic oscillator length along $x$ and $\lambda_\mathrm{photon}$ is the photon wavelength; (iii) the reduced electron--hole wave function overlap originating from the electric dipole moment induced by the in-plane electric field.

Using the above framework we can now proceed to trace the resonance center frequencies of the discrete states as a function of top gate voltage $\vtg$, and thus evolving confinement strength. However, owing to the intricate structure of our reflectance data, this fitting procedure cannot be performed in an automated way for the entire range of gate voltages. Therefore, we extract these quantities rather by selecting a narrow spectral range, which only contains the confined states, and fitting a superposition of pure Lorentzians $\sum_i L_0(E;E_{0,i},\Gamma_i,A_i)$ i.e.\,the phase factor $\zeta_\mathrm{QC}$ is set to $0^\circ$. This assumption likely introduces a systematic offset in the estimation of the resonance energy. However, this is not a big hindrance since our focus is on the evolution of transition energies as a function of $\vtg$. Furthermore, we restrict this analysis only to states which are depicted in color in Fig.\,\ref{fig:3_6_fits}\,{\bfC}. The evolution of states shown in gray is not traced, since their amplitude decays rapidly for decreasing $\vtg$.

\begin{figure}[htbp]
    \centering
	\includegraphics[width=\textwidth]{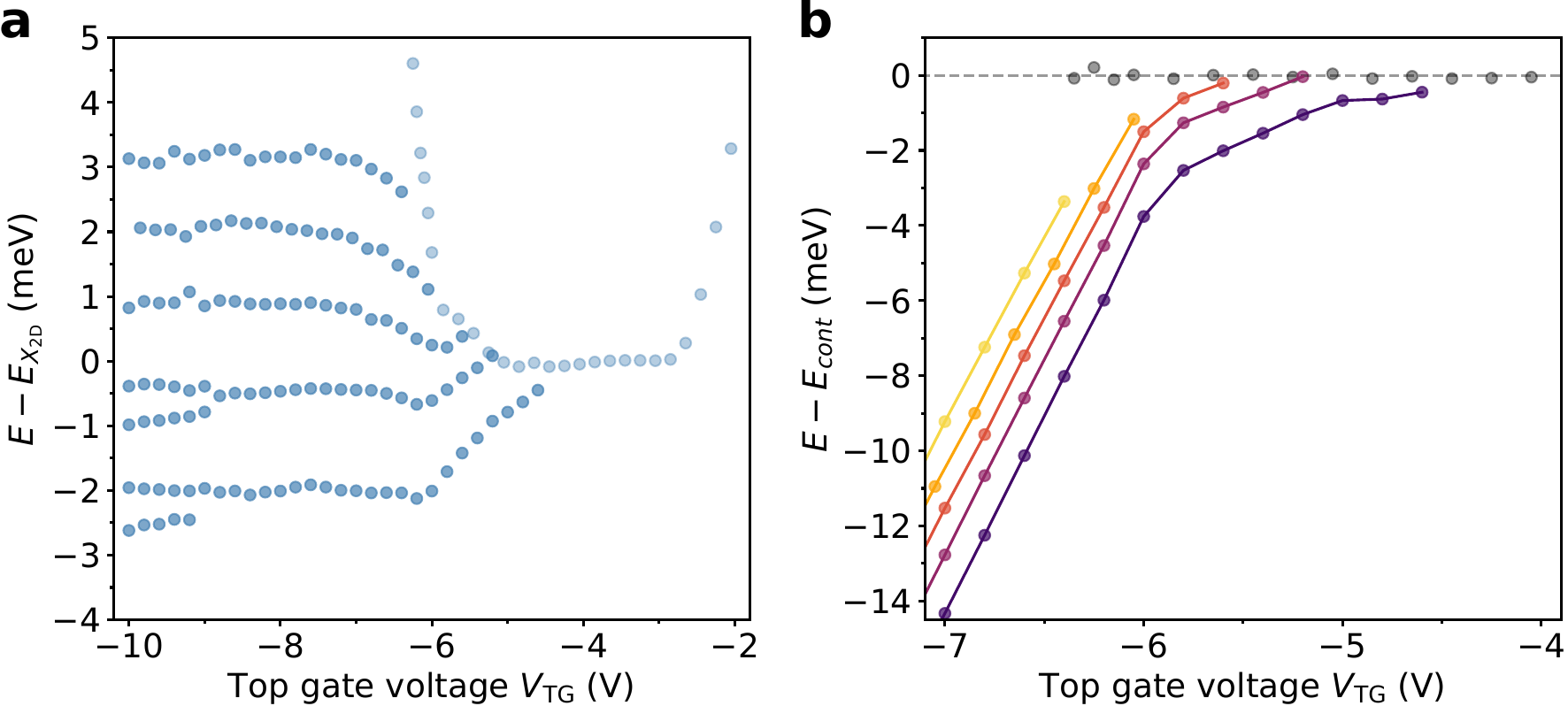}
	\caption{\textbf{Evolution of resonance energies with gate voltage $\mathbf{\vtg}$.}
    (\bfA) Resonance energies ($E$) obtained from fits of reflectance spectra with respect to the 2D exciton energy. We observe a maximum level spacing of $\hbar\omega \approx 1.5\,$meV between the lowest two confined states. This is accompanied by a narrowing of the linewidth from $\Gamma_\mathrm{2D} \approx 2\,$meV for free excitons to $\Gamma_{\mathrm{QC}} \approx 300\,\mu$eV for confined excitons. (\bfB) Energy of discrete resonances with respect to the free exciton continuum $E_{\mathrm{cont}}$. The continuum energy is the 2D exciton energy in the neutral regime, and the repulsive polaron energy in the doped regime.
    }  
	\label{fig:3_7_fit_evolution}
\end{figure}

In the voltage regime $-6\,\mathrm{V} < \vtg < -4\,\mathrm{V}$, which corresponds to the i-i-n charging configuration, we expect the redshift of the narrow resonances with decreasing $\vtg$ to arise mainly from the field-induced confinement (Fig.\,\ref{fig:3_2_Confinement_scheme}\,{\bfC}) mechanism. As can be seen in Fig.\,\ref{fig:3_7_fit_evolution}\,{\bfA}, here the dc Stark shift lowers the energy below the continuum given by the 2D exciton energy $E_\mathrm{X,2D}$. Furthermore, in the p-i-n regime ($\vtg < -6\,$V), as region II becomes hole-doped, we observe further quantized modes with energy higher than the $E_\mathrm{X,2D}$ that split off from the repulsive polaron branch RP$^+$. In this regime, the confinement potential is the sum of the field-induced and interaction-induced contributions. Therefore, the free-particle continuum is no longer the 2D exciton state but the blueshifted repulsive polaron (RP$^+$) in region II. In other words, to escape the confinement, an exciton in the lowest state must pay not only the dc Stark energy shift, but a further repulsive polaron energy, $E = \beta \cdot \sigma_\mathrm{max}$, in which $\sigma_\mathrm{max}$ denotes the maximum charge density. The energy of confined resonances with respect to the continuum energy ($E_\mathrm{cont}$) at each $\vtg$ is shown in Fig.\,\ref{fig:3_7_fit_evolution}\,{\bfB}. This clearly demonstrates the successive emergence of discrete states below the free-particle continuum as the potential is made deeper. Concurrently, we observe a broadening and loss of oscillator strength for the higher-lying discrete states. 

\begin{figure}[htbp]
    \centering
	\includegraphics[width=8.5cm]{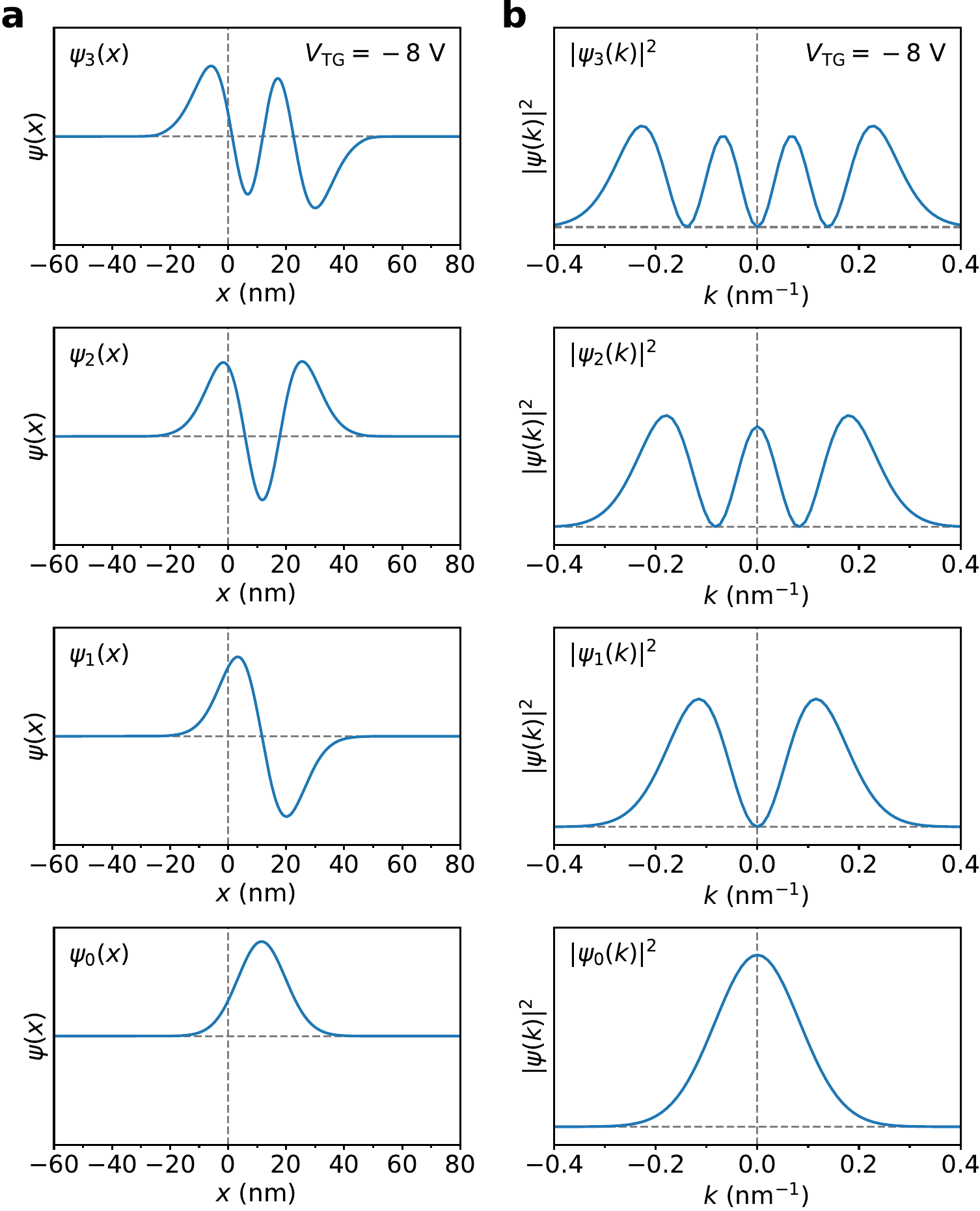}
	\caption{\textbf{Eigenstates of the confining potential.}
    Exciton center-of-mass wave functions corresponding to the energetically lowest four states in the confining potential realized at $\vtg=-8$\,V and $\vbg=4$\,V (Fig.\,\ref{fig:3_3_Electrostatic_sims}\,{\bfF}) in real- (\bfA) and k-space (\bfB) coordinates. In real space, the center of the wave function is offset away from $x=0$\,nm since the potential minimum occurs at finite values of $x$.
    }  
	\label{fig:3_8_wfs}
\end{figure}

In addition to the features mentioned above, we also observe that the level separations deduced from the reflectance spectra match well with those obtained from the electrostatic simulations of the device. An energy separation $\Delta E$ of around $1.5\,$meV between the lowest two resonances at $\vtg = -8\,$V is observed, suggesting a harmonic oscillator confinement length $\ell_x$ of approximately $6\,$nm. This value can be contrasted with the confinement length scale extracted from the numerically determined eigenfunctions of the confinement potential. Fig.\,\ref{fig:3_8_wfs} depicts the exciton COM wave functions associated with the energetically lowest four states of the potential in Fig.\,\ref{fig:3_3_Electrostatic_sims}\,{\bfF}. They are offset away from $x=0$\,nm since the electric field maximum and thus the potential minimum occur at finite $x$. Furthermore, they exhibit a similar shape as those of a quantum harmonic oscillator, such that the region around the minimum of the confining potential in Fig.\,\ref{fig:3_3_Electrostatic_sims}\,{\bfF} can be approximated by a harmonic trap $V(x)=kx^2$ with an effective spring constant $k \sim 10^6\,$keV\,cm$^{-2}$. In accordance with this value, we estimate a confinement length from the spatial extent of $\psi_0(x)$ (lowest panel in Fig.\,\ref{fig:3_8_wfs}\,\bfA) of $\ell_x \approx 8$\,nm, given by the relation $\mathrm{FWHM}/2\sqrt{2 \ln(2)}$, where FWHM is the full width at half maximum of the wave function.

\begin{figure}[htbp]
    \centering
	\includegraphics[width=7cm]{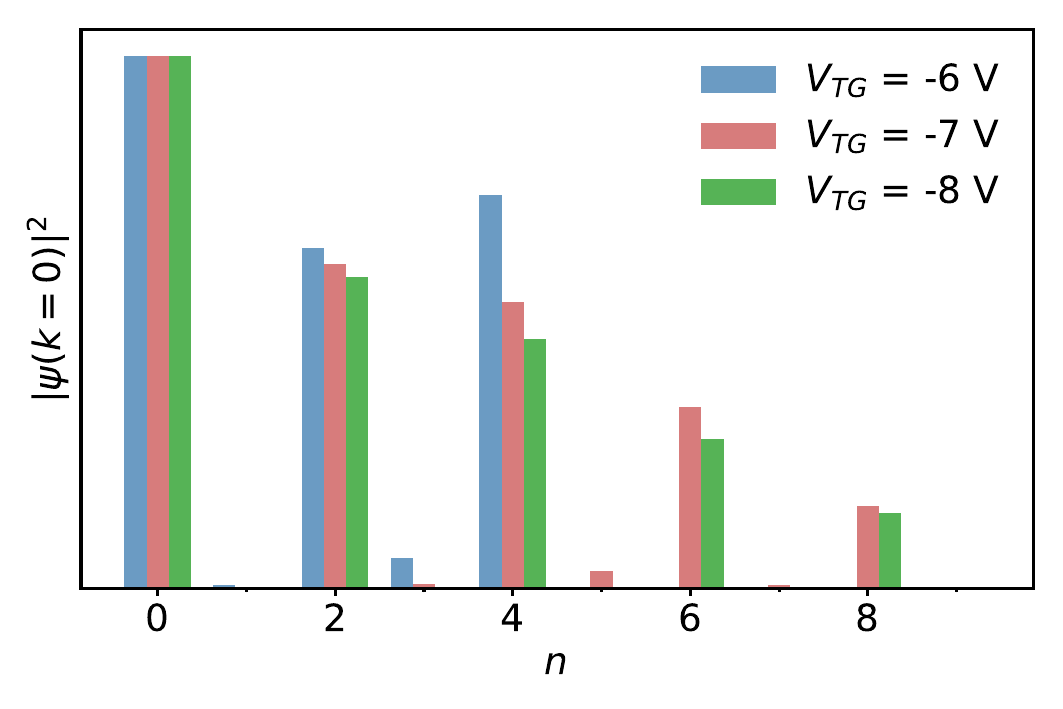}
	\caption{\textbf{Zero-COM-content of the eigenstates in the confining potential as a function of the quantum number $\boldsymbol{n}$ and for different applied gate voltages $\boldsymbol{\vtg}$.}
    Odd states have a negligible value at $k=0$, and thus should not lead to emission. For even states the zero-momentum content decreases with increasing $n$. Hence, the oscillator strength of higher lying confined states should be smaller.
    }  
	\label{fig:3_9_parity}
\end{figure}

The structure of these wave functions also allows to understand the observation that the number of predicted bound states determined from our simulations typically exceeds the number of discrete resonances seen experimentally in the differential reflection spectra. We hypothesize that the origin of this behavior could be rooted in the parity of these wave functions. This can be seen most clearly by Fourier-transforming the real-space probability amplitudes, through which we obtain the corresponding probability densities in k-space (Fig.\,\ref{fig:3_8_wfs}\,\bfB). Since the radiative coupling of a particular state should be given by its zero-COM-momentum content, which becomes negligible for odd states (Fig.\,\ref{fig:3_9_parity}), they should not lead to radiative emission. Hence, only the even states should have finite oscillator strength, which still decreases with increasing quantum number $n$. On a qualitative level, this trend is also observed in our experimental data.

\begin{figure}[htb]
    \centering
	\includegraphics[width=12cm]{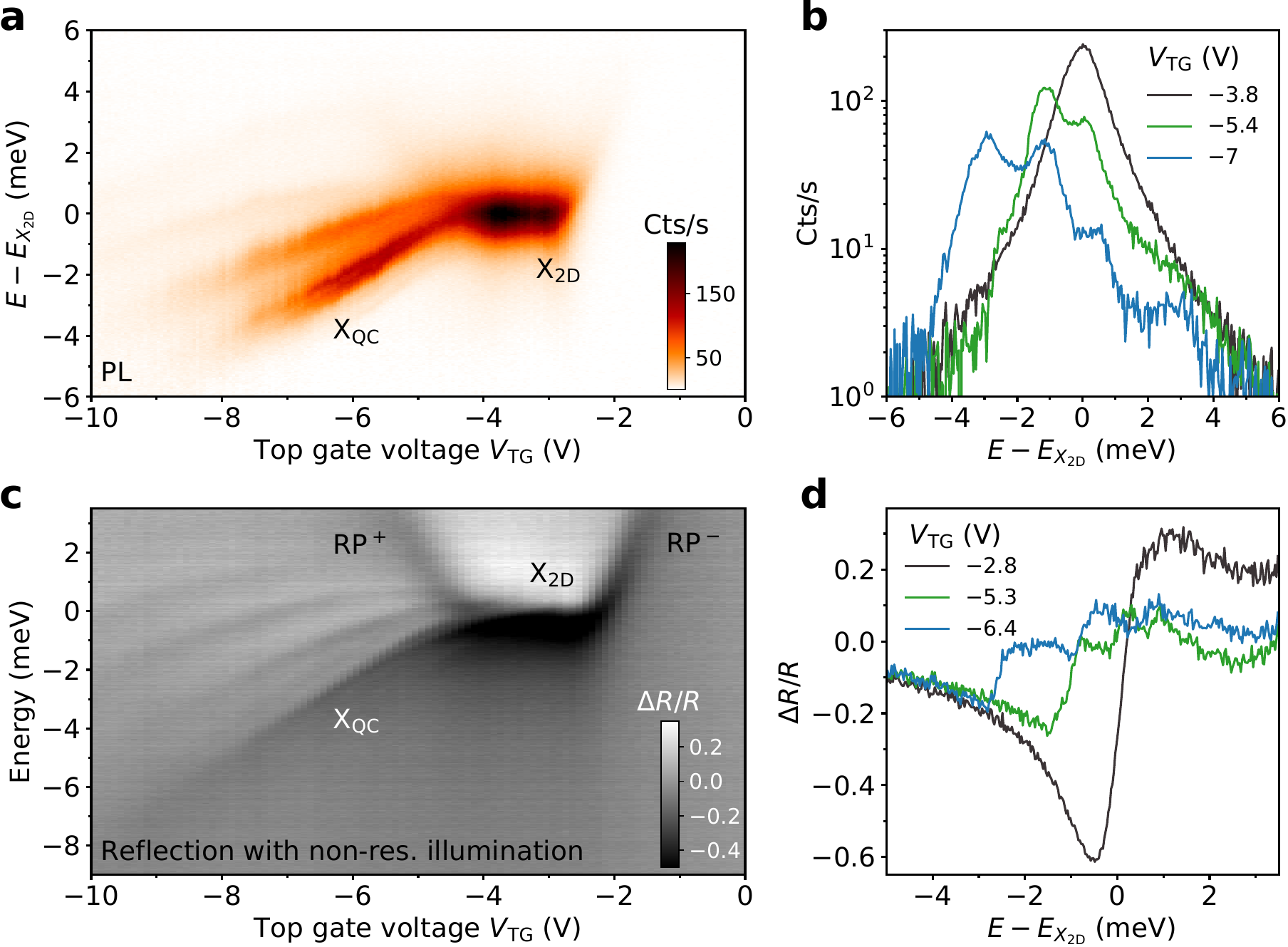}
	\caption{\textbf{Signatures of quantum confined excitons under non-resonant illumination.}
    (\bfA) Photoluminescence spectrum measured on the TG edge at $\vbg = 4\,$V as a function of $\vtg$ shows discrete red shifting emission lines associated with quantum confined excitons $\xqc$. Their emission energy $E$ is depicted relative to the resonance energy of the 2D exciton ($\xfree$). (\bfB) PL spectra on a logarithmic scale at $\vtg=-3.8\,$V (black), $-5.4\,$V (green), and $-7\,$V (blue). At the lowest $\vtg$ four resonances can be distinguished, two of which appear higher in energy than the 2D exciton energy $E_\mathrm{X,2D}$. (\bfC) Normalized differential reflectance $\Delta R/R$ acquired on the TG edge at $\vbg = 4\,$V as a function of $\vtg$, under simultaneous non-resonant PL illumination. In contrast to the reflectance data shown in Fig.\,\ref{fig:3_5_WL}, the $\xqc$ resonances evolve with decreasing $\vtg$ in a similar manner as in the PL measurement. (\bfD) Reflectance spectra at $\vtg=-2.8\,$V (black), $-5.3\,$V (green), and $-6.4\,$V (blue).
    }  
	\label{fig:3_10_PL_WL}
\end{figure}

In addition to characterizing the optical response in the vicinity of the TG edge by means of differential reflectance (Fig.\,\ref{fig:3_5_WL}), we also measure PL while varying $\vtg$ and keeping $\vbg$ fixed at $4$\,V (Fig.\,\ref{fig:3_10_PL_WL} {\bfA}). Owing to the heavy electron-doping in the bottom-gated area (region I) away from the TG, no optical signature associated with this region is observed in the plotted spectral range. Similar to the reflectance data, the existence of a prolonged exciton resonance ($\xfree$: $-4.5\,\mathrm{V} \lesssim \vtg \lesssim -3\,\mathrm{V}$) indicates charge neutrality in the dual-gated region II. A further decrease in $\vtg$ gives rise to multiple narrow resonances $\xqc$ which redshift monotonously and originate in the intermediate region III. In Fig.\,\ref{fig:3_10_PL_WL} {\bfB} we show individual PL spectra at fixed $\vtg$, showing how a single exciton resonance $\xfree$ at $\vtg = -3.8$\,V (black curve) evolves into four distinct resonances at $\vtg = -7$\,V (blue curve). Furthermore, while the emission intensity of these states decreases with increasing energy, by plotting the PL spectra on a logarithmic scale we can clearly discern the existence of two $\xqc$ resonances at an energy higher than the 2D exciton energy $E_\mathrm{X,2D}$. This observation further substantiates our claim of a joint interaction-induced and field-induced confinement mechanism acting together.

However, a direct comparison of the reflectance data (Fig.\,\ref{fig:3_5_WL}) with the PL measurement (Fig.\,\ref{fig:3_10_PL_WL}\,{\bfA} and {\bfB}) reveals a pronounced discrepancy in the voltage dependence of confined exciton resonances for $\vtg \lesssim -6\,$V. In the regime $-6\,\mathrm{V} < \vtg < -4\,\mathrm{V}$, where the area under the TG is still mostly insulating, the reflectance and PL lines shift with the same slope. This changes at the onset of strong hole-doping under the TG at $\vtg \approx -6\,$V, which is evidenced by a change in the slope of the repulsive polaron resonance in Fig.\,\ref{fig:3_5_WL} \bfA. Whereas the PL lines continue to redshift almost linearly with $\vtg$, the reflectance lines show an abrupt change in slope particularly at the lowest energies. To elucidate the origin of this discrepancy, we probe differential reflectance at the same optical spot under simultaneous illumination by a broadband LED and a non-resonant PL laser. As depicted in Fig.\,\ref{fig:3_10_PL_WL} {\bfC} and {\bfD}, in this case the $\xqc$ resonances evolve with decreasing $\vtg$ in a similar manner as in the PL measurement. We suspect that due to the much higher excitation intensity of the PL illumination laser (few $\mu$W) as compared to the broadband LED (few nW), optically-induced doping effects \cite{Ju2014,Seo2020} can play a significant role in determining the actual electrostatic landscape in such device architectures. As the excitation intensity is increased, excess charges might be introduced in the p-doped and n-doped regions, leading to an overall modification of the confinement potential and thus altering the evolution of the $\xqc$ resonances with gate voltage. This suggests that optical excitation can act as an additional control parameter for defining the precise shape of the confinement potential, an interesting feature which could be exploited in future studies.

%%%%%%%%%%%%%%%%%%%%%%%%%%%%%%%%%%%%%%%%%%%%%%%%%%%%%%%%%%%%%%%%%%%%%%%%
\section{Quantum confinement in alternative configurations}
\label{chap:1D:sec:alternative_configs}
%%%%%%%%%%%%%%%%%%%%%%%%%%%%%%%%%%%%%%%%%%%%%%%%%%%%%%%%%%%%%%%%%%%%%%%%

\begin{figure}[htb]
    \centering
	\includegraphics[width=14cm]{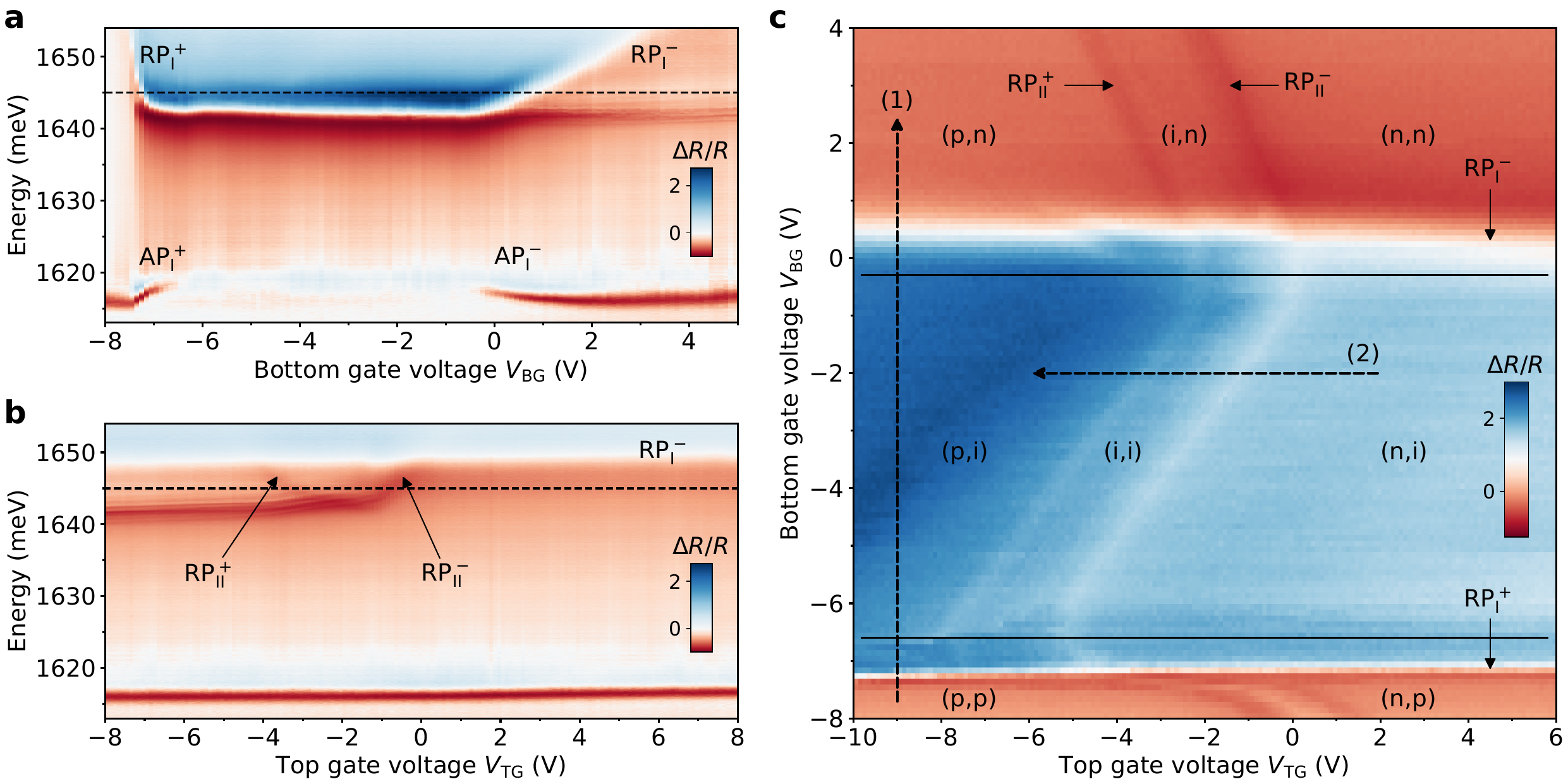}
	\caption{\textbf{Doping characteristics of Device 1.}
    We use the charge-density-dependent energy shifts of the repulsive polaron states to obtain the doping configuration of the device. Because the RP states exist in regions I and II of the device and on electron and hole sides, we use the notation RP$^\mathrm{charge}_\mathrm{region}$ to denote the different states. (\bfA) Normalized reflectance $\Delta R/R$ measured for fixed $\vtg = -6\,$V as a function of $\vbg$, in which we mainly observe features from region I. (\bfB) $\Delta R/R$ measured for fixed $\vbg = 1.8\,$V as a function of $\vtg$, in which we observe the spectral changes in region II. In addition to the features from regions II, we see the repulsive polaron resonance from region I at $E \approx 1648\,$meV, which is not affected by $\vtg$. (\bfC) Normalized reflectance taken at a fixed energy $E_\mathrm{X,2D} + \Gamma_\mathrm{2D}/2 = 1645\,$meV, in which $\Gamma_\mathrm{2D}$ is the bare 2D exciton linewidth. The horizontal solid lines demarcate the n-doped, neutral and p-doped regimes in region I. The doping configurations in regions I and II in different voltage regimes are identified. For example, in the top left corner, which we have mainly focused our attention on so far, the doping configuration is $(\mathrm{II, I}) \equiv (\mathrm{p,n})$.
    }  
	\label{fig:3_11_Doping}
\end{figure}

So far we restricted our analysis of the quantum confined exciton states $\xqc$ to a very specific set of gate voltages ($\vbg = 4$\,V, $-10\,\mathrm{V} < \vtg < 0\,\mathrm{V}$) in Device 1. However, the existence of such states should be guaranteed also at other gate voltage settings -- as long as inhomogeneous in-plane electric fields persist in the device. To verify this claim, we explore the charging behaviour of Device 1 to determine all the different doping configurations that can be achieved and identify in this manner the specific set of gate voltages, which will allow creating inhomogeneous in-plane electric fields. To this end, we again turn to measuring normalized differential reflectance $\Delta R/R$ in the vicinity of the TG edge and perform a dual-gate $\vbg$-$\vtg$ scan. Owing to the diffraction-limited spot size, we are capable of simultaneously measuring the combined optical response of the three distinct regions of the device: the bottom-gated region I, the dual-gated region II and the intermediate region III (Fig.\,\ref{fig:3_2_Confinement_scheme} \bfA). We note that the energy of the repulsive polaron state is a sensitive detector of the local charge density. Therefore, by measuring the reflectance at fixed energy $E = E_\mathrm{X,2D} + \Gamma_\mathrm{2D}/2 =  1645\,$meV (dashed line in Fig.\,\ref{fig:3_11_Doping} {\bfA} and {\bfB}) we can precisely deduce the device doping configuration.

In Fig.\,\ref{fig:3_11_Doping} {\bfA}, we show the reflectance spectra as a function of $\vbg$ (for fixed $\vtg$), which shows the typical charging behavior that exhibits neutral exciton, repulsive and attractive polaron branches on the n-doping and p-doping regimes in region I. Moving from right to left along the horizontal dashed line takes us from the n-doped to insulating to p-doped regimes in region I. Similarly, in Fig.\,\ref{fig:3_11_Doping} {\bfB}, we show the reflectance as a function of $\vtg$ (for fixed $\vbg$), which also exhibits the neutral exciton and polaron branches. Once again, moving from right to left along the dashed line takes us from n-doped to insulating to p-doped regimes in region II. We note that, in both cases, as we move along the dashed lines, we observe two repulsive states: RP$^-$ and RP$^+$. These resonances demarcate regions of electron and hole doping. On the right of RP$^-$, the corresponding region is always electron-doped, whereas on the left of RP$^+$, the region is always hole-doped. In between, the region is in the neutral regime.

Using this method, we can monitor the charge configurations in both regions I and II by measuring the reflectance at fixed energy as a function of $\vtg$ and $\vbg$, as shown in Fig.\,\ref{fig:3_11_Doping} {\bfC}. Following this procedure, we observe that the RP$^+$ and RP$^-$ resonance exhibit a zigzag shape in the $\vbg$-$\vtg$ plane. As the energy of the repulsive polaron state is proportional to charge density, moving on the zigzag line amounts to moving along a path of constant charge density. To identify the charging configurations, we use the notation (II,I). For example, (p, n) refers to p-doping in region II and n-doping in region I. In general, moving vertically at fixed $\vtg$ tunes the doping in region I. For example, moving along path 1 takes the doping configuration from (p,p) to (p,i) to (p,n). Similarly, moving horizontally while keeping $\vbg$ fixed tunes the doping in region II. Hence, moving along path 2 takes the doping configuration from (n,i) to (i,i) to (p,i).

We observe in Fig.\,\ref{fig:3_11_Doping} {\bfC} that all charging configurations are possible in our device. This is surprising considering that the electrical contact to the TMD monolayer exists only in region I and not in region II. Therefore, any charges that have to reach region II from the contacts (in region I) have to necessarily pass through region I in an equilibrium setting. For example, in the top left corner we observe the p-n regime, which we have mainly focused our attention on so far. In this voltage range, it is energetically unfavorable for holes to reach region II by traversing region I, which is n-doped. To explain this, we consider a charge injection mechanism based on optical doping, which is a local effect.

When region I is neutral, a voltage bias between TG and BG results in electric fields ($F \propto \vbg - \vtg$) which can lead to a trapping potential for charges under the TG. However, because the Fermi energy is in the gap, in thermodynamic equilibrium, it is not energetically favorable for electrons or holes to populate this trap. Nevertheless, on optical excitation, excitons can dissociate under the strong in-plane electric field generated at the edges of the TG. Thus, individual charge carriers are left behind which are accelerated towards region II, in which they ultimately get trapped. On the other hand, the generated field between TG and BG also leads to a finite rate of charges tunnelling out of the TMD in the out-of-plane direction. We suspect that this interplay between charge injection through exciton dissociation and charge removal by means of leakage currents defines the doping configuration in our device. The impact of both mechanisms is influenced by the electric field magnitude. Correspondingly, the observed charge density in region II also scales linearly, $\sigma \propto \vbg - \vtg$.

The doping situation is different when the device is globally n-doped or p-doped. In order to reach charge neutrality, $\vtg$ now needs to be tuned to a voltage that removes the charges under the TG, therefore at the equal density line $\sigma \propto \vtg + \vbg$. We expect that, in this scenario, doping of opposite charge carriers in region II also arises from the aforementioned optical doping mechanism. The validity of this claim can be further demonstrated by making use of the fact that the TG in Device 1 can also serve as a quantum point contact. By monitoring the TG voltage necessary to pinch off the source-drain current under different conditions, it is possible to deduce that without any optical illumination charges of opposite polarity cannot be injected in the region under the TG (see Appendix section \ref{appendix:qpc}).

\begin{figure}[htb]
    \centering
	\includegraphics[width=13cm]{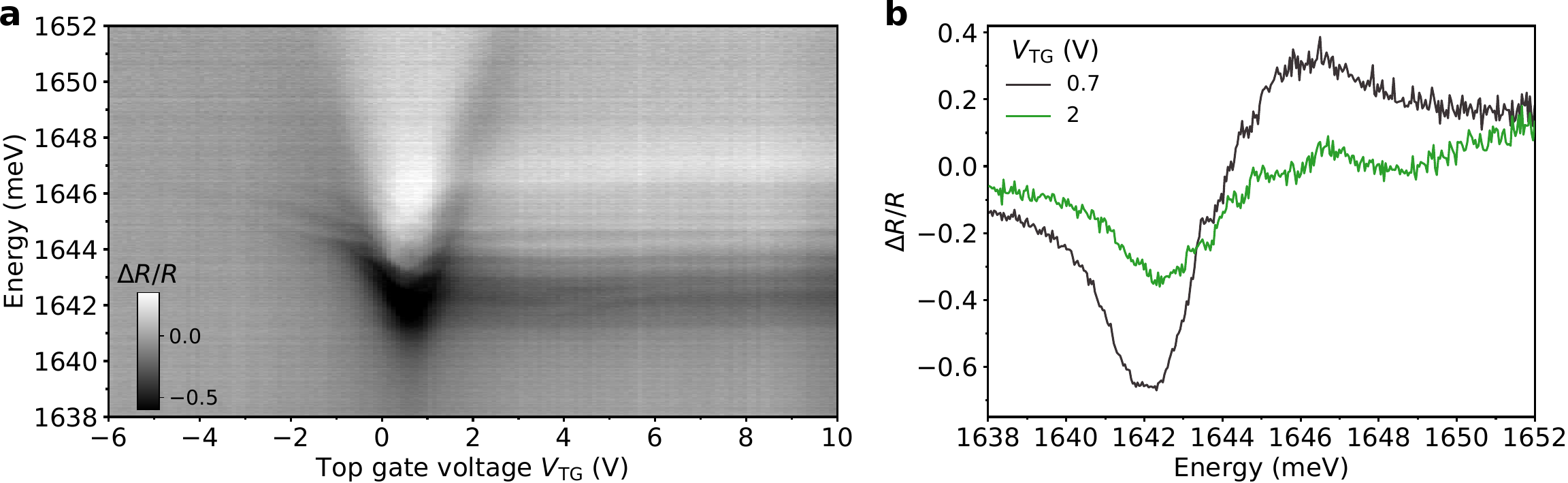}
	\caption{\textbf{Quantum confinement in the n-i-p regime.}
    (\bfA) Normalized reflectance $\Delta R/R$ from Device 2 as a function of $\vtg$ for fixed $\vbg = -8\,$V, corresponding to hole doping in region I. We observe qualitatively similar signatures of quantum confinement as the p-i-n regime, which includes narrow discrete states emerging out of the repulsive polaron (RP$^-$) continuum for $\vtg \gtrsim 0.5\,$V. (\bfB) Reflectance spectra at $\vtg=0.7\,$V (black) and $2\,$V (green).
    }  
	\label{fig:3_12_n-i-p}
\end{figure}

Having mainly focused on the p-i-n regime of Device 1 so far, based on Fig.\ref{fig:3_11_Doping} {\bfC}, we note that the n-i-p regime can also be reached, in which region I is p-doped and region II is n-doped. In this setting the mechanism of quantum confinement in the i-region should work in the same way. In Fig.\,\ref{fig:3_12_n-i-p}, we show the normalized reflectance measured as a function of $\vtg$ for fixed $\vbg = -8\,$V. We observe similar qualitative signatures of quantum confinement, i.e.\,narrow discrete lines emerging out of the repulsive polaron continuum. As expected, these lines now appear at positive $\vtg$. However, we do observe conspicuous quantitative differences between the p-i-n and n-i-p settings. We find that a prolonged neutral region, that is seen in Fig.\,\ref{fig:3_5_WL} \bfA, is absent in the data shown in Fig.\,\ref{fig:3_12_n-i-p}. Because of this we suspect that the discrete resonances in the n-i-p scenario do not exhibit the sharp initial redshift as observed in the voltage range $-6\,\mathrm{V} < \vtg < -4\,\mathrm{V}$ in Fig.\,\ref{fig:3_5_WL} \bfA.

We suspect that the origin of this discrepancy could be rooted in the interplay of optical doping effects and leakage currents in our devices. Due to the small valence band offset between h-BN and MoSe$_2$ the tunnelling rate for holes is expected to be considerably larger than for electrons, causing a substantially larger leakage current for the former. This gives rise to a long lifetime of trapped electrons. Hence, while increasing $\vtg$, a non-equilibrium electron population in region II can be sustained immediately after the hole population has been fully depleted.

\begin{figure}[htb]
    \centering
	\includegraphics[width=13cm]{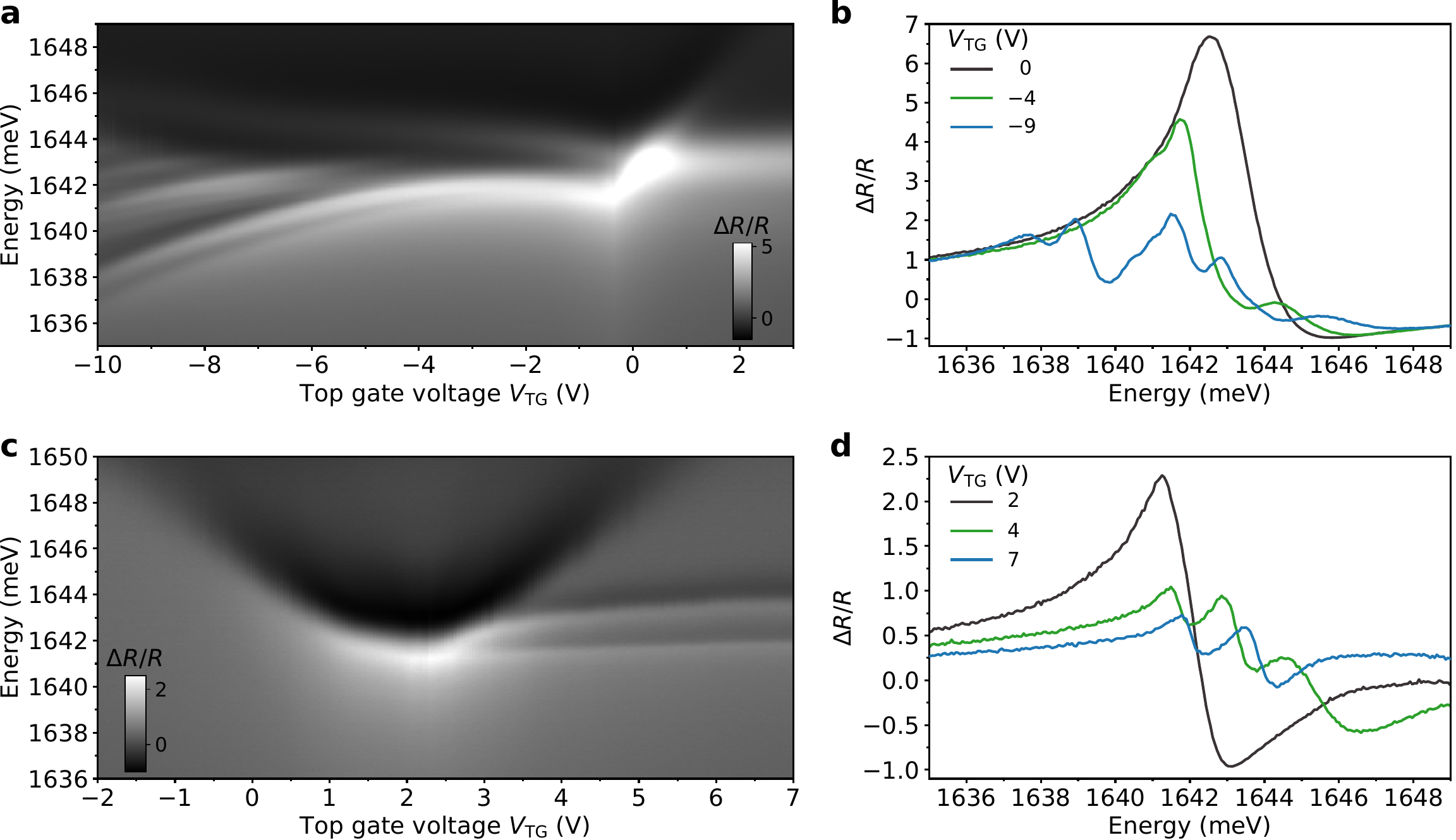}
	\caption{\textbf{Signatures of quantum confined excitons in differential reflectance in Device 2.}
    (\bfA) Normalized differential reflectance $\Delta R/R$ acquired on the edge of the few-layer graphene TG at $\vbg = 1\,$V as a function of $\vtg$, in full agreement with the observations in Device 1. For $\vtg\lesssim-2\,$V, a p-i-n doping configuration is realized. (\bfB) The corresponding reflectance spectra at $\vtg=0\,$V (black), $-4\,$V (green), and $-9\,$V (blue). (\bfC) $\Delta R/R$ acquired on the TG edge at $\vbg = -7\,$V as a function of $\vtg$. For $\vtg\gtrsim2\,$V, a n-i-p doping configuration is realized. (\bfD) The corresponding reflectance spectra at $\vtg=2\,$V (black), $4\,$V (green), and $7\,$V (blue).
    }  
	\label{fig:3_13_WL_Dev2}
\end{figure}

In Device 1 we realized tunable exciton quantum confinement using top and bottom gates that are defined using electron beam lithography and subsequent gold evaporation. However, given that our confinement method solely relies on a device structure featuring gates with partial spatial overlap, in principle, we can completely forego the need for nanoscale lithographic patterning. Therefore, to demonstrate the simplicity of our confinement scheme we fabricate a second device, Device 2 (Fig.\,\ref{fig:3_4_Devices} \bfB), in which all gate electrodes are made of few-layer graphene (see section \ref{chap:1D:sec:devices}). As such, if pre-patterned contact electrodes are utilized, fabrication of the entire device only consists of stacking the respective materials into a van der Waals heterostructure. Analogous to Figs.\,\ref{fig:3_10_PL_WL} {\bfC} and \ref{fig:3_12_n-i-p}, we depict in Fig.\,\ref{fig:3_13_WL_Dev2} the normalized differential reflectance from Device 2, measured on the edge of its TG. Qualitatively, similar signatures are observed, both in the p-i-n and n-i-p doping configurations, further emphasizing the versatility of our method.

  %%%%%%%%%%%%%%%%%%%%%%%%%%%%%%%%%%%%%%%%%%%%%%%%%%%%%%%%%%%%%%%%%%%%%%%%
\chapter[Center-of-mass exciton wave function]{Center-of-mass exciton wave function\footnote{This chapter is adapted from the following publication \cite{Thureja2022}:\\ \\
\textbf{Electrically tunable quantum confinement of neutral excitons}\\
Thureja, D.; Imamoglu, A.; Smole\'{n}ski, T.; Amelio, I.; Popert, A.; Chervy, T.; Lu, X.; Liu, S.; Barmak, K.; \mbox{Watanabe}, K.; Taniguchi, T.; Norris, D. J.; Kroner, M.; Murthy, P. A.\\
\textit{Nature}\textbf{ 606}, 298 (2022)
}}
\label{chap:COM}
%%%%%%%%%%%%%%%%%%%%%%%%%%%%%%%%%%%%%%%%%%%%%%%%%%%%%%%%%%%%%%%%%%%%%%%%

\begin{figure}[htb]
    \centering
	\includegraphics[width=12cm]{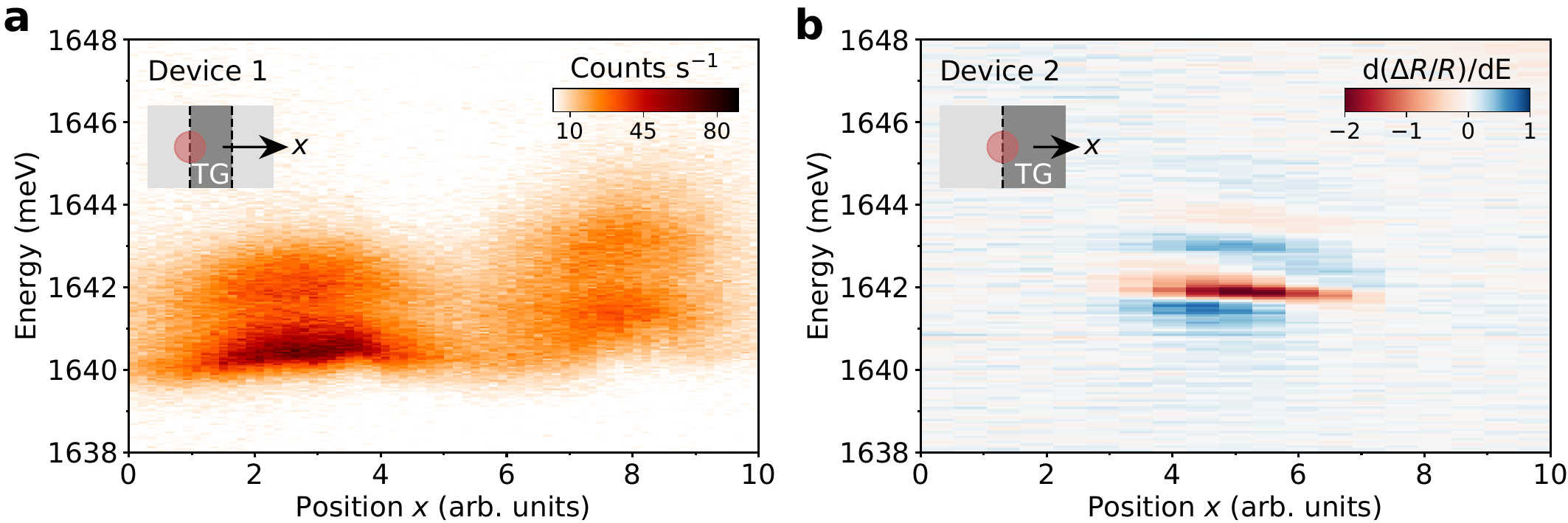}
	\caption{\textbf{Position dependence of 1D quantum confined excitons.}
    Position-dependent spectra as the diffraction-limited optical spot is scanned across the TG edge in Device 1 and Device 2. (\bfA) PL measurement at $\vbg=5\,$V and $\vtg=-6\,$V in Device 1 as the optical spot is scanned across the two edges of the TG, which leads to the discrete states appearing twice. (\bfB) In Device 2, we set $\vbg=-6\,$V and $\vtg=6.5\,$V, and show the derivative of the reflection contrast as we scan across a single edge.
    }  
	\label{fig:4_1_pos_dependence}
\end{figure}

The strong in-plane confinement perpendicular to the TG edge implies that exciton confinement in our system is effectively 1D in nature. In Fig.\,\ref{fig:4_1_pos_dependence}, we show the position-dependent optical spectra of confined exciton states in Device 1 (PL) and Device 2 (derivative of reflection contrast), respectively. These measurements are performed by scanning the optical spot across the sample using nanopositioners, while keeping $\vbg$ and $\vtg$ fixed. In Device 1, the TG consists of two edges separated by roughly $1\,\mu$m (see inset of Fig.\,\ref{fig:4_1_pos_dependence} \bfA) and hence the discrete states appear at two spatial locations, while vanishing in the intermediate region. In Device 2, because the graphene TG has a single edge (inset of Fig.\,\ref{fig:4_1_pos_dependence} \bfB), the $\xqc$ states appear once as we scan across the edge and vanish on either side.

\begin{figure}[htb]
    \centering
	\includegraphics[width=14cm]{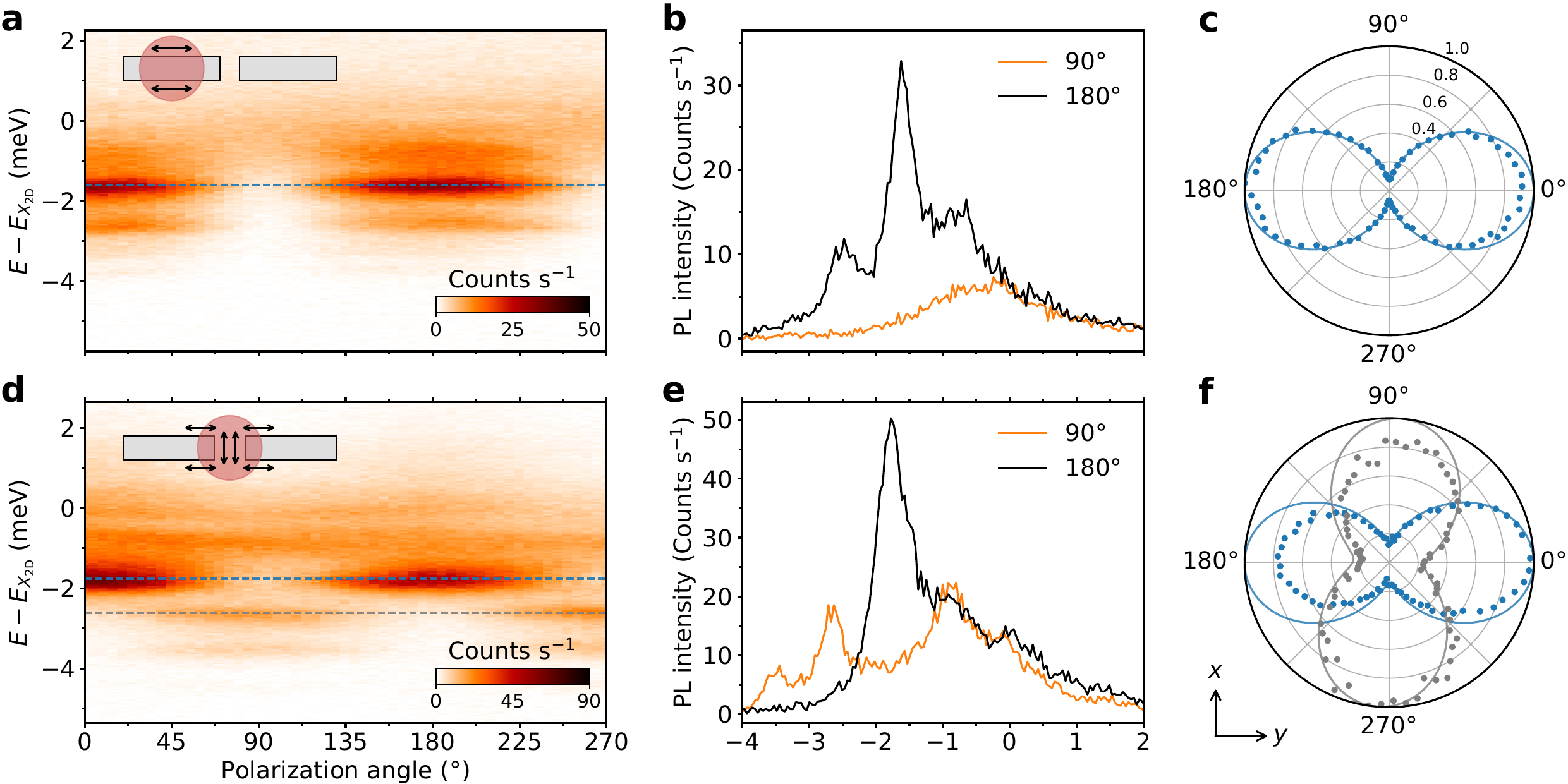}
	\caption{\textbf{Polarization dependence of 1D quantum confined excitons in Device 1.}
    PL measurements as a function of linear polarization angle taken at $\vbg = 5\,$V and $\vtg=-5.5\,$V at two different positions (see insets): (\bfA) where the spot encompasses two parallel edges, and (\bfD) where the optical spot is aligned on the split-gate, and we expect to have two sets of orthogonally oriented wires. The polarization axes in these two cases are illustrated by black arrows in the insets. (\bfB) and (\bfE) show linecuts from these datasets at orthogonal detection angles. (\bfC) and (\bfF) depict polar plots of normalized emission intensity associated with confined states for data shown in (\bfA) and (\bfD), respectively. The PL emission from confined modes exhibit a high degree of linear polarization ($\xi$ around $80\,\%$) in the longitudinal direction of the wire. This is clearly evidenced by the fact that, whereas in (\bfA) all states are $y$-polarized, we observe additional $x$-polarized confined states in (\bfD) arising from the orthogonally oriented wires. 
    }  
	\label{fig:4_2_polarization_dev1}
\end{figure}

To show the effect of such an anisotropic potential on the COM motion of excitons, we turn our attention to the polarization properties of emission from the confined states. Fig.\,\ref{fig:4_2_polarization_dev1} shows linear polarization-resolved PL measurements taken at two positions on Device 1, which are illustrated in the respective insets: Fig.\,\ref{fig:4_2_polarization_dev1} {\bfA} depicts PL from the $200$-nm-wide region which encompasses two parallel edges. Fig.\,\ref{fig:4_2_polarization_dev1} {\bfD} on the other hand, shows PL obtained by aligning the optical spot on the split-gate region, which contains two sets of orthogonal wire segments. In both measurements, the PL emission from confined exciton states exhibit a high degree of linear polarization $\xi = (I_{\mathrm{max}} - I_{\mathrm{min}})/(I_{\mathrm{max}} + I_{\mathrm{min}})$, where $I_{\mathrm{max}}$ and $I_{\mathrm{min}}$ are  maximum and minimum intensities in the polarization scan. These are obtained by fitting the function $A\cdot\mathrm{cos}^2(\theta-\theta_0)+C$ to the normalized PL intensities as function of the detection angle $\theta$. At $\vtg$ of around $-6\,$V, we measure typical values of $\xi \approx 0.8$, and a maximum value at some positions of $\xi_{\mathrm{max}} \approx 0.96$. This high degree of linear polarization is comparable to what was previously reported in other 1D systems, such as semiconductor nanowires \cite{Wang2001}, carbon nanotubes \cite{Lefebvre2004}, and 1D moir\'{e} excitons \cite{Bai2020}. Whereas in Fig.\,\ref{fig:4_2_polarization_dev1} {\bfA} the PL lines are polarized only along the $y$-direction, we observe additional sets of $x$-polarized lines in Fig.\,\ref{fig:4_2_polarization_dev1} {\bfD}, which arise from the 1D exciton wires in the gap region that are oriented along $x$. This can also be seen in linecuts of the PL spectra taken along the $x$- and $y$-directions, corresponding to a polarization angle of $90^\circ$ and $180^\circ$, respectively (Fig.\,\ref{fig:4_2_polarization_dev1} {\bfB} and {\bfE}). In addition, we depict the normalized PL emission intensity at fixed energy for both spots in Fig.\,\ref{fig:4_2_polarization_dev1} {\bfC} and {\bfF}. The $0^\circ$ angle of the polar plots is set to the primary polarization axis, corresponding to the $y$-direction. As shown in Fig.\,\ref{fig:4_2_polarization_dev1} {\bfF}, the intensity maxima for the states originating in the gap region of the split-gate appear close to $90^\circ$, within our experimental uncertainties of $\pm 5^\circ$.

\begin{figure}[htb]
    \centering
	\includegraphics[width=11.5cm]{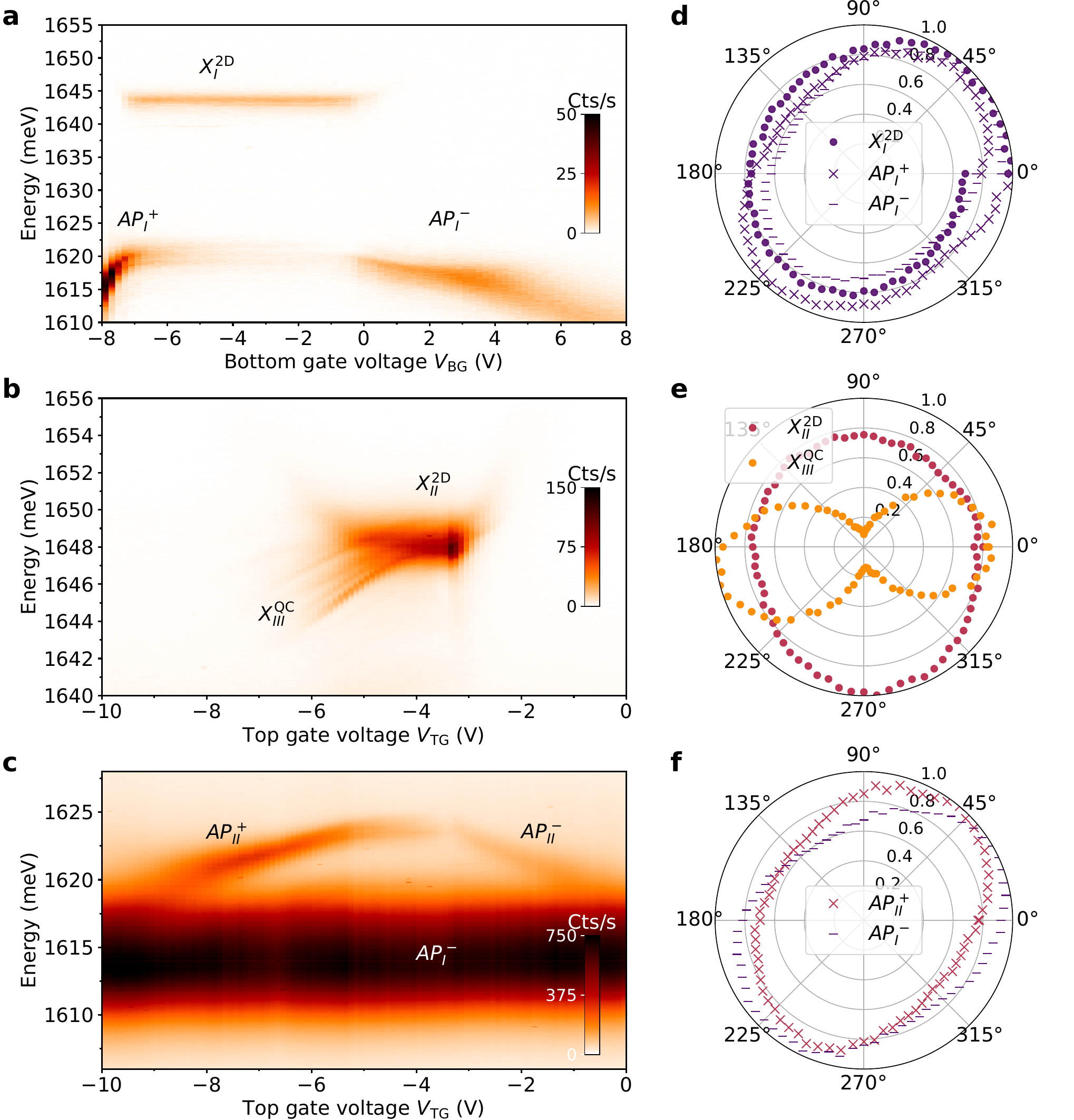}
	\caption{\textbf{Linear polarization anisotropy of different optical resonances.}
    (\bfA) PL BG ($\vbg$) scan conducted on bare MoSe$_2$ away from TG region. (\bfB), (\bfC) PL TG ($\vtg$) scan performed on the edge of the TG. (\bfD)-(\bfF) Polarization dependence of optical transitions in regions I, II and III, which are represented in purple, magenta and orange, respectively. The exciton, hole-side attractive polaron and electron-side attractive polaron is marked with a circle, cross and dash, respectively.
    }  
	\label{fig:4_3_polarization_anisotropy}
\end{figure}

To verify that the strong polarization anisotropy we report here for Device 1 is indeed associated with the confined states of the exciton, we perform polarization-resolved PL measurements also for various other optical transitions observed in the device. Fig.\,\ref{fig:4_3_polarization_anisotropy} {\bfA} demonstrates a reference BG ($\vbg$) scan conducted at a position away from the TG region. Also shown in Fig.\,\ref{fig:4_3_polarization_anisotropy} {\bfB}, {\bfC} is a $\vtg$ scan performed on the TG edge. We reiterate that, because the spot size of our excitation beam is diffraction-limited, PL emission from three distinct spatial regions is measured: (1) The region away from the TG is electron-doped and thus gives rise to a broad attractive polaron resonance ($\mathrm{AP}^-_\mathrm{I}$) centered around $1.615$ eV, which remains unaffected as a function of $\vtg$; (2) the region underneath the TG leads to neutral exciton ($\mathrm{X}^\mathrm{2D}_\mathrm{II}$) and attractive polaron resonances ($\mathrm{AP}^-_\mathrm{II}$ and $\mathrm{AP}^+_\mathrm{II}$) as it changes from being electron-doped, neutral and finally hole-doped when $\vtg$ is lowered; (3) the neutral intermediate region at the edge of the TG gives rise to redshifting confined exciton lines ($\mathrm{X}^\mathrm{QC}_\mathrm{III}$) as $\vtg$ is lowered.

In Fig.\,\ref{fig:4_3_polarization_anisotropy} {\bfD}-{\bfF}, we show the polarization dependence of these optical transitions by plotting the normalized PL emission as a function of linear polarization detection angle. $0^{\circ}$ thereby constitutes the direction along the quantum wire and thus the TG edge. As shown in Fig.\,\ref{fig:4_3_polarization_anisotropy} {\bfF}, the attractive polaron resonances originating from underneath the TG ($\mathrm{AP}^+_\mathrm{II}$) and away from the TG ($\mathrm{AP}^-_\mathrm{I}$) have a low degree of linear polarization $\xi \lesssim 20\%$. The neutral exciton originating from underneath the TG ($\mathrm{X}^\mathrm{2D}_\mathrm{II}$) exhibits a similar behavior ($\xi \approx 10\%$, see Fig.\,\ref{fig:4_3_polarization_anisotropy} {\bfE}). Furthermore, the primary polarization axis of these resonances has varying orientation and does not align with the TG edge. As a reference, we also compute $\xi$ for these optical transitions away from the TG and find that it is in the same range (Fig.\,\ref{fig:4_3_polarization_anisotropy} {\bfD}). In stark contrast to this behavior, the confined exciton states exhibit $\xi=96\%$ along with a primary polarization axis oriented within $\pm 5^{\circ}$ along the TG edge.

Owing to the similarity in polarization properties of exciton and polaron resonances from region I and II, we conclude that the strongly polarized emission of the confined excitonic states has its origin in the 1D confinement rather than a screening by the metallic TG, the effect of which cannot be distinguished from a typical strain-induced polarization dependence. If the geometry of the metallic TG had an impact on the polarization properties of the emission, also other optical transitions would exhibit an enhanced linear emission.

Excitons in 1D semiconductor nanowires have been previously reported to emit photons that are linearly polarized along the wire axis \cite{Wang2001,Akiyama1996,Lefebvre2004,Bai2020,Wang2021}. Linearly polarized emission originates from long-range electron--hole exchange interaction, which couples the valley degrees of freedom and the COM motion of excitons \cite{Glazov2015,Yu2015} (see section \ref{chap:theory:sec:fine_structure}). In fact, this coupling can be orders of magnitude greater in TMD monolayers than in III-V semiconductors due to the much larger exciton binding energy in the former. For finite exciton COM momenta, the valley-COM coupling leads to an energy splitting between longitudinal and transverse electromagnetic modes. Introducing spatial anisotropy, for instance in the form of a 1D confinement potential, breaks the rotational symmetry of the system, thus opening a gap between orthogonal linear polarization states at momentum $\mathbf{k}=0$. The magnitude of the $x$-$y$ polarization splitting $\delta$ depends on both the COM $k$-space wave function $\psi_\mathrm{COM}(\mathbf{k})$  and the relative wave function $|\psi_\mathrm{rel}(r=0)|^2$ according to (\cite{Yu2015, Glazov2015,Bai2020}):
\begin{equation}
\delta \propto |\psi_\mathrm{rel}(r=0)|^2 \int \psi_\mathrm{COM}^2(\mathbf{k})|\mathbf{k}|d\mathbf{k}
\label{eqn:polarization_splitting}
\end{equation}

\begin{figure}[htb]
    \centering
	\includegraphics[width=14cm]{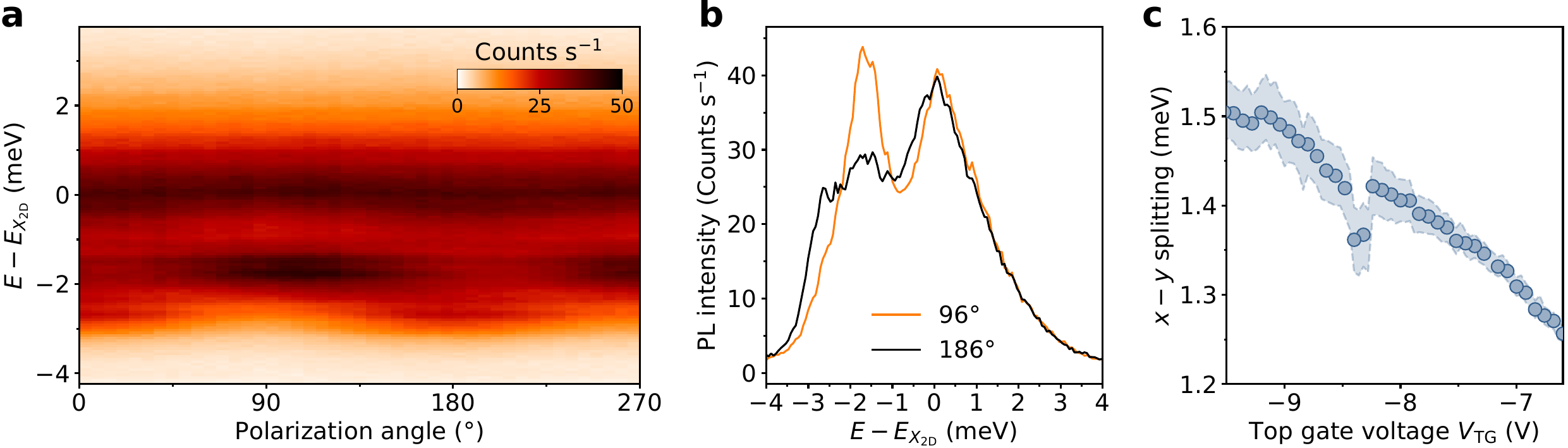}
	\caption{\textbf{Polarization splitting in Device 2.}
    (\bfA) PL spectra as a function of linear polarization angle in Device 2. (\bfB) PL spectra taken at orthogonal detection angles. The thicker h-BN spacer layers in Device 2 lead to a weaker confinement as compared to Device 1, giving rise to both $x$-polarized and $y$-polarized states with a finite energy splitting $\delta$ of around $1$ meV. The unmodified resonance at $E-E_\mathrm{X,2D}$ of around $0.5$ meV corresponds to the repulsive polaron emission from the region under the TG. (\bfC) The polarization splitting $\delta$ increases with decreasing $\vtg$, corresponding to tighter confinement. The shaded blue area represents fitting errors.
    }  
	\label{fig:4_4_polarization_splitting}
\end{figure}

In Device 1 (Fig.\,\ref{fig:4_2_polarization_dev1} \bfA-\bfC), we have shown all discrete states to be linearly polarized parallel to the TG edge (that is, $y$-polarized). On the other hand, while measuring linear polarization-resolved PL spectra in Device 2 (Fig.\,\ref{fig:4_4_polarization_splitting} \bfA), we observe a polarization doublet comprising both parallel ($y$-polarized) and perpendicular ($x$-polarized) components with an energy splitting of $\delta = E_\parallel - E_\perp \approx 1\,$meV (Fig.\,\ref{fig:4_4_polarization_splitting} \bfB). The discrepancy between the two devices can be explained by the thicker h-BN spacers in Device 2 compared with Device 1, which consequently leads to weaker confinement in the former. In Device 1, the confinement is strong enough that $x$-$y$ polarization splitting exceeds the confinement energy, whereas the weaker confinement in Device 2 allows to observe both polarization states and their relative splitting $\delta$. The dependence of polarization splitting on COM confinement can be seen clearly in Fig.\,\ref{fig:4_4_polarization_splitting} \bfC, in which we observe that $\delta$ increases with decreasing $\vtg$, which corresponds to stronger confinement.

We can understand this behavior by considering Eqn.\,\ref{eqn:polarization_splitting}. In applying this relation to our confinement scheme, we note that with increasing in-plane electric field, the confinement potential will get deeper and stiffer. On the one hand, this will lead to a decrease in $|\psi_\mathrm{rel}(r=0)|^2$ owing to an increased relative distance between the electron and hole due to the presence of in-plane electric fields. On the other hand, due to a simultaneous reduction in COM confinement length scale, $\psi_\mathrm{COM}$ will feature contributions from a larger range of COM momenta, thus causing an overall increase in polarization splitting $\delta$. Based on our observation in Fig.\,\ref{fig:4_4_polarization_splitting} {\bfC}, it is evident that the latter contribution is the dominant one. We further confirm that the modification in  $|\psi_\mathrm{rel}(r=0)|^2$ over the range of accessible electric fields in our devices is small in section \ref{chap:B:sec:lowest_state}.

A quantitative estimate of the polarization splitting can be obtained by approximating the trapping potential for the lowest confined state as a harmonic oscillator. This assumption is well justified in view of the numerical calculations performed in the previous section to determine the eigenfunctions of the confinement potential (Fig.\,\ref{fig:3_8_wfs}). In such a scenario, the expression in Eqn.\,\ref{eqn:polarization_splitting} can be evaluated analytically, and results in a polarization splitting
\begin{equation}
    \delta = \frac{2}{\sqrt{\pi}}\cdot\frac{J}{K \ell_\mathrm{x}} = \frac{3}{2\pi^{3/2}}\cdot\frac{a}{\ell_\mathrm{x}}\cdot J
\end{equation}
Here, we have encapsulated the term $|\psi_\mathrm{rel}(r=0)|^2$ in the exchange coupling strength $J$. In accordance with Eqn.\,\ref{eqn:exchange_hamiltonian}, $K=\frac{4\pi}{3a}$ is the distance from the $K^\pm$ points to the $\Gamma$ point in the Brillouin zone with a lattice constant $a$, and $\ell_\mathrm{x}$ denotes the harmonic oscillator confinement length. Considering changes in $|\psi_\mathrm{rel}(r=0)|^2$ to be small, we set $J$ to a fixed value of approximately $300\,$meV \cite{Shimazaki2021,Qiu2015}. We also assume $a=0.3\,$nm and $\ell_\mathrm{x}\approx6\,$nm, as established in the previous section for Device 1. These parameters yield a polarization splitting $\delta$ of around $4\,$meV, which is comparable to the confinement potential depth in Device 1, as can be seen in Fig.\,\ref{fig:3_7_fit_evolution}. This corroborates our assumption that the observed polarization signatures in a given device are significantly determined by an interplay between the depth and stiffness of the excitonic confinement potential.

\begin{figure}[htb]
    \centering
	\includegraphics[width=10cm]{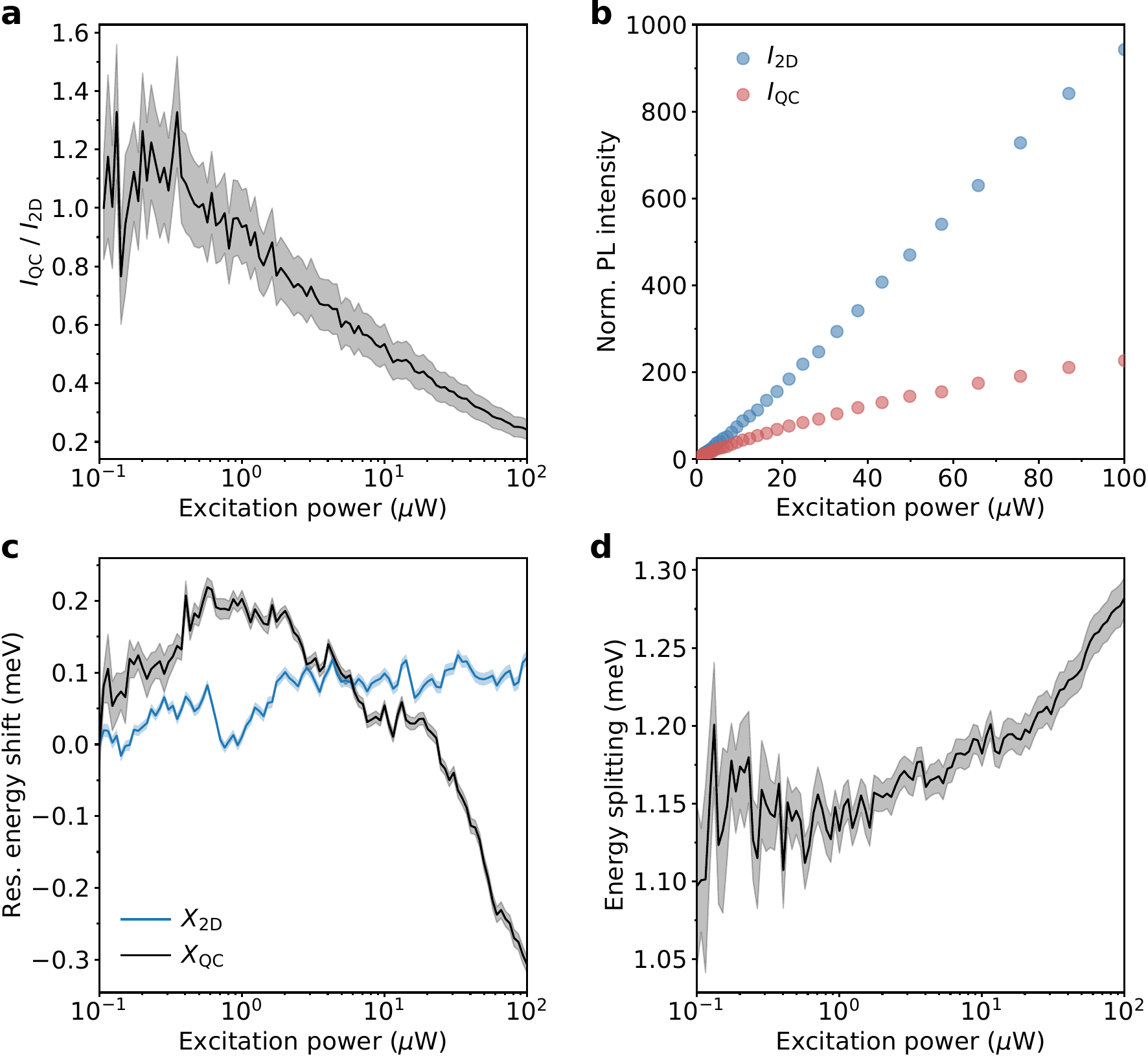}
	\caption{\textbf{Excitation power dependence of resonance energy in PL.}
    (\bfA) Power dependence of normalized PL emission intensity from quantum confined 1D excitons ($I_\mathrm{QC}$) relative to emission from free 2D excitons ($I_\mathrm{2D}$). (\bfB) $I_\mathrm{2D}$ scales linearly with power (blue points), whereas $I_\mathrm{QC}$ exhibits sub-linear behavior (red points). The PL intensities for $\xfree$ and $\xqc$ are normalized by their corresponding values at the lowest power. (\bfC) Shift in resonance energy of $\xfree$ (blue) and $\xqc$ (black) states relative to their energy measured at the lowest excitation power. (\bfD) Evolution of energy splitting between two successive quantum confined states with power. The shaded areas represent errors from fitting.
    }  
	\label{fig:4_5_power_dependence}
\end{figure}

Further indication of strong COM confinement is provided by comparing the excitation power dependence of PL emission of $\xfree$ and $\xqc$ states. In Fig.\,\ref{fig:4_5_power_dependence} \bfA, we show the ratio $I_\mathrm{QC}/I_\mathrm{2D}$ of the normalized PL intensity of $\xqc$ and $\xfree$ states. Whereas at low powers $I_\mathrm{QC}/I_\mathrm{2D}$ is relatively insensitive to power, the ratio decreases sharply in the high-power regime $>0.5\,\mu$W. The power dependence of the normalized PL emission for the two states is also shown separately in Fig.\,\ref{fig:4_5_power_dependence} \bfB. As expected for free 2D excitons, $I_\mathrm{2D}$ scales linearly with increasing power (blue points). On the other hand, the PL emission intensity from quantum-confined 1D excitons $I_\mathrm{QC}$ grows sublinearly (red points) with power. Such saturation behavior typically arises in confined systems owing to repulsive exciton-exciton interactions which prohibit high-density occupation of excitons within the optical spot.

Furthermore, we inspect the evolution of the resonance energy of these states with increasing excitation power (Fig.\,\ref{fig:4_5_power_dependence} \bfC). The free 2D exciton states only exhibit a slight variation in energy of approximately $0.1$\,meV blue shift as the excitation power is increased from $100$\,nW to $100$\,$\mu$W. For the analysis of $\xqc$ excitons, we track the power dependence of the two lowest quantum-confined states. For the lowest confined state, we observe that the resonance energy first exhibits a blue shift by approximately $0.2$\,meV at low powers. This trend changes sign to a red shift at high powers ($P > 1\,\mu$W) (Fig.\,\ref{fig:4_5_power_dependence} \bfC).

In an ideal scenario, the interaction between dipolar excitons, particularly in strongly confined systems, is expected to give rise to a shift of the resonance energy. However, it can be difficult to isolate such interaction effects from other effects that can arise from the continuous wave (CW) PL excitation \cite{Scuri2018}. Specifically, in our system, we suspect that the observed redshift occurs due to optically induced doping that introduces charges to the p-doped and n-doped regions, and thus modifies the spatial in-plane electric field distribution. This can be seen most clearly by extracting the energy splitting between the lowest quantum confined eigenstates as a function of excitation power, as shown in Fig.\,\ref{fig:4_5_power_dependence} \bfD. Here, we observe a clear increase in the level splitting, indicating tighter confinement at higher power.

If this picture is correct, it would be a very interesting feature of our scheme, as it potentially suggests a control of the confinement using optical excitation. However, we hesitate to make definite claims before carrying out time-resolved pump-probe measurements, which should help distinguish between contributions of interaction-induced blue shift and a possible enhancement of the in-plane electric field.

  %%%%%%%%%%%%%%%%%%%%%%%%%%%%%%%%%%%%%%%%%%%%%%%%%%%%%%%%%%%%%%%%%%%%%%%%
\chapter[Relative exciton wave function]{Relative exciton wave function\footnote{This chapter is adapted from the following publication \cite{Thureja2022}:\\ \\
\textbf{Electrically tunable quantum confinement of neutral excitons}\\
Thureja, D.; Imamoglu, A.; Smole\'{n}ski, T.; Amelio, I.; Popert, A.; Chervy, T.; Lu, X.; Liu, S.; Barmak, K.; \mbox{Watanabe}, K.; Taniguchi, T.; Norris, D. J.; Kroner, M.; Murthy, P. A.\\
\textit{Nature}\textbf{ 606}, 298 (2022)
}}
\label{chap:relative}
%%%%%%%%%%%%%%%%%%%%%%%%%%%%%%%%%%%%%%%%%%%%%%%%%%%%%%%%%%%%%%%%%%%%%%%%

%%%%%%%%%%%%%%%%%%%%%%%%%%%%%%%%%%%%%%%%%%%%%%%%%%%%%%%%%%%%%%%%%%%%%%%%
\section{Magnetic field dependence of lowest confined state}
\label{chap:B:sec:lowest_state}
%%%%%%%%%%%%%%%%%%%%%%%%%%%%%%%%%%%%%%%%%%%%%%%%%%%%%%%%%%%%%%%%%%%%%%%%

The presence of strong inhomogeneous in-plane electric fields influences not only the COM motion, but also the relative wave function of the confined exciton. To shed light on the modification of the internal structure of the $\xqc$ states, we perform circular-polarization-resolved measurements with an external out-of-plane magnetic field $B$ on Device 2. These measurements are performed at around $4$ K in a cryostat with fibre-optical access and a superconducting coil that can generate $B$-fields of up to $16$\,T in the direction perpendicular to the sample surface. Using nanopositioners, we can position the optical spot on the edge of the graphene TG in Device 2. As we ramp the magnetic field, we observe spatial drift of the sample. Hence, the optical spot must be readjusted to measure on the same position at every magnetic field. We achieve this by using the bright 2D exciton state as a reference spectrum and scanning the sample position such that the average energy of the Zeeman-split 2D exciton branches matches the reference exciton energy at charge neutrality and $B= 0\,$T. In addition to realigning the position, we also recalibrate the optical setup at every magnetic field to compensate for any Faraday rotation resulting from the strong $B$-fields. In this way, we are reliably able to perform circular-polarization-resolved measurements from $B= 0-16\,$T. At every value of the magnetic field, we perform measurements at different $\vtg$ and $\vbg$.

\begin{figure}[htb]
    \centering
	\includegraphics[width=10cm]{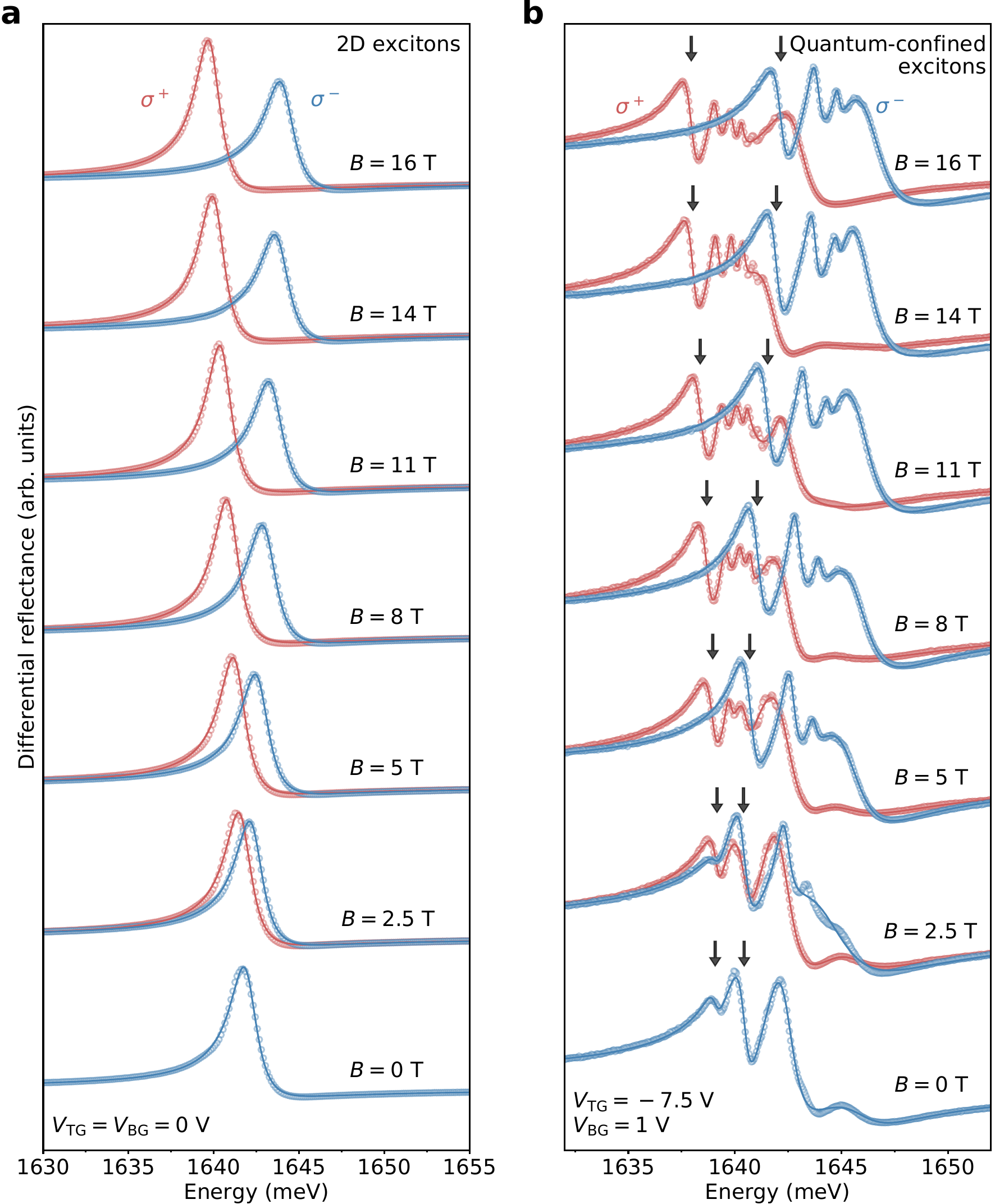}
	\caption{\textbf{Evolution of differential reflectance spectra with magnetic field.}
    Reflectance spectra as a function of magnetic fields ($B$) shown for 2D excitons (\bfA) and confined excitons (\bfB) in Device 2. The experimental data points are indicated with circles, and fits to the spectra are shown in solid lines.
    }  
	\label{fig:B-spectra}
\end{figure}

In Fig.\,\ref{fig:B-spectra}, we show the measured spectra at different $B$-fields, from $B=0\,$T to $B=16\,$T, for the free 1s excitons $\xfree$ and confined excitons $\xqc$, respectively. The $\xqc$ spectra are measured for fixed $\vbg = 1\,$V and $\vtg = -7.5\,$V. For both $\xfree$ and $\xqc$ states, we observe a splitting of spectral lines into two orthogonally polarized states. The corresponding energies, $E^+$ and $E^-$, are determined by performing a spectral analysis as described in section \ref{chap:1D:sec:signatures}. The measured spectra are fitted with the reflectance spectral profile $S(E)$ (Eqn.\,\ref{eqn:spec_func}) for $\xfree$, and a superposition of such spectral profiles $\sum_i S(E;E_{0,i},\Gamma_i,A_i,\zeta_i)$ for the quantum confined states. For $\xqc$, we focus on the $B$-dependent behaviour of the lowest-energy confined state (denoted with black arrows in Fig.\,\ref{fig:B-spectra} \bfB). Only at $B=0$\,T are both the two \emph{linear}-polarization-split quantum-confined ground states visible in the same \emph{circular} polarization. Hence, for the zero-field case, their phase factor $\zeta$ is assumed to be the same.

\begin{figure}[htb]
    \centering
	\includegraphics[width=11cm]{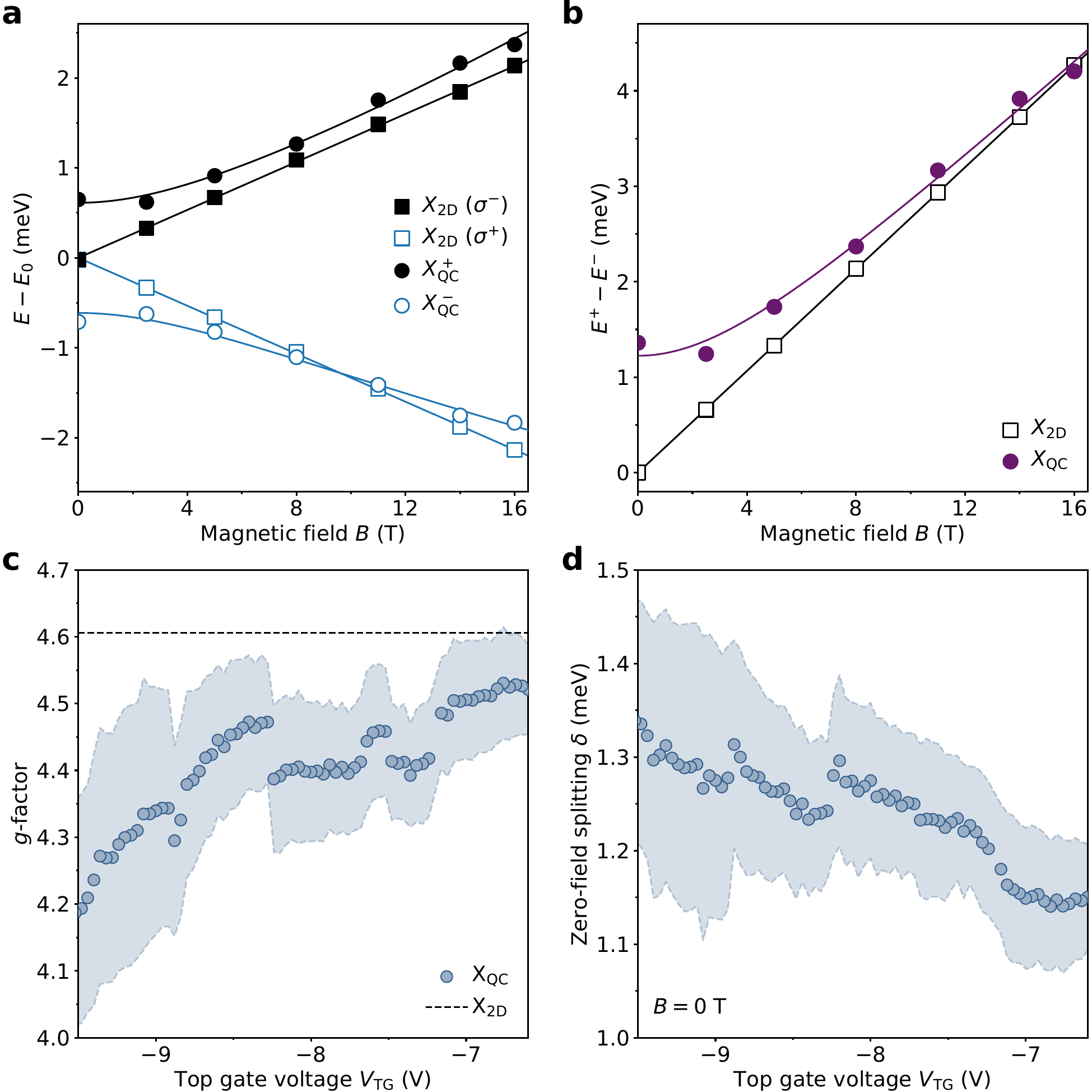}
	\caption{\textbf{Energy shifts of 2D and quantum-confined excitons.}
    (\bfA) Evolution of $E^+$ and $E^-$ as a function of $B$ for $\xfree$ (black closed squares and blue open squares) and $\xqc$ (black closed circles and blue open circles). The uncertainties in energies $E^+$ and $E^-$ are $\lesssim 15\,\mu$eV. All energies are plotted with respect to the average energy $\bar{E}(B = 0\,\mathrm{T})$. (\bfB) The Zeeman splitting $\Delta E = E^+ - E^-$ scales linearly with $B$ for $\xfree$ (black open squares), whereas for $\xqc$, polarization splitting is non-zero at $B=0\,$T and asymptotically approaches linear scaling for large $B$ (purple closed circles). (\bfC) $g-$factor and (\bfD) zero-field splitting $\delta$ as a function of top gate voltage $\vtg$, determined by fitting the Zeeman splitting $E^+ - E^-$ as a function of the magnetic field. The shaded areas represent fitting errors of around $100\,\mu$eV.
    }  
	\label{fig:B_Zeeman}
\end{figure}

From the resulting fits (shown in Fig.\,\ref{fig:B-spectra}), we extract the resonance energies $E^+$ and $E^-$, which are plotted as a function of $B$ in Fig.\,\ref{fig:B_Zeeman} \bfA. The corresponding fitting error is approximately $\lesssim 15\,\mu$eV. The Zeeman splitting $E^+ - E^-$ for both $\xfree$ and $\xqc$ can be obtained with a similar precision and is shown in Fig.\,\ref{fig:B_Zeeman} \bfB. The 2D exciton shows the expected linear-in-B splitting with perfect valley degeneracy at $B=0\,$T, leading to a $g$-factor of about $4.6$ (Fig.\,\ref{fig:B_Zeeman} \bfC). The confined excitons exhibit a finite polarization splitting $\delta$ of approximately $1\,$meV at $B=0\,$T and approach linear behavior asymptotically at high $B$-fields. This can be modelled with the $B$-field dependence
\begin{equation}
    \Delta E = \sqrt{\delta^2 + (g\mu_\mathrm{B} B)^2}
\end{equation}
in which $\delta$ is the zero-field splitting and $\mu_\mathrm{B}$ is the Bohr magneton. The fit results are shown in Fig.\,\ref{fig:B_Zeeman} \bfB, along with $\Delta E$ determined from experimental data points. We observe that, at low fields, the model appears to have a systematic error. The fitted values for $\delta$ systematically underestimate the value obtained by directly taking the difference of the fitted $E^+$ and $E^-$ resonance energy at $B=0$\,T. Therefore, the fitted values of $\delta$ exhibit larger fitting errors of around $100\,\mu$eV (Fig.\,\ref{fig:B_Zeeman} \bfD). The $g$-factor of $\xqc$ states as a function of $\vtg$ is shown in Fig.\,\ref{fig:B_Zeeman} \bfC. We observe a decrease of the $g$-factor below the value for 2D excitons as $\vtg$ is reduced. We do not have an explanation for this voltage-dependent $g$-factor. We speculate that strong in-plane electric fields may modify the Bloch states of free electrons and holes, as well as the relative motion of bound electrons and holes in excitons, thus leading to the observed voltage-dependent shifts of the $g$-factor.

\begin{figure}[htb]
    \centering
	\includegraphics[width=8cm]{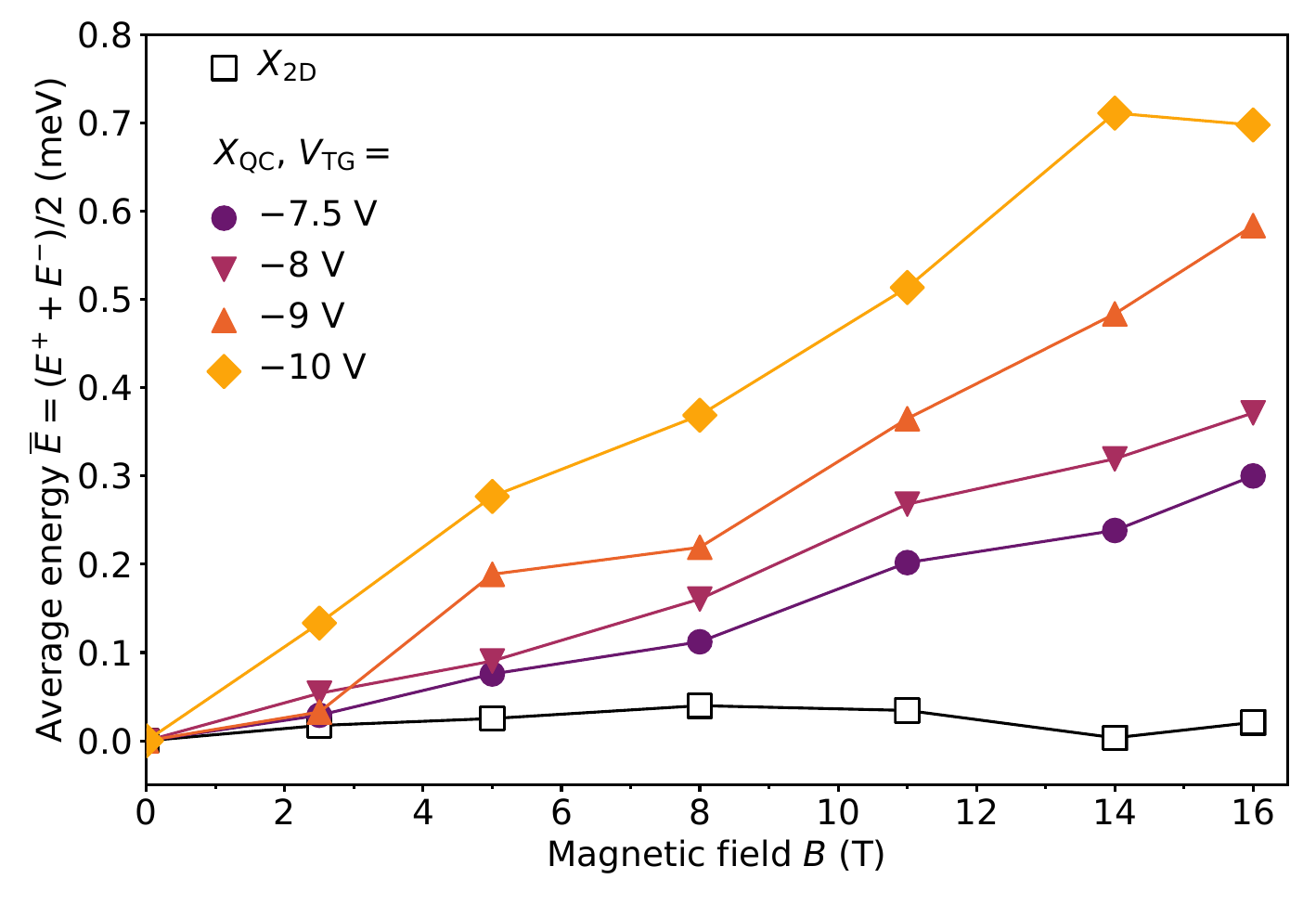}
	\caption{\textbf{Diamagnetic energy shift of 2D and quantum-confined excitons.}
    The average energy $\bar{E} = (E^+ + E^-)/2$ for $\xfree$ and $\xqc$ at different values of $\vtg$. The 2D exciton shows no measurable diamagnetic shift, even at 16 T (black open squares), consistent with the small Bohr radius. On the other hand, the confined dipolar excitons exhibit a sizeable shift, which also increases with decreasing $\vtg$. All energies are plotted with respect to the average energy $\bar{E}(B = 0\,\mathrm{T})$.
    }  
	\label{fig:B_diamag}
\end{figure}

Information on the relative wave function of excitons is encoded in the diamagnetic energy shifts (see section \ref{chap:theory:sec:B-field}), which can be extracted from the average energy $\bar{E} = (E^+ + E^-)/2$ of the two polarization states, which can be quantified with an experimental uncertainty of approximately $\lesssim 20\,\mu$eV. For the 1s exciton, the root mean square size $\braket{r^2}$ is related to the diamagnetic shift according to $E_\mathrm{dia} = e^2 \braket{r^2} B^2/8m_\mathrm{r}$, in which $m_\mathrm{r} = m^*_\mathrm{n} m^*_\mathrm{p}/(m^*_\mathrm{n} + m^*_\mathrm{p})$ is the reduced mass. In Fig.\,\ref{fig:B_diamag}, we show $\bar{E} = (E^+ + E^-)/2$ as a function of $B$, measured for both $\xfree$ (open black squares) and $\xqc$ states. The energy shifts of $\xqc$ are shown for four $\vtg$ values: $\vtg = -7.5\,$V, $-8\,$V, $-9\,$V and $-10\,$V. As 2D excitons (in the 1s manifold) in TMD monolayers are strongly bound and have small Bohr radii ($a_\mathrm{B}$ around $1\,$nm for MoSe$_2$ \cite{Goryca2019}), we do not observe a shift in $\bar{E}$ for $\xfree$ beyond our experimental uncertainties, which is also a consistency check for the position alignment procedure described above. By contrast, we observe strikingly enhanced shift of $\bar{E}$ for $\xqc$ states with increasing $B$-field. Moreover, we find a clear dependence of $\bar{E}$ on $\vtg$, that is, a stronger in-plane electric field corresponds to larger blue shift of the average energy.

These observations are surprising for several reasons. First, we note that, in conventional quantum-confined systems such as cleaved-edge-overgrowth quantum wires or self-assembled quantum dots \cite{Nagamune1992}, the confinement has the effect of increasing the electron--hole overlap and hence reducing the diamagnetic shift compared with bulk 2D systems. This is in strong contrast to our scenario, in which the confined states show much larger energy shifts than $\xfree$. Furthermore, the fact that the absolute energy of the $\xqc$ is only slightly redshifted from $\xfree$ by a few meV suggests that confined excitons have comparable binding energies as their free 2D counterparts. Furthermore, the high reflection contrast of $\xqc$ is also consistent with an exciton with a small Bohr radius. On the other hand, the substantially larger $\bar{E}$ of $\xqc$ indicates an enhanced spatial extent. For example, in the case of a free 2D exciton, an energy shift $\bar{E}$ of around $600\,\mu$eV at $B = 16\,$T would correspond to a root mean square size $\sqrt{\braket{r^2}}$ of about $6\,$nm, which would -- correspondingly -- have a much smaller oscillator strength and a different energy than the observed values. Taken together, our measurements indicate that the inhomogeneous in-plane $E$-field gives rise to a non-trivial spatially extended relative wave function of  $\xqc$ that is not captured by a simple hydrogenic picture of exciton states.

To better understand the physics of confined excitons in in-plane electric and out-of-plane magnetic fields and the resulting large shifts of $\bar{E}$, we exactly diagonalize the following Hamiltonian in relative coordinates $(r,\theta)$ between the electron and the hole comprising the exciton \cite{Lozovik2002,Chestnov2021}:
\begin{equation}
    H = -\frac{\hbar^2}{2\mu_+} \nabla^2  + \hbar\frac{e B}{2\mu_-}(-i\partial_\theta) + \frac{e^2B^2}{8\mu_+}r^2 + V_K(r) - eF_x x,
\end{equation}
where $\mu_\pm = \frac{m_\mathrm{n}^* m_\mathrm{p}^*}{m_\mathrm{n}^* \pm m_\mathrm{p}^*}$ and $V_\mathrm{K}$ is the Keldysh potential
\begin{equation}
    V_\mathrm{K}(r) = -\frac{e^2}{4\pi\varepsilon_0}\frac{\pi}{2\varepsilon r_0}[H_0(r/r_0) - Y_0(r/r_0)]
\end{equation}
We use $m_\mathrm{n}^* = 0.7m_\mathrm{e}$ \cite{Larentis2018}, $m_\mathrm{p}^* = 0.6m_\mathrm{e}$ \cite{Zhang2014} for effective electron and hole masses, $\varepsilon = 4.4$ for the dielectric constant, and $r_0 = 3.9/\varepsilon\,$nm for the screening length \cite{Goryca2019}. In the absence of any applied electric or magnetic field, we obtain with this procedure a binding energy of 1s excitons $E_\mathrm{B} = 217\,$meV, which is comparable with experimental values.

We exactly diagonalize the above Hamiltonian on the basis of Bessel functions with Dirichlet boundary conditions on a disk of radius $R$. The Hilbert space of a 2D disk of radius $R$ admits the basis $\ket{m,n}$ for $m \in \mathbb{Z}, n=1,2,..$, whose elements can be expressed in polar coordinates as
\begin{equation}
    \braket{r,\theta|m,n} = \frac{1}{Z_{m,n}}J_{m}(k_{m,n}r)e^{im\theta},
\end{equation}
in which $k_{m,n} = z_{m,n}/R$, with $z_{m,n}$ being the n-th zeros of the Bessel function $J_{m}(z)$. Typically, we consider up to $n=250$ modes and $m \leq 10$ angular modes. This basis is particularly convenient to treat a particle in an electric field, as only few angular modes are required. The difficulty in dealing with an external uniform in-plane electric field is that the Hamiltonian spectrum is not bounded from below, and -- hence -- the ground state of the system depends on the system size $R$. To mitigate this problem, we monitor the energy variation as a function of $B$ and $F_\mathrm{x}$, not of the ground state, but of the \emph{Most Localized Eigenstate at the Origin}, which we abbreviate as MLESO.

\begin{figure}[htb]
    \centering
	\includegraphics[width=14cm]{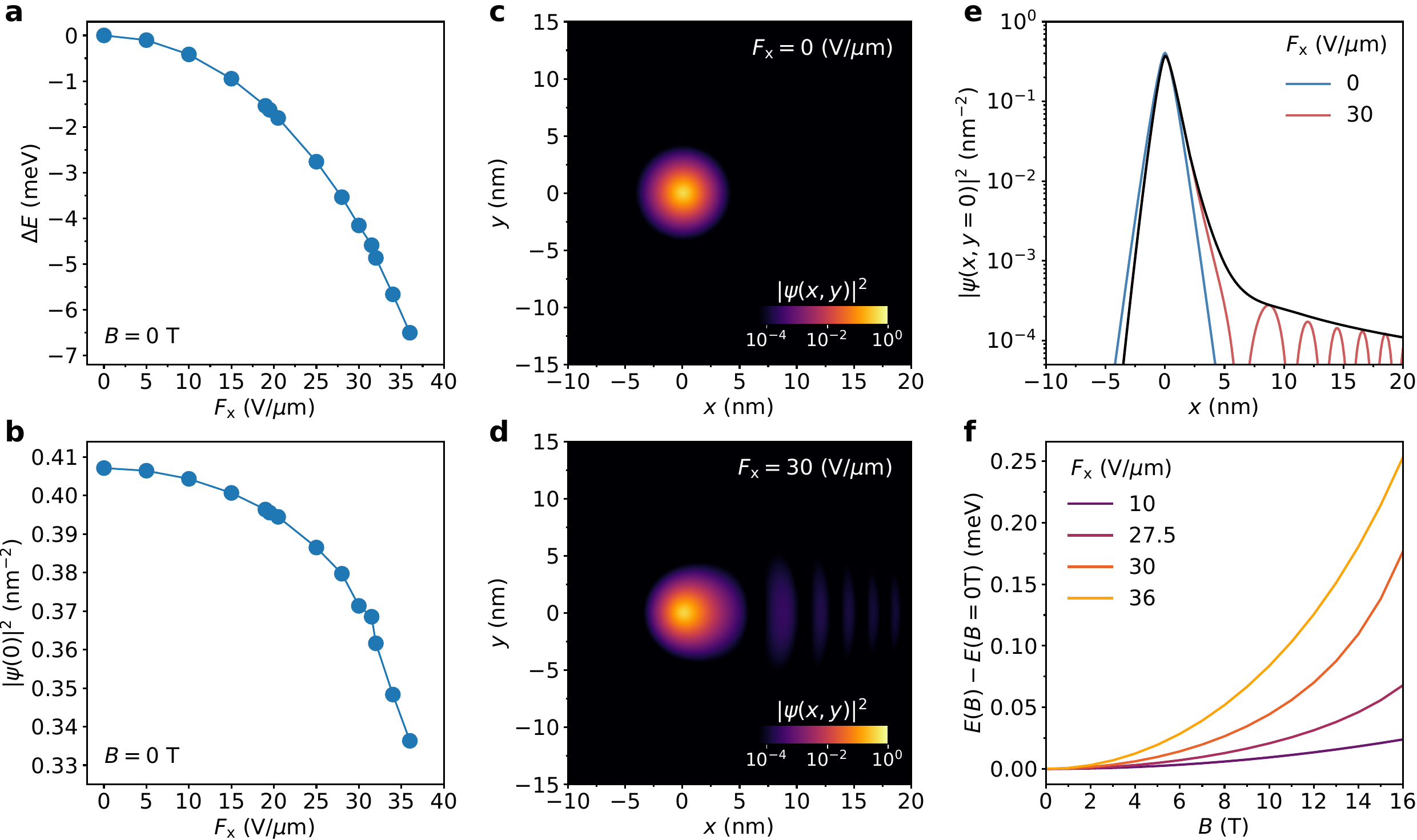}
	\caption{\textbf{Exact diagonalization of exciton relative motion in crossed electric and magnetic fields.}
    (\bfA) Energy of the \emph{Most Localized Eigenstate at the Origin} (MLESO) as a function of the in-plane electric field strength $F_\mathrm{x}$ at $B=0$\,T. A quadratic dependence with the electric field is observed, as expected from the dc Stark effect. (\bfB) The probability density $|\psi(r = 0)|^2$, proportional to the exciton oscillator strength, decreases by about $10-15\,\%$ with increasing electric field $F_\mathrm{x}$. Hence, a decrease in oscillator strength, as seen in the experiment, should primarily arise from center-of-mass quantum confinement of excitons. (\bfC, \bfD) Calculated probability density $|\psi(r)|^2$ for $F_\mathrm{x} = 0$ and $F_\mathrm{x} = 30\,\mathrm{V}/\mu\mathrm{m}$, respectively. (\bfE) Probability density $|\psi(x,y=0)|^2$ in the direction of the electric field for $F_\mathrm{x} = 0$ (blue) and $F_\mathrm{x} = 30\,\mathrm{V}/\mu\mathrm{m}$ (red). The oscillations seen at finite electric field arise from reflections from the finite-sized box assumed in the calculations. The black curve is a guide to the eye, to show the small but finite component of the wave function the exists outside the Coulomb potential. (\bfF) MLESO energy shift with respect to the energy $E(B=0\mathrm{\,T})$ as a function of $B$, for different values of $F_\mathrm{x}$. The magnitude of the predicted shift is of the same order as in the experimental observation.
    }  
	\label{fig:B_ED}
\end{figure}

Moreover, we focus on $R$ values in the range $25-40$ nm, comparable with the size of the intrinsic region of the experiment. Within this range, the results at different sizes are in agreement provided one avoids values of $R$ and $F_x$ for which an unbound state is exactly resonant with the bound state. We note that the oscillations seen in Fig.\,\ref{fig:B_ED} {\bfD} and {\bfE} are an artifact of the finite-sized box, whereas one would have a plane wave for an ionized state in a real continuum. A mathematical treatment of the limit $R\to\infty$ is the subject of current theoretical research.

In Fig.\,\ref{fig:B_ED} \bfA, we show the MLESO energy as a function of the in-plane electric field strength $|F_\mathrm{x}|$ at $B=0$\,T, which shows a quadratic $F_\mathrm{x}$ dependence, as expected from the dc Stark effect. The order of magnitude of the calculated shift is comparable with the measured Stark shifts in our experiment. In Fig.\,\ref{fig:B_ED} \bfB, we show $|\psi(r = 0)|^2$, which is proportional to the exciton oscillator strength, as a function of electric field, approximately spanning the range of field magnitudes achievable in the experiment. We observe that the wave function overlap in the Coulomb well, which directly corresponds to the oscillator strength, decreases with increasing $F_\mathrm{x}$, but only by about $10-15\,\%$. In the experiment, the overall oscillator strength is strongly reduced owing to COM confinement, but does not vary appreciably as a function of applied gate voltage. In Fig.\,\ref{fig:B_ED} {\bfC} and {\bfD}, we show the 2D probability density $|\psi(x,y)|^2$ in the relative frame for in-plane electric field values of $|F_\mathrm{x}| = 0$ and $|F_\mathrm{x}| = 30\,V/\mu$m, respectively. The probability density along $x$ at $y=0$ for these fields is shown in Fig.\,\ref{fig:B_ED} \bfE. Whereas at $|F_\mathrm{x}| = 0$ we obtain the expected decay of a tightly bound 2D exciton, the case for high electric fields indicates a modified wave function with an additional longer decay length scale. The spatial oscillations observed in Fig.\,\ref{fig:B_ED} {\bfE} are artifacts arising from a finite-sized box. In Fig.\,\ref{fig:B_ED} \bfF, we show the MLESO energy shift as a function of $B$ for different values of $F_\mathrm{x}$ and $R=40$ nm. The energy shift is shown with respect to the energy $E(B=0\mathrm{\,T})$.

These calculations indicate that, at strong enough in-plane electric fields, the electron (or hole) wave function in the relative frame of the exciton leaks out of the attractive Coulomb potential, and hybridizes with the quasi-continuum states. Furthermore, even though only a small fraction of the wave function leaks out, the state exhibits a substantial blue shift at finite $B$-fields, on the same order of magnitude as observed in our experiment. Moreover, we observe that the shift is larger for larger applied electric field, which is also qualitatively in agreement with experimental observation (Fig.\,\ref{fig:B_diamag}). Our observations and calculations therefore suggest that the quantum-confined exciton states have a modified internal structure compared with 2D excitons, or even in-plane dipolar excitons at weak fields.  This feature could potentially have novel applications, for instance towards enhancing exciton-exciton interactions \cite{Togan2018} and realizing effective gauge potentials \cite{Lim2017,Chestnov2021}.

%%%%%%%%%%%%%%%%%%%%%%%%%%%%%%%%%%%%%%%%%%%%%%%%%%%%%%%%%%%%%%%%%%%%%%%
\section{Magnetic field dependence of excited states}
%%%%%%%%%%%%%%%%%%%%%%%%%%%%%%%%%%%%%%%%%%%%%%%%%%%%%%%%%%%%%%%%%%%%%%%

\begin{figure}[htbp]
    \centering
	\includegraphics[width=14cm]{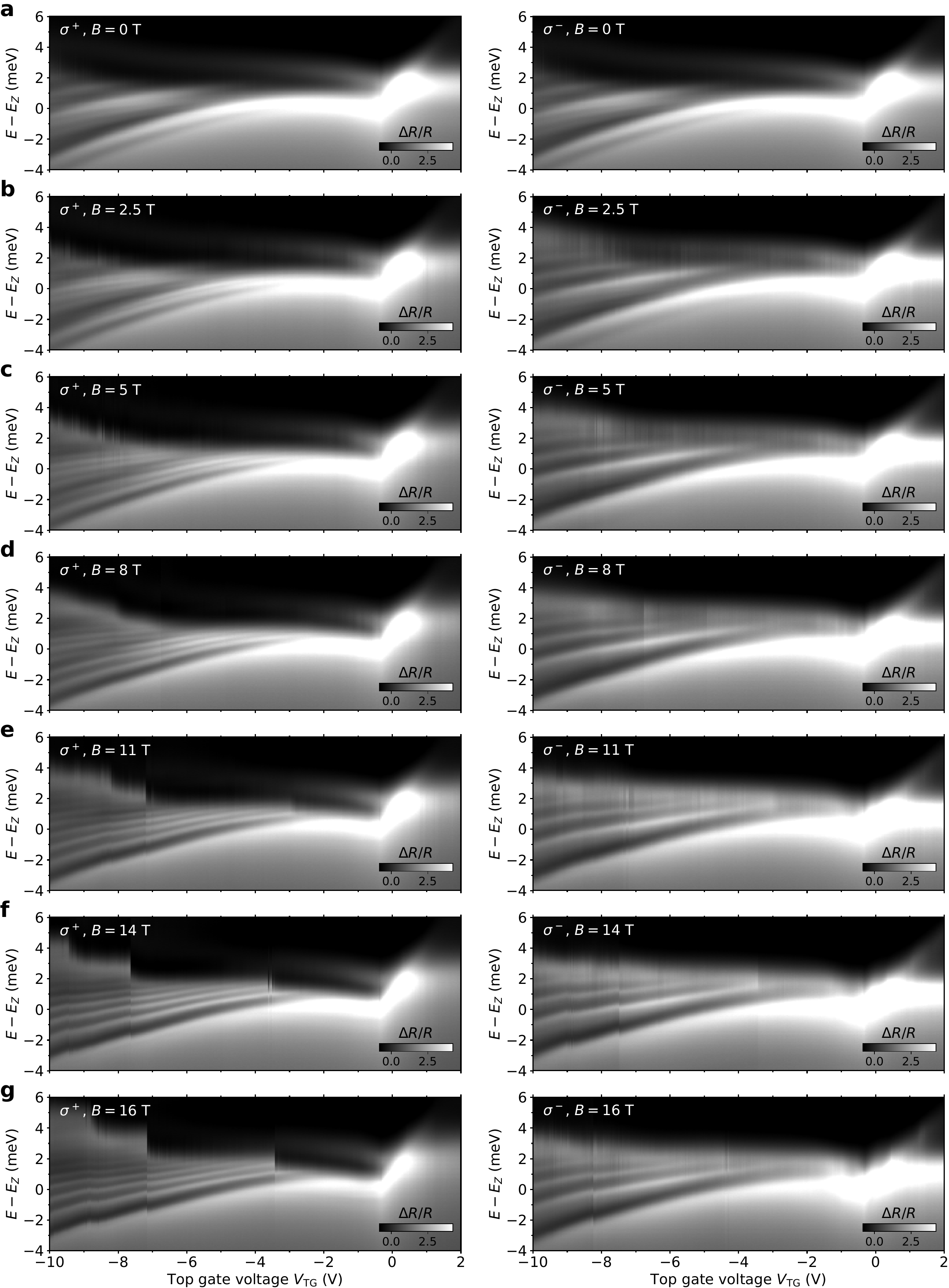}
	\caption{\textbf{Evolution of gate-dependent differential reflectance spectra with magnetic field.}
    % The left/right panels depict reflectance spectra in Device 2, acquired in $\sigma^+$/$\sigma^-$ measurement basis, respectively, at various magnetic fields. All spectra are plotted with respect to the free exciton energy $E_\mathrm{Z}$ at a given magnetic field.
    $\Delta R/R$ spectra in Device 2, acquired in $\sigma^+$ (left) and $\sigma^-$ (right) measurement basis for $\vbg=1\,$V at various magnetic fields. All spectra are shown with respect to the free exciton energy $E_\mathrm{Z}$ at a given magnetic field.
    }  
	\label{fig:B_excited}
\end{figure}

In the preceding section, we focused our attention solely on the magnetic field dependence of the lowest quantum confined state. However, a closer inspection of the individual spectra shown in Fig.\,\ref{fig:B-spectra} {\bfB} reveals intriguing modifications also for the excited states. Notably, at high magnetic fields, such as $B=16\,$T, the number of confined states appears to be greater in the $\sigma^+$ polarization basis than in the $\sigma^-$ basis. To more carefully examine this behavior, we present in Fig.\,\ref{fig:B_excited} the complete $\vtg$-dependent differential reflectance spectra at $\vbg=1\,$V and their evolution with increasing magnetic fields. To facilitate a comparison across the different gate scans, we plot all spectra with respect to the free exciton energy $E_\mathrm{Z}$ at a given magnetic field.

A striking difference between the two measurement bases can be discerned. While the total redshift of the lowest confined state remains at around $4\,$meV for all $B$-fields, the number of confined states in $\sigma^+$, corresponding to the lower Zeeman-split branch, clearly exceeds the number in $\sigma^-$. Qualitatively, this behavior can be interpreted as excitons in $K^+$ and $K^-$ valleys experiencing confinement of equal depth, albeit with a potential that is notably stiffer for the $K^-$ valley. Nevertheless, considering that the energies of the lower confined states are predominantly dictated by the dc Stark shift of a spatially varying in-plane electric field, no discernable valley-selective behavior should emerge, given that the device electrostatics are independent of the spin and valley index.

This reasoning does not hold for the interaction-induced confinement mechanism. Since the formation of polaron states is dominated by inter-valley exciton-charge interactions \cite{Efimkin2017,Back2017,Smolenski2018}, excitons in one valley should experience strong repulsive interactions only with carriers in the opposite valley. Since we expect all itinerant charge carriers in the device to be spin-polarized at $B=16\,$T, one may argue that a $K^+$ exciton will face a repulsive potential barrier only from charges in the $K^-$ valley, which at a positive $B$-field will correspond to electrons. We emphasize, however, that our confinement mechanism relies on excitons being confined in the neutral region of a \mbox{p-i-n} junction. Consequently, the reverse scenario, wherein $K^-$ excitons interact repulsively with spin-up polarized holes, is equally valid, which again establishes full symmetry with respect to spin and valley indices.

\begin{figure}[htb]
    \centering
	\includegraphics[width=10cm]{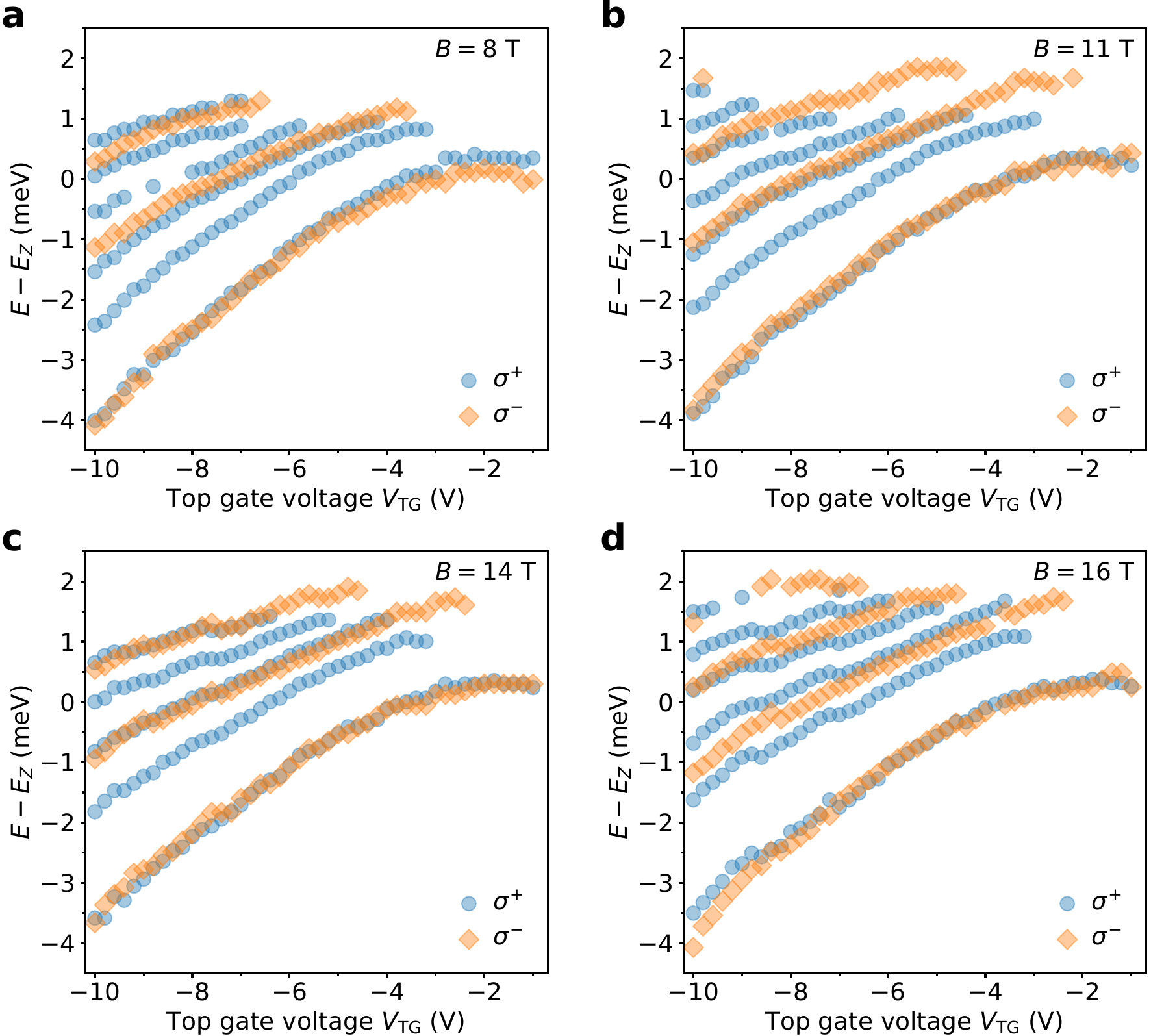}
	\caption{\textbf{Resonance energies of quantum-confined excitons at different magnetic fields.}
    Gate-dependent energy shifts of $\xqc$ states measured in $\sigma^+$ (blue) and $\sigma^-$ (orange) polarization basis for $\vbg=1\,$V at (\bfA) $B=8\,$T, (\bfB) $B=11\,$T, (\bfC) $B=14\,$T and (\bfD) $B=16\,$T. All energies are plotted with respect to the free exciton energy $E_\mathrm{Z}$ at a given magnetic field.
    }  
	\label{fig:B_excited_energies}
\end{figure}

To enable a more direct comparison of the confined states in $\sigma^+$ and $\sigma^-$, we overlay their extracted resonance energies in Fig.\,\ref{fig:B_excited_energies}. While the observed trend is not perfect, it appears that the $\sigma^-$ confined states overlap with every other state in $\sigma^+$ polarization, which suggests a possible relation to an effect rooted in the parity of the confined state wave function. Indeed, in section \ref{chap:1D:sec:signatures} (see also Fig.\,\ref{fig:3_9_parity}), we have argued that only even states should be bright, since their COM wave function is finite at $\mathbf{k}=0$. Nonetheless, it remains unclear how the application of a $B$-field could lead to brightening of an odd state in a valley-selective manner. Moreover, the evolution of gate scans for increasing $B$-fields shows a uniformity in oscillator strength across all confined states, especially evident in panels {\bfB} and {\bfC} of Fig.\,\ref{fig:B_excited}. Were there a brightening of a previously dark state, one might anticipate discernible variations in the oscillator strength between even and odd states -- however, such differences remain absent in experimental observations. Another possibility is to consider an additional lifting of degeneracy with application of a $B$-field, though the mechanism enabling such a valley-selective effect remains elusive. Therefore, additional experiments are needed to clarify these empirical findings. Specifically, it would be helpful to confirm the same behavior for $K^-$ excitons under the influence of a reversed magnetic field. In addition, exploring the $B$-field dependence of such confined excitons in devices with varying h-BN thicknesses, and those featuring gate electrodes made of gold as opposed to few-layer graphene, would further improve our understanding.

  %%%%%%%%%%%%%%%%%%%%%%%%%%%%%%%%%%%%%%%%%%%%%%%%%%%%%%%%%%%%%%%%%%%%%%%%
\chapter{Tunable exciton quantum confinement in 0D}
%%%%%%%%%%%%%%%%%%%%%%%%%%%%%%%%%%%%%%%%%%%%%%%%%%%%%%%%%%%%%%%%%%%%%%%%

%%%%%%%%%%%%%%%%%%%%%%%%%%%%%%%%%%%%%%%%%%%%%%%%%%%%%%%%%%%%%%%%%%%%%%%%
\section{Introduction}
\label{chap:0d:sec:intro}
%%%%%%%%%%%%%%%%%%%%%%%%%%%%%%%%%%%%%%%%%%%%%%%%%%%%%%%%%%%%%%%%%%%%%%%%

\begin{figure}[htb]
    \centering
	\includegraphics[width=12cm]{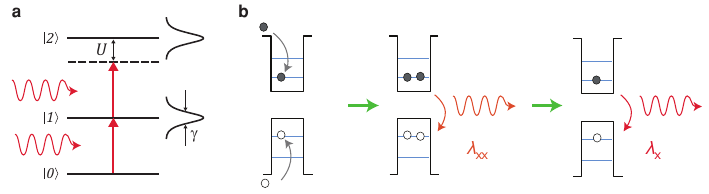}
	\caption{\textbf{Photon blockade due to inter-particle interactions.}
    ({\bfA}) An incoming photon in resonance with the system's fundamental transition will efficiently drive it from $\ket{0}$ to $\ket{1}$. However, inter-particle interactions cause a shift in energy of the doubly occupied state $\ket{2}$ by an amount $U$. Therefore, subsequent photons at the same wavelength will fail to drive the system further and ascend the ladder of states, causing the system to spontaneously emit one photon at a time. In addition, to achieve an efficient blocking, the magnitude of the interaction energy $U$ should exceed the decay rate $\gamma$ of a given state. ({\bfB}) Radiative cascade in a QD: Optically generated charge carriers in the barrier relax through the QD states via carrier collisions or phonon interactions (grey arrows). Their radiative recombination leads to sequential emission of two photons at separate wavelengths ($\lambda_\mathrm{XX}$, $\lambda_\mathrm{X}$) due to Coulomb interactions. The first photon is emitted as the biexciton $\ket{XX}$ relaxes to the exciton $\ket{X}$. Subsequently, relaxation into the QD ground state $\ket{0}$ leads to emission of the second photon. Reproduced with permission from Springer Nature \cite{Senellart2017}.
    }  
	\label{fig:0d_blockade}
\end{figure}

The observation of emergent phenomena in a many-body system crucially relies on the presence of interactions among its constituent elements. These interactions can manifest as a ``blockade'' where the occupancy of a state by one particle prevents concurrent occupation by another—a concept akin to Pauli's exclusion principle. And indeed, the first demonstration of such an effect involved electrons and was termed ``Coulomb blockade'' \cite{Fulton1987, Kastner1993}. While Pauli's exclusion principle is fundamentally grounded in the exchange antisymmetry of fermionic particles, a blockade effect can also be realized by leveraging repulsive interactions among excitations generated in a driven system. A notable instance occurs in ensembles of neutral atoms in highly excited electronic states, where long-range dipolar interactions give rise to a ``Rydberg blockade'' \cite{Urban2009,Gaetan2009}. As such, even in bosonic media blockade phenomena can arise, provided their energy level spectrum exhibits sufficient anharmonicity. This scenario is depicted in Fig.\,\ref{fig:0d_blockade} \bfA. When a single incoming photon is in resonance with a system's fundamental transition, it efficiently drives the system from $\ket{0}$ to $\ket{1}$. However, subsequent photons with the same energy become detuned from the \mbox{$\ket{1}\rightarrow\ket{2}$} transition due to inter-particle interactions, preventing them from ascending the ladder of states and driving the system further. Consequently, the system exclusively emits one photon at a time, thereby transforming a coherent stream of incoming photons, characterized by Poissonian fluctuations, into a sequence of individual photons, in essence mimicking a ``photon turnstile'' \cite{Kim1999,Michler2000}. To characterize the effectiveness of this process, the interaction energy $U$ needs to be put in comparison with the decay rate $\gamma$ of the respective state. Efficient blocking is thereby achieved when the ratio $U/\gamma$ exceeds unity, i.e.\,interactions start to dominate over dissipation. Evidently, in the limit where the nonlinearity $U$ is infinitely large, an ideal two-level system is realized.

It is therefore not surprising that pioneering experimental demonstrations of such single-photon emission were based on atoms \cite{Kimble1977}, ions \cite{Diedrich1987} and molecules \cite{Basche1992}. Currently, substantial efforts are also directed towards solid-state material systems \cite{Aharonovich2016}, with self-assembled quantum dots (QDs) earning considerable attention and emerging as a potential key enabler towards the realization of distributed quantum systems \cite{Kimble2008,Lodahl2015,Uppu2021}. The quantum confinement of electrons and holes within a QD leads to discrete energy levels for both charge carriers, effectively allowing the QD to function as an ``artificial atom''. Under non-resonant excitation, the radiative cascade within a QD can be succinctly described as follows (Fig.\,\ref{fig:0d_blockade} \bfB): Optically generated charge carriers in the barrier relax through the QD states via carrier collisions or phonon interactions, eventually resulting in radiative recombination and photon emission. The decay from the biexciton state $\ket{\mathrm{XX}}$ to the exciton state $\ket{\mathrm{X}}$ leads to the emission of a single photon. Subsequently, another photon is emitted during the transition from $\ket{\mathrm{X}}$ to the QD ground state $\ket{\mathrm{0}}$ \cite{Moreau2001,Santori2002}. Notably, the wavelength of the $\ket{\mathrm{XX}}\rightarrow\ket{\mathrm{X}}$ transition differs from that of the $\ket{\mathrm{X}}\rightarrow\ket{\mathrm{0}}$ transition due to strong Coulomb interaction between carriers. This distinctive feature enables spectral filtering of a single emission line, thus facilitating the realization of a single-photon source. 

\begin{figure}[htb]
    \centering
	\includegraphics[width=14cm]{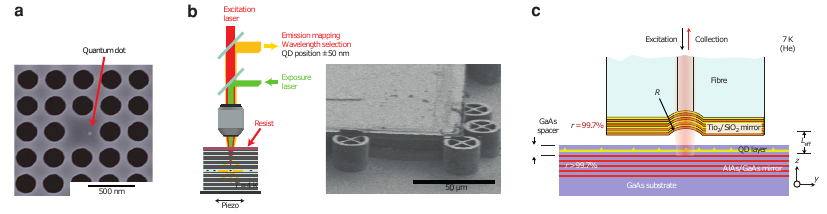}
	\caption{\textbf{Integration of self-assembled QDs in photonic cavities.}
    ({\bfA}) SEM image of a self-assembled QD placed at region of electric field maximum of a photonic crystal cavity. The QD location is identified using AFM / SEM, after which the cavity is patterned \cite{Badolato2005}. ({\bfB}) Left: Experimental scheme for cryogenic in-situ lithography \cite{Dousse2008}, in which first the QD position is optically identified by characterizing the spatial emission intensity. A second green laser line, aligned to the first, exposes the photoresist and allows patterning the micropillar cavity around the QD. Right: SEM image of four micropillar cavities containing a QD at their center \cite{Somaschi2016}. ({\bfC}) Schematic of open microcavity \cite{Muller2009}. One side of the cavity is formed by a QD layer grown atop a DBR mirror, while the opposing side consists of an external optical fiber facet housing a dimple, coated by a DBR. The QD-cavity detuning can be changed by modifying the fiber-substrate distance. $R$ is the mirror radius of curvature, $r$ is the mirror reflectivity, $L_\mathrm{eff}$ is the effective cavity length. Reproduced with permission: ({\bfA}) Ref.\,\cite{Badolato2005}, AAAS; ({\bfB}) Ref.\,\cite{Senellart2017}, Springer Nature; ({\bfC}) Ref.\,\cite{Muller2009}, AIP.
    }  
	\label{fig:0d_dot_cavity}
\end{figure}

Self-assembled QDs have yielded remarkable advances in quantum optics \cite{Gao2012,DeGreve2012,Schaibley2013,Delteil2015,Najer2019}. Nonetheless, the scalability of these QDs towards forming larger quantum systems faces numerous hurdles. Since self-assembled QDs are naturally formed by strain relaxation mechanisms, they occur at random spatial positions on the sample. Furthermore, since no two QDs are identical, slight variations in the QD geometry can lead to significant spectral inhomogeneous broadening. Both these effects pose a great challenge in interfacing QDs with photonic nanostructures, for example to improve photon extraction efficiencies through increased light--matter interaction. For accurate positioning of the photonic structure around the QD (Fig.\,\ref{fig:0d_dot_cavity} \bfA), its position needs to be precisely predetermined, for instance using scanning electron microscopy (SEM) or atomic force microscopy (AFM) \cite{Badolato2005,Hennessy2007}. Subsequently, multiple digital etching steps might be required to tune the cavity in resonance with the QD. Alternatively, cryogenic in-situ lithography techniques have been devised, where first the QD position is optically determined by generating a spatial emission profile using a laser. A second laser line, aligned to the first, then patterns the photonic structure around the QD (Fig.\,\ref{fig:0d_dot_cavity} \bfB) \cite{Dousse2008}. Yet another approach makes use of open microcavities based on distributed Bragg reflectors (DBRs) \cite{Muller2009,MiguelSanchez2013}. One side of the cavity comprises a QD layer grown atop a DBR mirror, while the other side consists of an external optical fiber facet housing a dimple, coated by a DBR (Fig.\,\ref{fig:0d_dot_cavity} \bfC). This configuration has the advantage of allowing the cavity resonance to be tuned by modifying the fiber-substrate distance.

In light of these developments, a major objective in solid-state quantum optics has been to devise methods for achieving individual electrical control of QDs to facilitate their mutual coupling. Concurrently, their deterministic positioning in relation to a photonic crystal cavity or waveguide should be ensured \cite{Lodahl2015,Tuerschmann2019,Montblanch2023}. In this chapter, we emphasize how our method of tunable quantum confinement of excitons meets these demands, while offering additional benefits. In the subsequent chapter, we will elaborate on potential routes for integration with photonic structures. As illustrated in chapter \ref{chap:1D}, using our methodology, exciton confinement potentials can be precisely positioned through the appropriate design of gate electrodes. Consequently, the confined exciton energy can be electrically tuned, allowing to correct for spatial disorder. It is worth noting that the energy levels in self-assembled QDs can also be influenced by electric fields using gates through the quantum-confined Stark effect \cite{Miller1984}. However, our confinement scheme goes beyond merely modifying the energy levels of a pre-existing confinement potential. By carefully shaping the electrostatic potential in the device, \emph{we can dynamically manipulate the exciton COM confinement length}. This has two significant implications.

Firstly, as the exciton confinement length scale $\ell_x$ is reduced from the unconfined 2D limit to below $10\,$nm, the excitonic light--matter coupling strength $g^\mathrm{exc}$ is reduced $\propto \sqrt{A_\mathrm{exc}}$, where $A_\mathrm{exc}$ is the excitonic area in the center-of-mass (COM). Correspondingly, the oscillator strength of the confined exciton resonance is modified $\propto |g^\mathrm{exc}|^2 \propto A_\mathrm{exc}$ (see also section \ref{chap:theory:sec:LM_coupling}). Moreover, as elaborated in section \ref{chap:1D:sec:signatures}, this reduction in COM extent also leads to a decrease in excitonic radiative decay, thus causing a narrowing of the excitonic linewidth. Other factors that contribute to this effect are a reduced electron--hole wave function overlap induced by the in-plane electric field, and a lower inhomogeneous broadening since the exciton COM motion is restricted to a smaller region in space. In addition, a stronger COM localization will also impact the absolute exciton--exciton interaction shift $U$, as it scales inversely with the excitonic area $A_\mathrm{exc}$. \emph{These modifications grant us the capability to dynamically traverse various regimes of the ratio $U/\gamma$, while simultaneously preserving the potential for brighter excitonic emission than conventional solid-state quantum emitters.}

In particular, we can estimate a minimum confinement length scale necessary for accessing the ``blockade'' regime i.e.\,$U/\gamma \approx 1$. Theoretical calculations predict an exciton--exciton interaction strength, $\overline{U}$, to be approximately $3 E_\mathrm{B} a_\mathrm{B}^2 \sim 1\,\mu$eV\,$\mu$m$^2$ (where $E_\mathrm{B}$ and $a_\mathrm{B}$ are the exciton binding energy and Bohr radius, respectively) \cite{Ciuti1998,Shahnazaryan2017}. However, recent experimental studies report a considerably lower value, around an order of magnitude less, or $\overline{U} \sim 0.1\,\mu$eV\,$\mu$m$^2$ \cite{Barachati2018,Scuri2018,Tan2020,Uto2023}. With this lower interaction strength alongside a conservative linewidth estimate of around $300\,\mu$eV for $\xqc$ states (see section \ref{chap:1D:sec:signatures}), confining the exciton to an area of $300\,$nm$^2$ should suffice for achieving sub-Poissonian light emission. In the case of an isotropic trap, this equates to a confinement length $\ell_x = \ell_y$ of around $15-20\,$nm. This requirement could be further relaxed by accounting for the dipolar nature of our confined excitons, which would potentially lead to an enhanced interaction strength $\overline{U}$.

%%%%%%%%%%%%%%%%%%%%%%%%%%%%%%%%%%%%%%%%%%%%%%%%%%%%%%%%%%%%%%%%%%%%%%%%
\section{Design considerations and preliminary results}
%%%%%%%%%%%%%%%%%%%%%%%%%%%%%%%%%%%%%%%%%%%%%%%%%%%%%%%%%%%%%%%%%%%%%%%%

Various possibilities exist for designing heterostructures that permit such excitonic confinement length scales. Based on our experimental findings on 1D quantum confined excitons (see section \ref{sec:1DX_sims}), a gate geometry which allows for 0D quantum confinement should be designed such that the electrostatic potential can be modulated on the order of the bandgap $E_\mathrm{g}$ (approximately $2\,$eV) over a length scale of few tens of nanometers along both $x$- and $y$-directions.

%%%%%%%%%%%%%%%%%%%%%%%%%%%%%%%%%%%%%%%%%%%%%%%%%%%%%%%%%%%%%%%%%%%%%%%%
\subsection{Exciton confinement on a ring}
%%%%%%%%%%%%%%%%%%%%%%%%%%%%%%%%%%%%%%%%%%%%%%%%%%%%%%%%%%%%%%%%%%%%%%%%

\begin{figure}[htb]
    \centering
	\includegraphics[width=12cm]{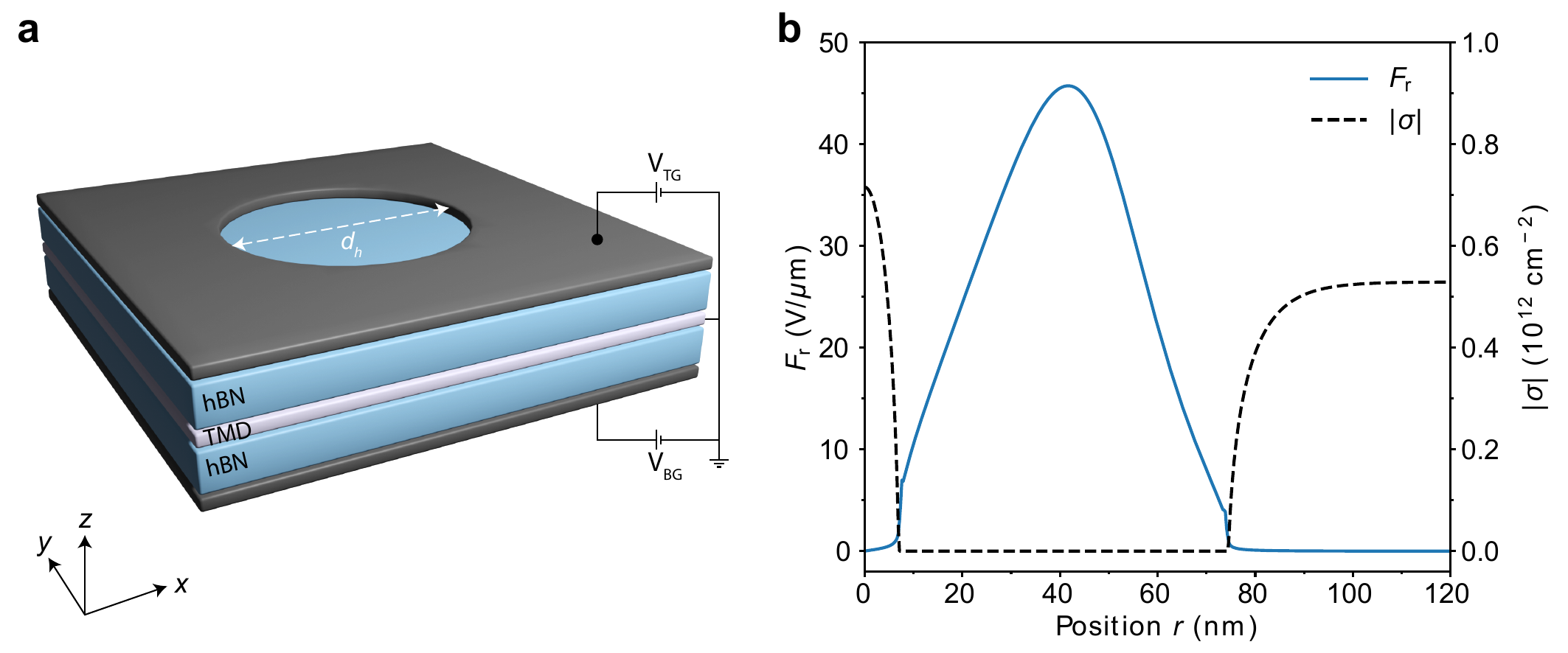}
	\caption{\textbf{Annular exciton confinement.}
    ({\bfA}) Schematic of a device architecture for achieving tunable exciton confinement in an annular geometry. The monolayer TMD is encapulated by h-BN. While the BG covers the entire TMD area, the TG contains an opening of diameter $d_h$. ({\bfB}) Magnitude of the in-plane electric field $|F_\mathrm{r}|$ and charge density $|\sigma|$ in the TMD layer as a function of the radial position coordinate $r$. The origin is set to be at the center of the opening. The diameter $d_h$ is set to $100\,$nm. The strongest in-plane electric fields result in a device doping configuration that leads to formation of a p-i-n junction, analogous to Fig.\,\ref{fig:3_3_Electrostatic_sims} ({\bfE}) and ({\bfF}). Therefore, the gate voltages are defined as $\vbg=-3\,$V and $\vtg=5\,$V. In this manner, a positively charged island is created in the center of the opening, surrounded by a globally electron-doped TMD. In the annulus demarcating the boundary between these two regions, radial in-plane electric fields $F_\mathrm{r}$ arise, leading to exciton confinement on a ring with a diameter $d_X$ of approximately $80\,$nm.
    }  
	\label{fig:confinement_annular}
\end{figure}

A straightforward device concept for achieving this consists of a heterostructure featuring a h-BN encapsulated TMD monolayer with a bottom gate (BG) that extends over the entire TMD area, and a top gate (TG) containing an opening of diameter $d_h$ (Fig.\,\ref{fig:confinement_annular} \bfA). To better comprehend the electrostatic landscape in such architectures we conduct electrostatic simulations, using the same procedure and material parameters as outlined in section \ref{sec:1DX_sims}. For the entire set of simulation results shown in this section the thickness of both the top and bottom h-BN dielectrics is assumed to be $20\,$nm. The diameter of the opening $d_h$ is set to $100\,$nm. Given that the geometry is rotationally symmetric, we can calculate the relevant electrostatic quantities, magnitude of the in-plane electric field and charge density, as a function of the radial position coordinate $r$. The origin is thereby defined as the center of the opening. To achieve strong in-plane electric fields the doping configuration should be such that a p-i-n junction is formed, analogous to Fig.\,\ref{fig:3_3_Electrostatic_sims} {\bfE} and {\bfF}. Therefore, we have set $\vbg=-3\,$V and $\vtg=5\,$V. The resulting magnitude of in-plane electric field $|F_\mathrm{r}|$ and charge density $|\sigma|$ distribution is depicted in Fig.\,\ref{fig:confinement_annular} \bfB. This voltage configuration leads to formation of a positively charged island centered around $r=0$, which is surrounded by a globally electron-doped TMD. The annular boundary between these two regions remains undoped and gives rise to radial in-plane electric fields $F_\mathrm{r}$. This constrains the excitonic motion to a ring with an approximate diameter $d_X$ of $80\,$nm.

One significant advantage of this device architecture, especially for the doping configuration illustrated in  Fig.\,\ref{fig:confinement_annular} \bfB, is that the regions surrounding the confined excitons are charged, and thus give rise to repulsive polaron (RP) states which exhibit a blueshift relative to excitonic resonances. Consequently, spectral signatures tied to confined excitons would appear unaffected by any broad background resonances. This greatly facilitates their investigation, either by enabling resonant excitation in a resonance fluorescence measurement, or through spectral filtering in a photoluminescence experiment. Essentially, in such a setting the spectral signatures would be similar to those displayed in Fig.\,\ref{fig:3_5_WL} {\bfA} for $\vtg$ below $-6.5\,$V, where only $\xqc$ states are observed devoid of any RP emission. An additional benefit of this architecture pertains to fabrication of such devices. Typically, when TMD heterostructures are stacked, local inhomogeneities in the device are unavoidable due to formation of bubbles or cracks. Given that the gate electrode on the top is patterned in this architecture, spatial regions with narrow exciton linewidths could be identified first, after which the desired feature is etched into the TG made of few-layer graphene. Various etch methods have been proposed, which have near-perfect etch selectivity between graphene and h-BN \cite{Son2018,Jessen2019}, such that only the TG is etched while leaving the h-BN underneath unaffected.

However, a disadvantage of this structure is that the confinement geometry is in the shape of a ring and thus not strictly zero-dimensional. The excitonic localization area approximately corresponds to $\pi\,d_X\cdot\ell_x\approx2500\,$nm$^2$, which greatly exceeds our imposed condition for reaching the blockade regime. Electrostatic simulations conducted in a wider parameter range have shown that a further reduction of the diameter $d_h$ in the TG results only in a marginal improvement. The diameter $d_X$ of the excitonic ring could certainly be reduced in this manner, however the range of modulation in electrostatic potential also decreases, thus effectively lowering the achievable magnitude of in-plane electric field strength. For this reason, it was decided to not pursue this device architecture further for experimental investigation. Nevertheless, we emphasize that such a device implementation could become interesting for realizing synthetic gauge fields for polaritonic systems \cite{Chestnov2021}.

%%%%%%%%%%%%%%%%%%%%%%%%%%%%%%%%%%%%%%%%%%%%%%%%%%%%%%%%%%%%%%%%%%%%%%%%
\subsection{Cross-gate confinement geometry}
%%%%%%%%%%%%%%%%%%%%%%%%%%%%%%%%%%%%%%%%%%%%%%%%%%%%%%%%%%%%%%%%%%%%%%%%

\begin{figure}[htbp]
    \centering
	\includegraphics[width=13.5cm]{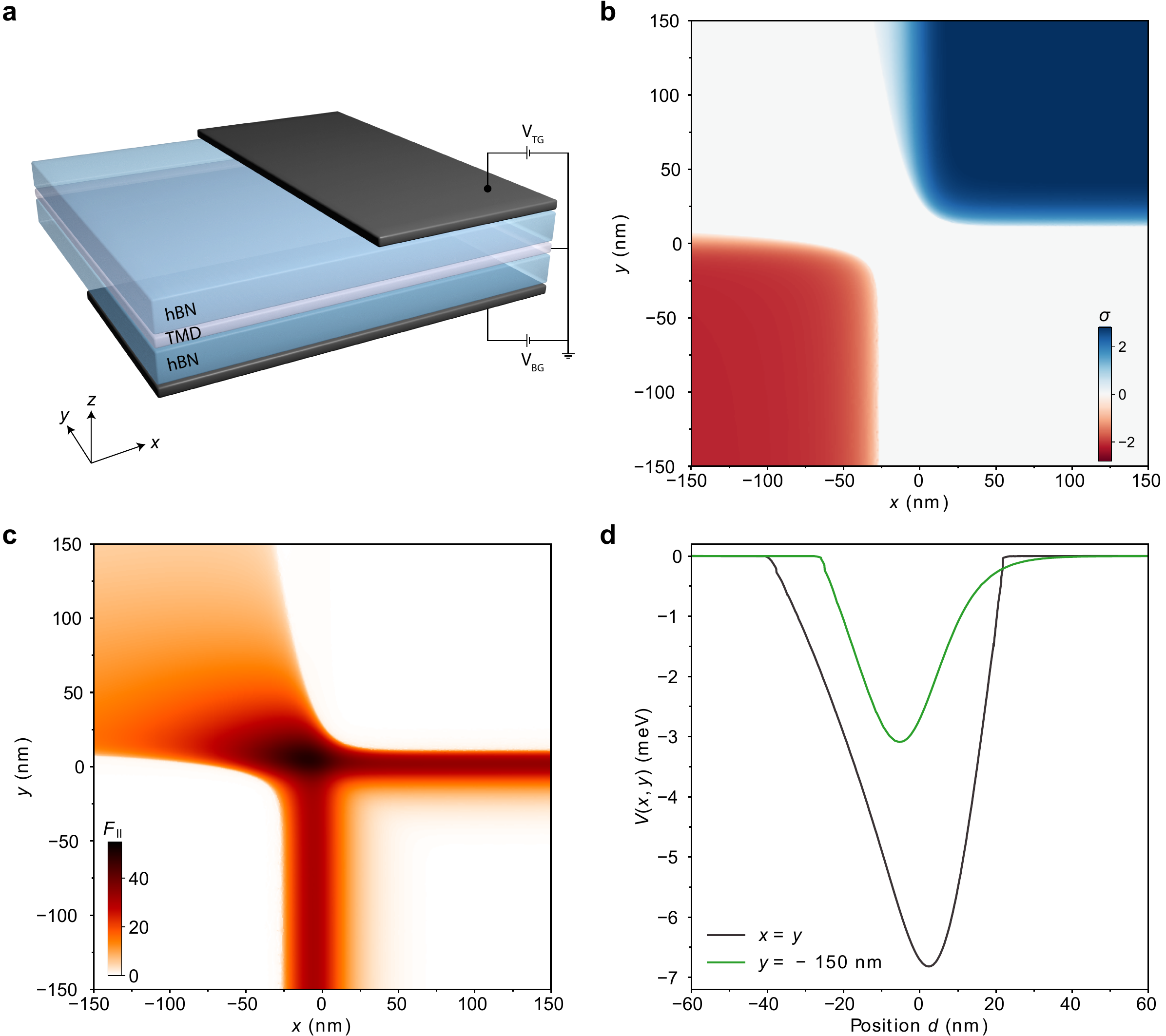}
	\caption{\textbf{Cross-gate confinement geometry.}
    A zero-dimensional quantum confining potential for excitons can be realized at the intersection of two one-dimensional confining potentials. ({\bfA}) This confinement scheme can be implemented by utilizing a TG covering only the right half of the device, and a BG covering only the bottom half. This results in four distinct areas: a dual-gated region I, a bottom-gated region II, a top-gated region III and an ungated region IV. The charge-density configuration of these regions (in units of $10^{12}\, \mathrm{cm}^{-2}$) is shown in (\bfB), for $\vbg=2\,$V and $\vtg=-2.5\,$V. Two regions remain neutral: region IV because it is ungated, and region I since TG and BG counteract each other in that area. Region II and III are electron-doped and hole-doped, respectively. The magnitude of the in-plane electric fields $F_{\parallel}$ (in units of V/$\mu$m) in this doping configuration is depicted in ({\bfC}). It clearly shows two one-dimensional regions of field maxima located at around $x=0$ and $y=0$, which coincide and lead to the strongest in-plane fields at $x=y=0$. (\bfD) Magnitude of the resulting field-induced confinement potential for excitons across the 1D and 0D traps, shown in green and black, respectively.
    }  
	\label{fig:confinement_cross-gate}
\end{figure}

Another possibility for realizing a zero-dimensional quantum confining potential for excitons is to form an intersection between two one-dimensional confining potentials. This strategy is reminiscent of the cleaved edge overgrowth method \cite{Pfeiffer1990,Goni1992,Wegscheider1997}, where quantum wires and dots can be fabricated using molecular beam epitaxy by controlled cleaving and regrowth of a quantum well on the cleaved facet. The proposed device architecture for implementing this confinement scheme is illustrated in Fig.\,\ref{fig:confinement_cross-gate} \bfA. The monolayer TMD semiconductor is thereby encapsulated by h-BN spacers of equal thickness. As we have established, a 1D quantum confining potential will form along the edge of the gate electrodes. Therefore, by employing a TG which extends over the right half of the TMD and a BG that solely spans the lower half, two such potentials can intersect at the device center. This arrangement of gate electrodes yields four distinct regions: a dual-gated region I, a bottom-gated region II, a top-gated region III, and an ungated region IV.

The resulting charge-density configuration of these regions, determined through electrostatic simulations, for an apt choice of gate voltages ($\vbg=2\,$V and $\vtg=-2.5\,$V) is shown in Fig.\,\ref{fig:confinement_cross-gate} \bfB. It clearly depicts the formation of a planar p-i-n junction, analogous to the electrostatic device configuration discussed in section \ref{sec:1DX_sims}. Intriguingly, the peculiar arrangement of gate electrodes results in a non-uniform depletion width, which tapers to its narrowest at a single point at the interface between regions II and III, corresponding to the center of the device. In accordance with this doping landscape, the analysis of the in-plane electric field magnitude $F_\parallel$ (Fig.\,\ref{fig:confinement_cross-gate} \bfC) reveals two prominent 1D regions of field maxima located at the boundary of region I. At their intersection point at $x=y=0$ the strongest in-plane fields are observed, thus giving rise to a highly localized confinement potential for excitons. A linecut of the field-induced confinement potential, determined according to Eqn.\,\ref{eqn:potential} (neglecting the interaction shift), across the device diagonal is shown in black in Fig.\,\ref{fig:confinement_cross-gate} \bfD. In comparison to the potential depth of the 1D trap (green), a sufficient energetic detuning should be achievable to distinguish the 0D from the 1D quantum confined states.

The strength of this device architecture lies in its simplicity. Following the same design principle as the devices described in chapter \ref{chap:1D}, the need for nanoscale patterning is completely mitigated in such a design, since both BG and TG are essentially extended gates. The sole requirement to be fulfilled is the perpendicular orientation of their respective edges, which can be easily accommodated in the device fabrication procedure. Furthermore, as evidenced by Fig.\,\ref{fig:confinement_cross-gate} {\bfC} and {\bfD}, the minimum of the confinement potential can be well approximated by a near-isotropic harmonic trap. Thus, achieving confinement lengths below $10\,$nm in both $x$- and $y$-directions appears feasible, promising a more straightforward access to the blockade regime than the previously discussed device architecture. A potential hindrance in this architecture is the excitonic background resonance associated with regions I and IV, which may obfuscate the detection of quantum-confined states.

\begin{figure}[htb]
    \centering
	\includegraphics[width=12.65cm]{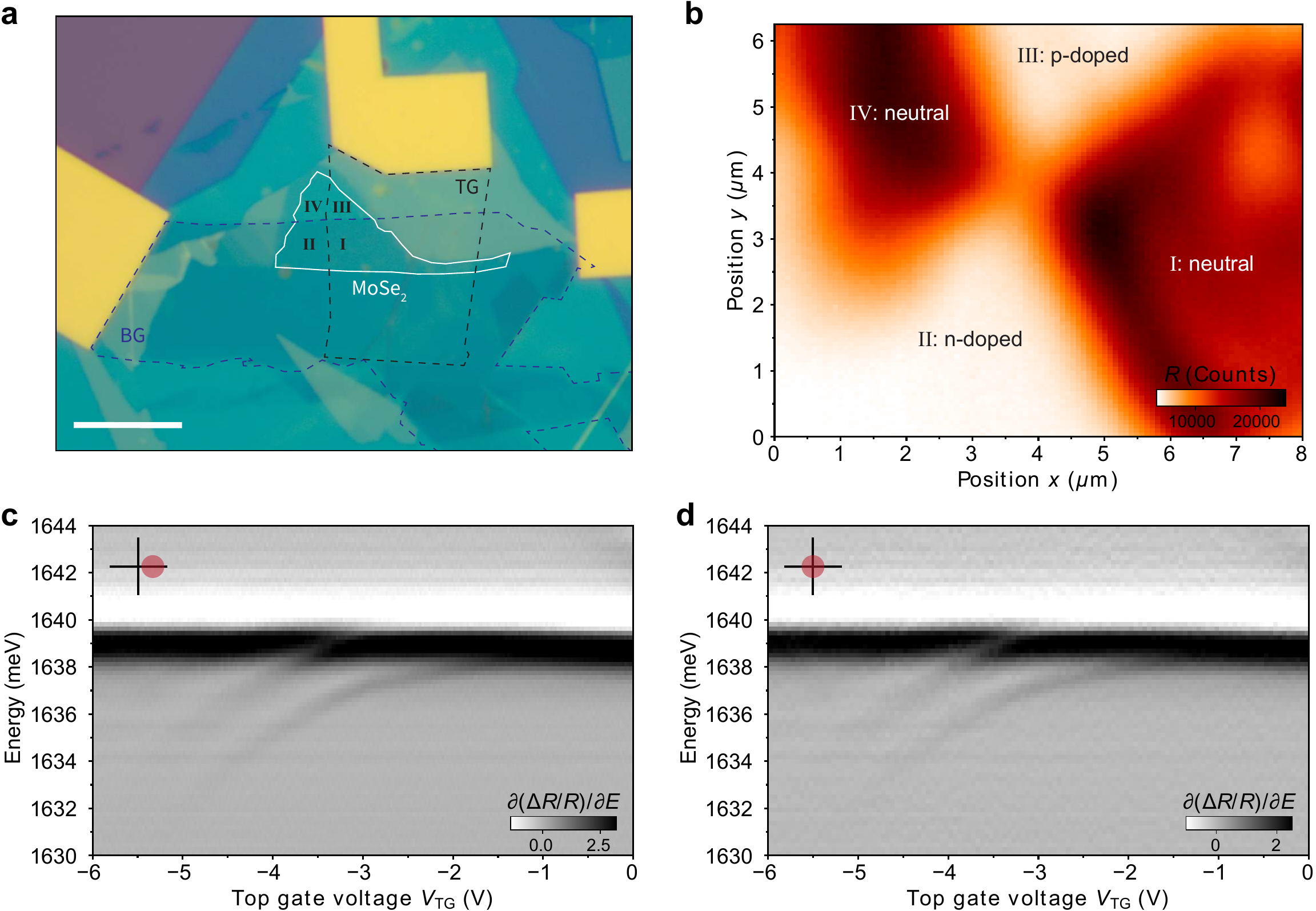}
	\caption{\textbf{Overview of of Device 3.}
    (\bfA) Optical micrograph of Device 3. The outline of the MoSe$_2$ monolayer is indicated by the white line. The BG and TG are made of few-layer graphene, and depicted with dashed blue and black lines. Their arrangement in the device allows creation of four distinct regions in the TMD monolayer: a dual-gated region I; a bottom-gated region II; region III, which is gateable through the TG and the global silicon BG; region IV, which is only gateable through the global silicon BG. The contact to the monolayer MoSe$_2$ is also formed by few-layer graphene (not shown). The scale bar denotes $10\,\mu$m. (\bfB) Position-dependent bare reflectance $R$ shown at the 2D exciton resonance energy $E_\mathrm{X,2D}$, for $\vbg=4\,$V and $\vtg=-4\,$V. The global silicon BG is grounded. This configuration allows doping only regions II and III, while maintaining charge neutrality in I and IV, analogous to the charge-density map shown in Fig.\,\ref{fig:confinement_cross-gate} (\bfB). (\bfC), (\bfD) Derivative of reflection contrast measured as a function of $\vtg$ for $\vbg=1.5\,$V at the boundary between region I and III, and at the intersection point at the device center as shown in the inset, respectively. While clear discrete redshifting resonances indicating quantum-confined states are visible in both cases, no clear distinction between the two can be made.
    }  
	\label{fig:device_cross-gate}
\end{figure}

Nevertheless, owing to the ease of fabrication, Device 3 was prepared with an arrangement of gates as depicted in Fig\,\ref{fig:confinement_cross-gate} \bfA. An optical micrograph of this device is shown in Fig.\,\ref{fig:device_cross-gate} \bfA. It consists of a MoSe$_2$ monolayer encapsulated by top and bottom h-BN dielectric spacers of thickness $20\,$nm and $13\,$nm, respectively. The gate electrodes along with the contact to the MoSe$_2$ monolayer are made of few-layer graphene, which are connected to metal electrodes formed with Ti/Au ($5\,$nm/$85\,$nm). Additional details about the fabrication procedure can be found in appendix section \ref{appendix:fabrication}. Fig.\,\ref{fig:device_cross-gate} {\bfA} clearly shows the formation of four different regions based on their overlap with the respective gate electrodes. While the BG enables electrostatic potential modulation in regions I and II, the TG affects regions I and III. In contrast to the device structure investigated by electrostatic simulations, Device 3 is placed on a Si/SiO$_2$ substrate, which allows modifying the doping configuration in regions III and IV through the global silicon BG.

To verify whether the ideal doping configuration, as depicted in Fig.\,\ref{fig:confinement_cross-gate} {\bfB}, can be physically realized, we perform a position-dependent reflectance measurement on Device 3, using an experimental setup as described in section \ref{chap:1D:sec:devices}. The resulting bare reflectance $R$ at the 2D exciton resonance energy $E_\mathrm{X,2D}$ is illustrated in Fig.\,\ref{fig:device_cross-gate} {\bfB}, for $\vbg=4\,$V, $\vtg=-4\,$V and a grounded global Si BG. Owing to the bright excitonic emission originating from regions I and IV, we conclude that charge neutrality is maintained in these areas. On the other hand, lack of emission from regions II and III indicates the presence of charges, since the resonance energy is blueshifted. Thus, a p-i-n junction with a point-like interface between the p-doped and n-doped domains can be formed.

Furthermore, this spatial map allows us to precisely position our optical spot for interrogating quantum-confined states. First, we investigate the boundary between region I and III to characterize 1D quantum-confined excitons. We set $\vbg=1.5\,$V and measure the normalized differential reflectance $\Delta R/R$ as a function of $\vtg$, the derivative of which with respect to energy is plotted in Fig.\,\ref{fig:device_cross-gate} \bfC. In line with previous observations, as $\vtg$ is lowered multiple spectral lines split off from a 2D excitonic continuum, centered around $1640\,$meV. A clear redshift of approximately $6\,$meV can be observed, which allows for a clear distinction between $\xfree$ and $\xqc$ states. In addition, an energy splitting $\Delta E$ of around $2-2.5\,$meV can be discerned, as opposed to $\Delta E \sim 1.5\,$meV seen in Device 1, indicative of a tighter confinement potential due to thinner h-BN spacer layers.

However, this device also exhibits features that are contrary to expectation. For example, the $\xqc$ states exhibit a linewidth of around $1\,$meV, which is broader than the corresponding spectral features in Device 1. A tighter confinement should lead to narrower lines. Based on this empirical evidence, combined with the observation of similarly broad $\xqc$ states in Device 2, we currently suspect that gate electrodes made of few-layer graphene could potentially be the reason for this effect. While a comprehensive microscopic understanding of this behavior is lacking, we hypothesize that the large intrinsic charge density in metallic gates, e.g.\,prepared with Au, could allow to more efficiently screen charge noise and thus help in suppressing electrostatic disorder. An additional concern associated with this device emerges when the same gate-dependent reflectance scan is conducted at the interface between the doped regions II and III (Fig.\,\ref{fig:device_cross-gate} \bfD). Notably, no pronounced differences can be identified when compared to 1D quantum-confined states seen at the boundary between regions I and III (Fig.\,\ref{fig:device_cross-gate} \bfC). To elucidate the underlying cause behind this problem, gathering an expanded dataset on such confinement geometries would be instructive. This will allow us to determine whether this issue is unique to the present device, possibly a consequence of local disorder, or conversely a fundamental flaw inherent to the proposed confinement scheme. Since in the current form each device only permits a single location for realizing 0D confinement, this would require fabricating a series of new devices.

Both the aforementioned points of concern can be addressed in a single device by employing gold-based gate electrodes patterned in an appropriate geometry. Nevertheless, an intrinsic issue with this design is that as the gate voltages are tuned, \emph{both} 1D and 0D quantum-confining potentials are affected simultaneously. The sole mechanism to distinguish optical signatures associated with each of them is based on an energetic detuning or a difference in oscillator strength. However, as evidenced by Figs.\,\ref{fig:device_cross-gate} {\bfC} and {\bfD}, if the 1D trap smoothly evolves into a more localized 0D trap centered at the crossing point, these factors might not be sufficient. A potential remedy to this predicament could be to rather probe the electroluminescence signal by driving a current through the doped regions. Since the p-i-n junction is at its narrowest only at a single point, the emission will predominantly originate from the crossing point and thus carry signatures only of the 0D trap. In terms of device fabrication, this would necessitate the incorporation of at least one additional contact to the TMD monolayer. Nonetheless, instead of pursuing this device concept further experimentally, we sought to devise a strategy which allows to more clearly distinguish signatures associated with both confining potentials directly in a reflectance or photoluminescence experiment.

%%%%%%%%%%%%%%%%%%%%%%%%%%%%%%%%%%%%%%%%%%%%%%%%%%%%%%%%%%%%%%%%%%%%%%%%
\subsection{Split-gate confinement geometry}
\label{chap:0d:sec:split-gate}
%%%%%%%%%%%%%%%%%%%%%%%%%%%%%%%%%%%%%%%%%%%%%%%%%%%%%%%%%%%%%%%%%%%%%%%%

\begin{figure}[htbp]
    \centering
	\includegraphics[width=13.5cm]{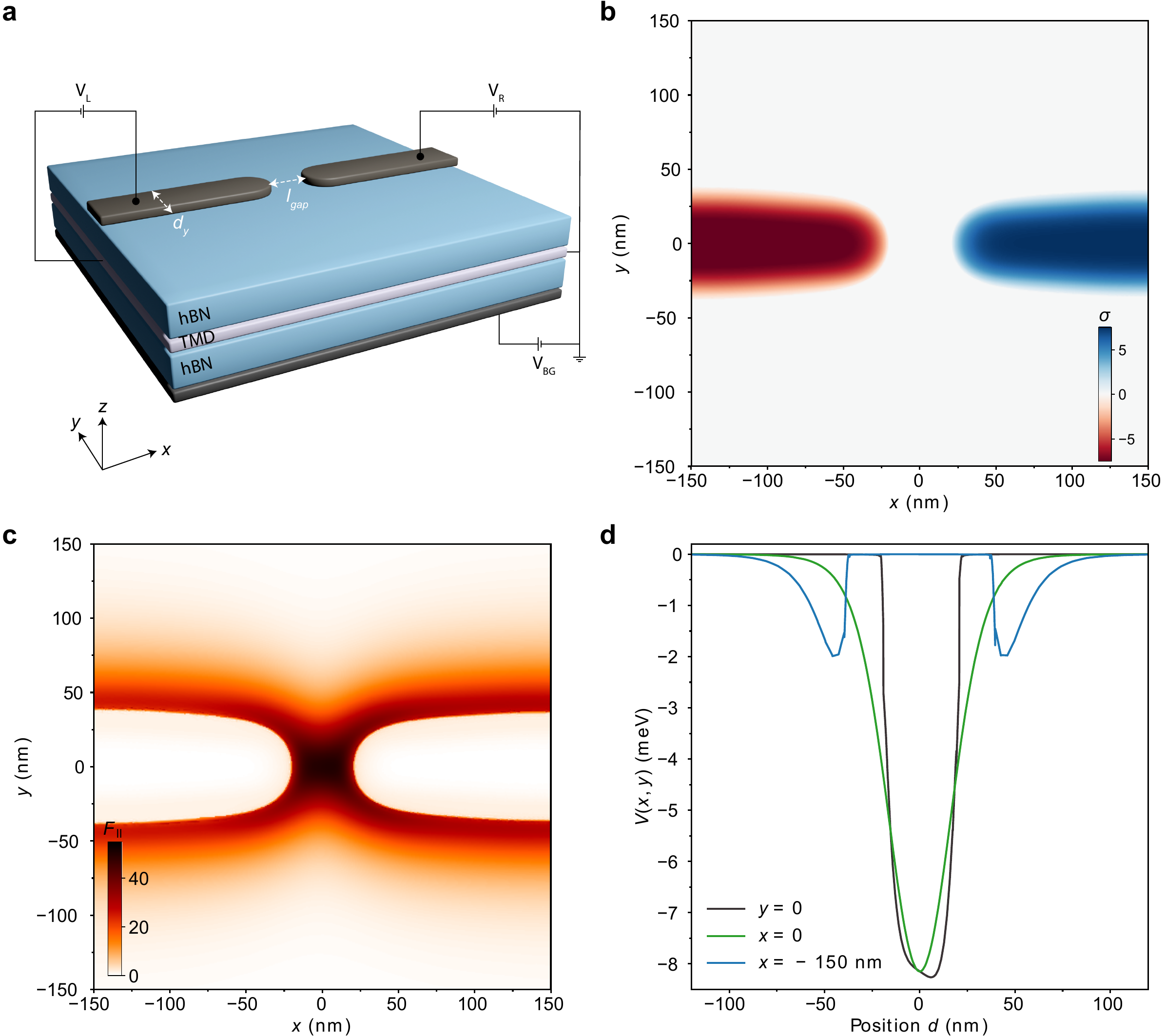}
	\caption{\textbf{Split-gate confinement geometry.}
    ({\bfA}) Proposed device architecture for realizing a zero-dimensional quantum confining potential for excitons. The monolayer TMD semiconductor is encapulated by h-BN. The BG extends across the entire TMD area, and the TG consists of a pair of split-gate electrodes, analogous to a quantum point contact, which are separated by a distance of $l_\mathrm{gap}$. The width of the TG electrodes is $d_y$. (\bfB) A desirable charge-density configuration can be achieved by creating a voltage asymmetry between the left and right TG electrodes, such that charges are injected only in the top-gated region. The extended TMD region remains charge neutral. Following parameters are assumed for the electrostatic simulation: $V_L = -V_R = 6\,$V, gap size $l_\mathrm{gap} = 50\,$nm, TG width $d_y=50\,$nm. The charge density is plotted in units of $10^{12}\, \mathrm{cm}^{-2}$. The magnitude of the in-plane electric fields $F_{\parallel}$ (in units of V/$\mu$m) in this nanoscale p-i-n junction is depicted in ({\bfC}). While the in-plane fields are finite along the gate electrode edges, they are the strongest in the gap region between the electrodes. (\bfD) Magnitude of the resulting 0D confinement potential for excitons along the $x$- and $y$-axis is shown in black and green, respectively. The 1D confining potential located at the gate electrode edge is also shown in blue for reference.
    }  
	\label{fig:confinement_split-gate}
\end{figure}

To decouple the 1D confinement occurring along the edge of gate electrodes from a 0D confinement an additional degree of freedom is required, such that both can be tuned independent of each other. This feat can be accomplished by adding a third gate electrode to the device architecture, as illustrated in Fig.\,\ref{fig:confinement_split-gate} \bfA. The base device structure is a monolayer TMD semiconductor encapsulated by h-BN dielectrics of equal thickness and features a BG that extends across the entire TMD region. The TG consists of a pair of split-gate electrodes, analogous to a quantum point contact, which are separated by a distance $l_\mathrm{gap}$ and denoted by $\vl$ and $\vr$.

To realize a 0D confinement potential in this geometry, a voltage asymmetry between the left and right TG electrodes is created, leading to electron and hole injection under the respective gates, while the extended TMD region remains charge neutral. The charge-density configuration, as determined through electrostatic simulations, for such a scenario is shown in Fig.\,\ref{fig:confinement_split-gate} \bfB. The voltages are defined as $V_L = -V_R = 6\,$V, along with a gap size $l_\mathrm{gap}$ of $50\,$nm and TG width $d_y$ of $50\,$nm. The resulting magnitude of the in-plane electric field distribution $F_{\parallel}$ in the TMD layer is plotted in Fig.\,\ref{fig:confinement_split-gate} \bfC. While finite in-plane fields persist along the edge of the TG electrodes, they are clearly the strongest in the gap region. The field-induced confinement potential for excitons along the $x$- and $y$-directions can be determined using Eqn.\,\ref{eqn:potential} (neglecting the interaction shift), and is illustrated in Fig.\,\ref{fig:confinement_split-gate} {\bfD} in black and green, respectively. A linecut across the 1D confinement potential along the left TG is shown in blue.

Evidently, the confinement length scales $\ell_x$ and $\ell_y$ in this device architecture are set by the gap size $l_\mathrm{gap}$ and the TG width $d_y$, respectively. Therefore, nanoscale patterning is unavoidable, which adds to the device fabrication complexity. However, for gold electrodes made with electron beam lithography (EBL) a value of sub-$50\,$nm for both these parameters is within reach. In this scenario, $\ell_x$ and $\ell_y$ can reach values under $10\,$nm, such that the excitonic confinement occurs on an area $<100\,$nm$^2$, which is an encouraging prospect towards reaching the blockade regime. 

Other than this experimental inconvenience, this device architecture holds several significant advantages, which allows to greatly facilitate the isolation of optical signatures associated with 0D excitons from all the other background resonances. While the charge neutral global TMD region will inevitably lead to a persistent excitonic resonance, the achievable electric field strengths are sufficient to attain a redshift greater than the 2D exciton linewidth (Fig.\,\ref{fig:confinement_split-gate} {\bfD}, black and green curves). The 0D states are also more naturally separated from the 1D states. As indicated by the blue curve in Fig.\,\ref{fig:confinement_split-gate} {\bfD}, the 1D confinement potential is shallower for the split-gate geometry than the cross-gate device geometry (Fig.\,\ref{fig:confinement_cross-gate} {\bfD}), even though the charge densities in the former are greater in magnitude than the latter. This is because charge neutrality in region I of the cross-gate design is achieved by the BG and TG counter-acting their electrostatic potential modulation, giving rise to stronger electric fields in general. Furthermore, the 1D edge confinement in the split-gate scenario is purely a result of the voltage asymmetry between the TGs and the global BG. Therefore, by incorporating a bottom h-BN spacer layer in the device stack, which is much thicker than the top spacer layer, the magnitude of the fringing fields at the gate edge can be substantially reduced (see also Eqns.\,\ref{eqn:depletion_width} and \ref{eqn:field_max}). Since the excitonic potential in the gap region is solely defined by the voltage asymmetry between the TG electrodes, it remains unaffected by this modification. In addition, an even greater sensitivity to the 0D states can be achieved by employing Stark-shift modulation spectroscopy \cite{Alen2003}, in which the left and right gate voltages are modulated at distinct frequencies, and the optical signal is demodulated at the sum- or difference-frequency using lock-in detection. Investigation of the electroluminescence signal originating from the gap region is also possible, in similar fashion to the case of cross-gate confinement described earlier. Lastly, the location of 0D confinement can be freely selected post-device preparation and scaled up to multiple sites with ease, constrained only by the available area of the TMD monolayer.

\begin{figure}[htbp]
    \centering
	\includegraphics[width=14cm]{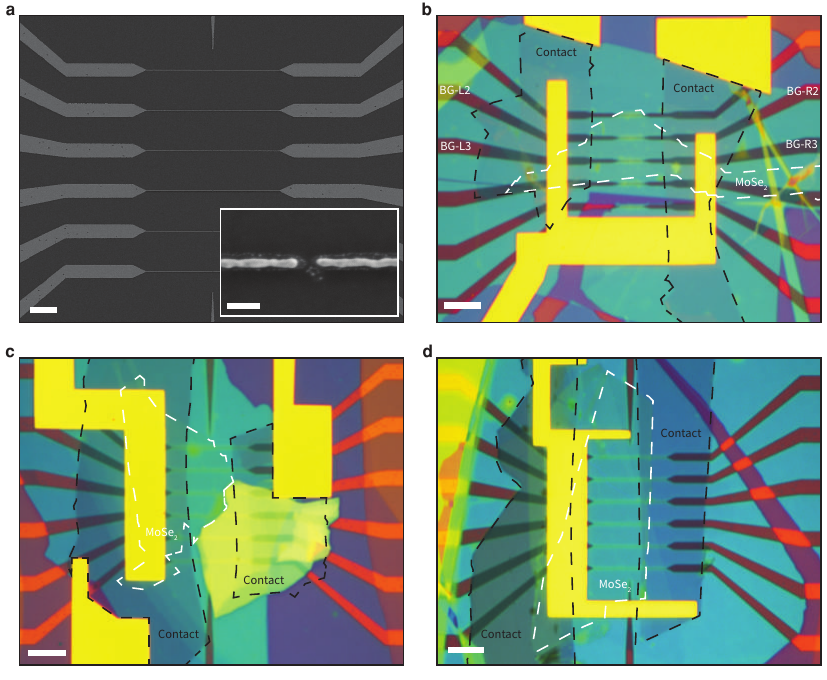}
	\caption{\textbf{Micrographs of Devices 4, 5 and 6.}
    ({\bfA}) SEM image of pairs of split-gate electrodes fabricated on a Si/SiO$_2$ substrate (scale bar: $2\,\mu$m). The inset shows a magnified image of an isolated set of split-gates with a width $d_y=25\,$nm and gap size $l_\mathrm{gap}=60\,$nm (scale bar: $100\,$nm). (\bfB)-(\bfD) Optical micrographs of Devices 4, 5 and 6, respectively. The white dashed line indicates the monolayer MoSe$_2$ flake, which is encapsulated in h-BN, and deposited on the bottom split-gate electrodes. Contacts to the TMD monolayer are formed by few-layer graphene, indicated in black dashed lines. All devices are capped with a global TG, also made of few-layer graphene (not shown). For Device 4 the split-gate electrodes used for device characterization are also labelled. The parameter $l_\mathrm{gap}$ is varied across the three devices and has a mean value of $80\,$nm for Device 4, $70\,$nm for Device 5 and $60\,$nm for Device 6. The scale bar denotes $5\,\mu$m.
    }  
	\label{fig:device_split-gate}
\end{figure}

Due to these advantageous features, we sought to fabricate multiple devices with a split-gate architecture for varying gap sizes $l_\mathrm{gap}$ and gate widths $d_y$ to realize different confinement length scales and thus determine the best tradeoff between oscillator strength and optical nonlinearity. Furthermore, for the first iteration of devices, split-gate electrodes were placed at the bottom of the device stack, rather than the top. This allowed us to perform a pre-characterization of the gate electrodes using SEM to determine $l_\mathrm{gap}$ and $d_y$ and overall ensure that none of the electrodes are electrically shorted. Fig.\,\ref{fig:device_split-gate} {\bfA} depicts a SEM image of multiple pairs of split-gate electrodes prepared using EBL and evaporation of Ti/Au ($3\,$nm/$10\,$nm) on a Si/SiO$_2$ substrate. The inset shows a magnified view of an isolated pair of split-gate electrodes with $l_\mathrm{gap}<100\,$nm. In total, three such pre-patterned substrates are fabricated with the following ($l_\mathrm{gap}$, $d_y$) configuration: ($80\,$nm, $70\,$nm), ($70\,$nm, $30\,$nm) and ($60\,$nm, $30\,$nm), to be used in Device 4, Device 5 and Device 6, respectively. Optical micrographs of these devices are shown in Fig.\,\ref{fig:device_split-gate} \bfB-\bfD. They consist of a monolayer MoSe$_2$ flake encapsulated by h-BN dielectric spacers with a thickness of approximately $20-25\,$nm. An extended TG that covers the entire TMD region is made with few-layer graphene. Additionally, all devices feature two contacts, also made with few-layer graphene, to the MoSe$_2$ monolayer, to potentially drive a current through the doped regions and probe excitonic states in the gap region by means of electroluminescence. Metal electrodes to both contacts and the TG are formed with Ti/Au ($5\,$nm/$85\,$nm). Additional fabrication details are provided in appendix section \ref{appendix:fabrication}.

\begin{figure}[htb]
    \centering
	\includegraphics[width=10.75cm]{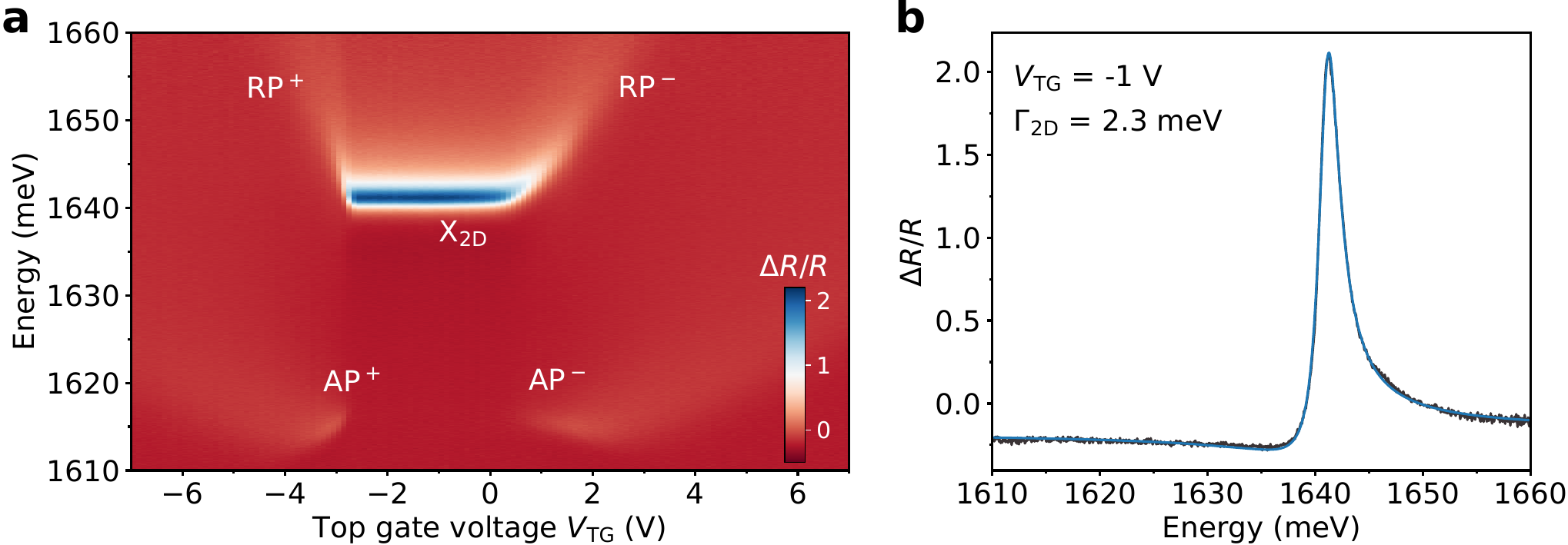}
	\caption{\textbf{Differential reflectance in Device 4.}
    ({\bfA}) The normalized differential reflectance $\Delta R/R$ as a function of $\vtg$, acquired in an extended TMD region away from the split-gate electrodes. The bare 2D exciton resonance appears during device charge neutrality and is indicated with X$_\mathrm{2D}$. The hole-side and electron-side repulsive polaron branches (RP$^+$ and RP$^-$), along with their attractive counterparts (AP$^+$ and AP$^-$) are also denoted. (\bfB) Spectrum taken at $\vtg=-1\,$V and fit to a line shape according Eqn.\,\ref{eqn:spec_func}. The 2D exciton resonance exhibits a linewidth of $2.3\,$meV.
    }  
	\label{fig:device4_reflectance}
\end{figure}

Among this series of devices, Device 4 will exhibit the largest confinement length scales and thus the greatest excitonic oscillator strength, facilitating the spectral characterization of quantum-confined states. Thus, we have performed preliminary experimental characterization of 0D quantum-confined states on this device and will discuss these results in the remainder of this section. To achieve the desired doping configuration (Fig.\,\ref{fig:confinement_split-gate} \bfB), charge neutrality should be maintained in the extended TMD region. An appropriate value for $\vtg$ can be determined by acquiring normalized differential reflectance $\Delta R/R$ on a region of the device away from the split-gate electrodes (experimental procedures as in section \ref{chap:1D:sec:devices}). For normalization we use the bare reflectance free of excitonic resonances, obtained by heavily doping the device. Fig.\,\ref{fig:device4_reflectance} {\bfA} shows the typical doping-dependent behaviour of a gated monolayer in reflectance, with the onset of doping characterized by emergence of repulsive polaron (RP) and attractive polaron (AP) branches. Charge neutrality, indicated by presence of the exciton resonance $\xfree$, is maintained for $-2.5\,\mathrm{V}\lesssim\vtg\lesssim0\,\mathrm{V}$. A narrow linewidth of $2.3\,$meV is extracted for $\xfree$ by fitting a line shape according to Eqn.\,\ref{eqn:spec_func} (Fig.\,\ref{fig:device4_reflectance} \bfB).

\begin{figure}[htb]
    \centering
	\includegraphics[width=13cm]{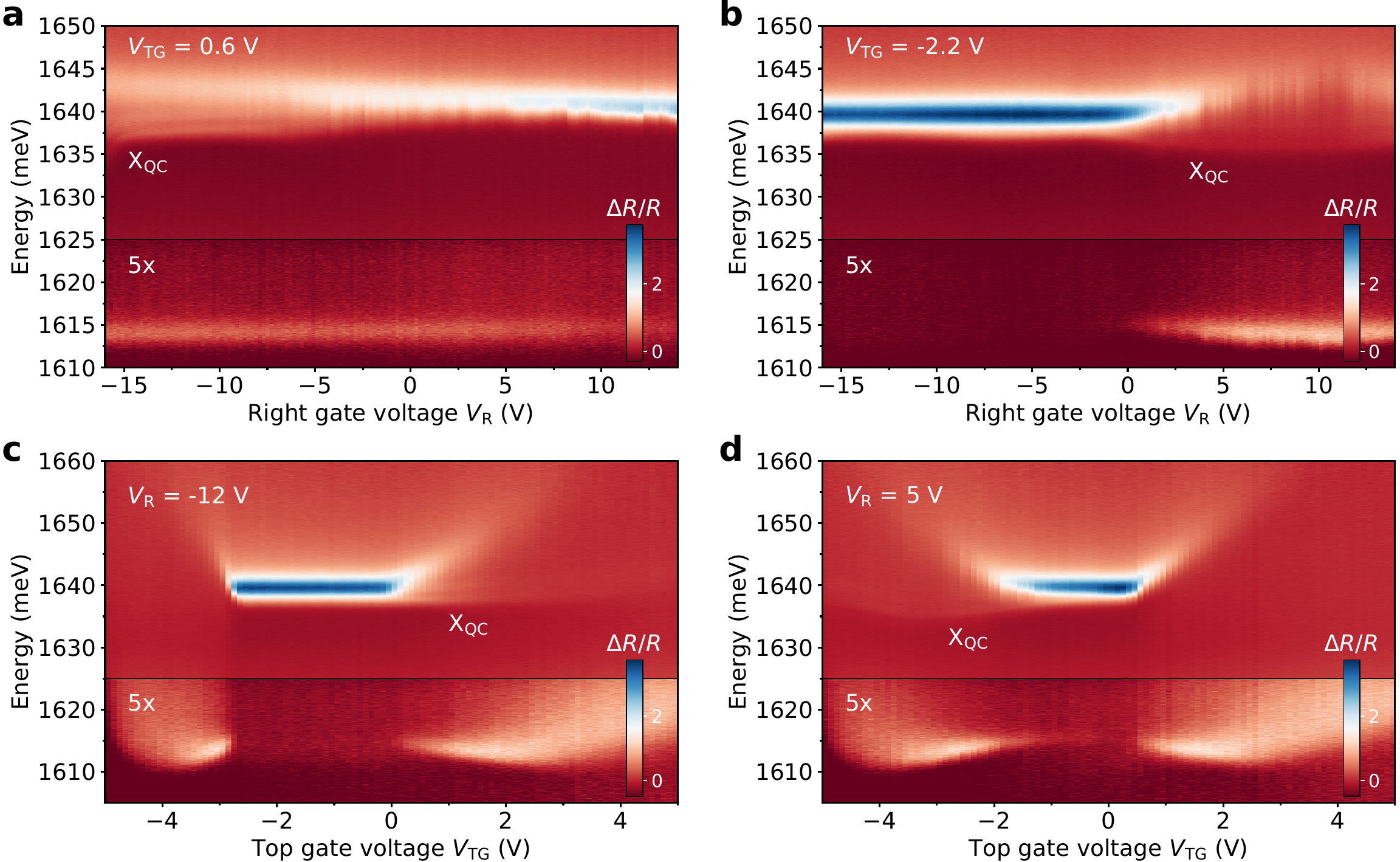}
	\caption{\textbf{Signatures of 1D quantum excitons in Device 4.}
    Reflectance measurement conducted on the central narrow part of electrode BG-R3 away from the gap region. Owing to the weak oscillator strength of spectral signatures in the low energy range, they are shown amplified by a factor 5. (\bfA), (\bfB) $\Delta R/R$ as a function of right split-gate voltage $\vr$, fixed $\vtg=0.6\,$V and $\vtg=-2.2\,$V, respectively. (\bfC), (\bfD) $\Delta R/R$ as a function of global TG voltage $\vtg$, fixed $\vr=-12\,$V and $\vr=5\,$V, respectively. In addition to the typical features associated with the neutral exciton and repulsive/attractive polaron branches, multiple narrow spectral lines ($\xqc$) are observed, which split off from the neutral exciton resonance. In line with our observations on 1D quantum-confined excitons in previous devices, these discrete states emerge for positive $\vtg$ at negative $\vr$, and vice versa.
    }  
	\label{fig:device4_1DX}
\end{figure}

To probe the optical response of the split-gate confinement geometry, we primarily focus on the gap regions formed by BG electrodes BG-L2 and BG-R2, and BG-L3 and BG-R3, hereinafter referred to as gap 2 and gap 3, respectively (Fig.\,\ref{fig:device_split-gate} \bfB). First, we identify signatures associated with 1D quantum-confined excitons to establish a reference point for the subsequent optical characterization of the gap region. To this end, we measure normalized differential reflectance $\Delta R/R$  on the central narrow part of BG-R3 away from the gap region, as shown in Fig.\,\ref{fig:device4_1DX}. In panels {\bfA} and {\bfB}, the right gate voltage $\vr$ is tuned for fixed $\vtg=0.6\,$V and $\vtg=-2.2\,$V, respectively. Similarly, panels {\bfC} and {\bfD} depict $\Delta R/R$ for varying $\vtg$ and fixed $\vr=-12\,$V and $\vr=5\,$V, respectively. In all cases the typical exciton and polaron resonances can be identified. In addition, multiple narrow $\xqc$ states are visible, which split off from the the neutral exciton resonance and redshift. In accordance with observations on previous devices, their appearance is exclusively tied to a voltage asymmetry between the scanned gate voltages, i.e.\,they are observable only for positive $\vtg$ and negative $\vr$ (Fig.\,\ref{fig:device4_1DX} {\bfA} and {\bfC}) or negative $\vtg$ and positive $\vr$ (Fig.\,\ref{fig:device4_1DX} {\bfB} and {\bfD}).

\begin{figure}[htb]
    \centering
	\includegraphics[width=7cm]{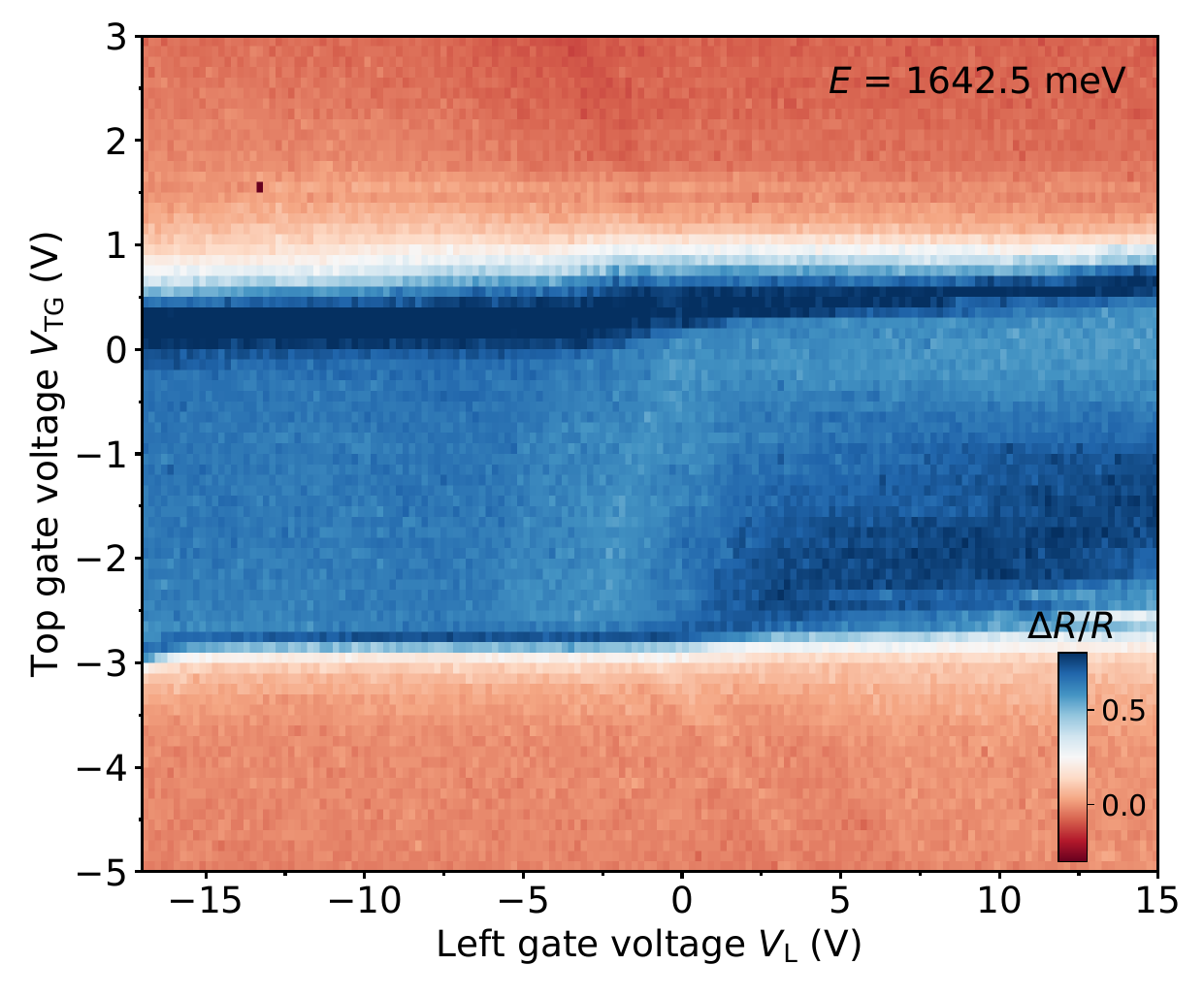}
	\caption{\textbf{Doping characteristics of Device 4.}
    We focus on charge-density-dependent energy shifts of the repulsive polaron to identify the doping configuration of Device 4. We measure normalized differential reflectance $\Delta R/R$, both as a function of $\vl$ (BG-L3) and $\vtg$, and plot the result at fixed energy, which is chosen slightly blueshifted to the bare 2D exciton i.e.\,at $E=1642.5\,$meV. Similar spectral features are observed as in Device 1 (Fig.\,\ref{fig:3_11_Doping} \bfC). The prominent horizontal lines demarcate the n-doped, neutral and p-doped regimes in the extended TMD region. The fainter set of resonances exhibiting a zigzag behaviour originate in the TMD region above BG-L3 and indicate onset of doping in this area. It can be clearly seen that for $\vtg$ ranging from approximately $-1\,$V to $0\,$V, $\vl$ can be tuned across its entire range without greatly affecting charge neutrality in the extended TMD region.
    }  
	\label{fig:device4_doping}
\end{figure}

We can verify whether these $\xqc$ states emerge as a result of a p-i-n junction that is formed at the interface between the back-gated and extended TMD region by performing a full dual-gate scan, where the TG and BG voltages are varied across their entire range. Then, by mapping the differential reflectance $\Delta R/R$ at fixed energy $E = E_\mathrm{X,2D} + \Gamma_\mathrm{2D}/2 = 1642.5\,$meV the charging behaviour of the device can be characterized more rigorously. Since the optical spot is diffraction-limited the measured optical response will contain information about the doping state of both the dual-gated and extended TMD regions. In similar manner as for Device 1 (Fig.\,\ref{fig:3_11_Doping} \bfC), the emergence of an RP resonance will serve as an indicator for the onset of doping. Fig.\,\ref{fig:device4_doping} depicts the result of such a dual-gate scan, where $\vtg$ and $\vl$ (BG-L3) are varied. Overall, the observed features match very well with the doping characteristics of Device 1. The prominent horizontal lines indicate onset of electron doping for $\vtg\gtrsim0.5\,$V and hole doping for $\vtg\lesssim-3\,$V in the extended TMD region. Concurrently, a fainter set of resonances exhibiting a zigzag behaviour is visible. These spectral lines point at the doping state in the TMD region above BG-L3. To their right, for increasing $\vl$, the back-gated area is electron-doped. To their left, for decreasing $\vl$, it becomes hole-doped. Based on this charging map, the most desirable conditions for probing 0D states are for $-1\,\mathrm{V}\lesssim\vtg\lesssim0\,$V. Within this range the back-gated area can be both p-doped and n-doped, while least affecting charge neutrality in the extended TMD region, in good agreement with the desired doping configuration shown in Fig.\,\ref{fig:confinement_split-gate} \bfB. Such charging maps were also acquired for other split-gate electrodes (BG-L2, BG-R2 and BG-R3) which showed similar doping behaviour.

\begin{figure}[htb]
    \centering
	\includegraphics[width=14cm]{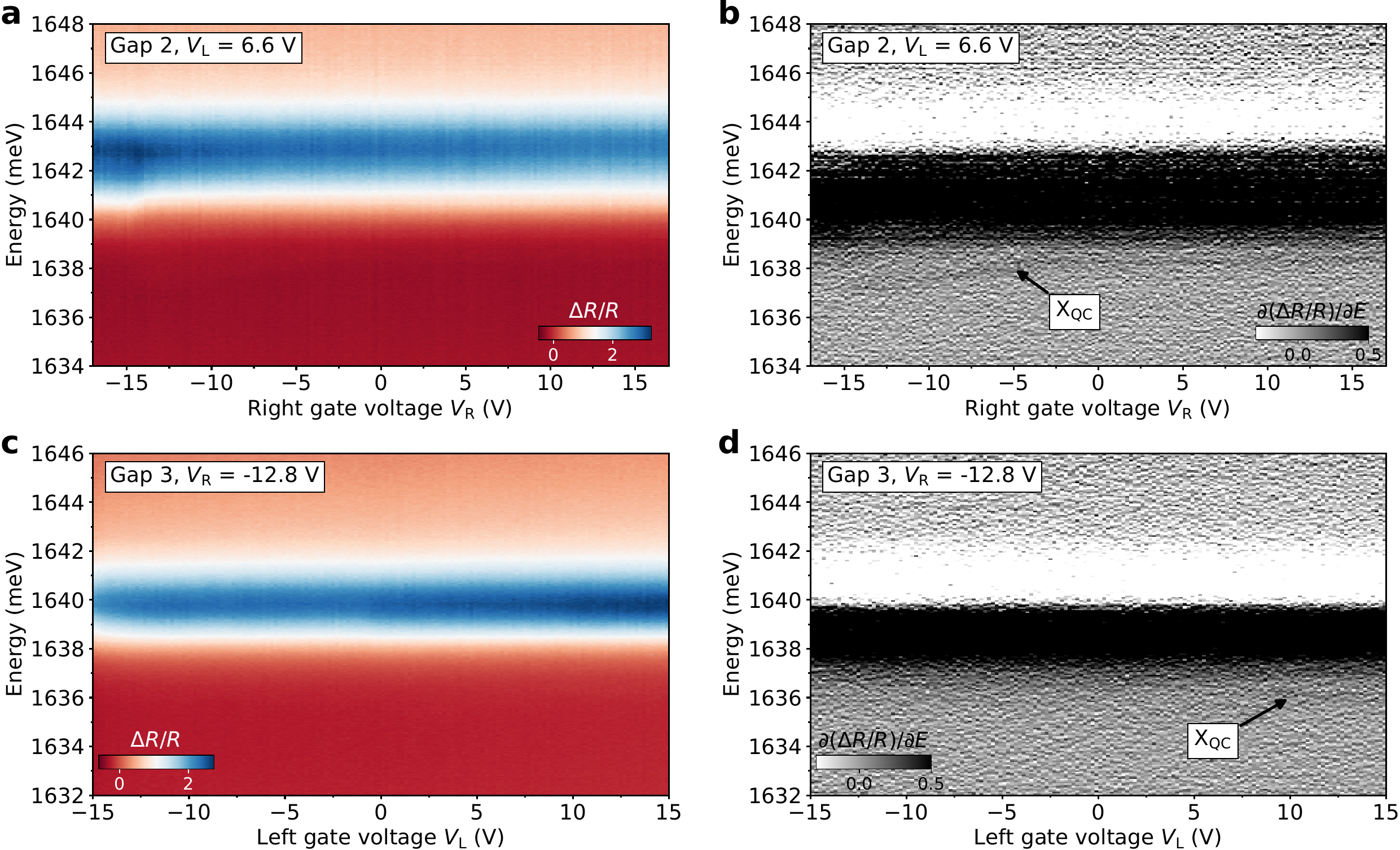}
	\caption{\textbf{Preliminary signatures of 0D quantum confined excitons.}
    (\bfA) Normalized reflection contrast $\Delta R/R$ as a function of $\vr$ for fixed $\vl=6.6\,$V and $\vtg=-0.4\,$V, acquired on gap 2 between BG-L2 and BG-R2. Other than a broad exciton resonance associated with the extended TMD region, no additional features are discernible. (\bfB) Derivative of normalized reflection contrast at the voltage configuration as in (\bfA) reveals an additional faint and narrow spectral line which redshifts for decreasing $\vr$. (\bfC), (\bfD) Normalized reflection contrast $\Delta R/R$ and its derivative, respectively, as a function of $\vl$ for fixed $\vr=-12.8\,$V and $\vtg=-0.4\,$V, acquired on gap 3 between BG-L3 and BG-R3. Similar behaviour is observed as on gap 2, however the faint resonance monotonously redshifts for increasing $\vl$. We tentatively attribute these faint resonances as stemming from quantum-confined excitons localized in a 0D trap.
    }  
	\label{fig:device4_0D}
\end{figure}

Having identified a suitable range of gate voltages for probing the optical response of 0D quantum-confined excitons, we monitor normalized differential reflectance $\Delta R/R$ on gap 2, while keeping $\vtg$ fixed at $-0.4\,$V and tuning $\vl$ and $\vr$. Fig.\,\ref{fig:device4_0D} {\bfA} depicts one particular instance of this $\vl$-$\vr$ dual-gate scan for varying $\vr$ and fixed $\vl=6.6\,$V. Across the entire range of BG voltages, a broad exciton resonance centered around $E_\mathrm{X,2D}$ of approximately $1643\,$meV can be identified, the energy of which remains unaffected as $\vl$ or $\vr$ are changed. Owing to its large oscillator strength and lack of shift in energy with gate voltage, we identify this resonance as stemming from the charge neutral extended TMD region. For energies lower than $E_\mathrm{X,2D}$, no clear resonance is discernible in the reflectance measurement, in stark contrast to spectral signatures observed for 1D quantum-confined excitons (e.g.\, see Fig.\,\ref{fig:3_5_WL}, Fig.\,\ref{fig:device_cross-gate} {\bfC} or Fig.\,\ref{fig:device4_1DX}). However, by taking the derivative of reflectance contrast with respect to energy a faint and narrow spectral line can be observed which redshifts for decreasing $\vr$ (Fig.\,\ref{fig:device4_0D} \bfB). Very similar behaviour is observed on gap 3: In the bare reflectance measurement as a function of $\vl$ and fixed $\vr=-12.8\,$V only the 2D exciton resonance associated with the extended TMD region is visible (Fig. Fig.\,\ref{fig:device4_0D} \bfC). On the other hand, in the reflectance contrast derivative, a second faint resonance appears which redshifts for increasing $\vl$ (Fig.\,\ref{fig:device4_0D} \bfD).

Combined together, these features strongly suggest that the additional resonances, visible only in the derivative of reflection contrast, are in fact associated with 0D quantum-confined excitons. Their substantially weaker oscillator strength, as compared to 1D quantum-confined excitons, is to be expected, owing to a confinement length of under $10\,$nm now also along an additional spatial dimension. Moreover, the redshift of these resonances is most pronounced, only when there is a voltage asymmetry between $\vl$ and $\vr$. While on gap 2 this case is depicted for positive $\vl$ and negative $\vr$ (Fig.\,\ref{fig:device4_0D} \bfB), on gap 3 the opposite scenario is shown (Fig.\,\ref{fig:device4_0D} \bfD). If these features were rather linked to 1D confinement potentials along the edge of BG electrodes, the redshift would have been symmetric around $\vl$ or $\vr$ approximately $0\,$V. This is because the fringing electric fields along the edge of the BG electrodes are solely a result of the voltage difference between BG and TG, which would change in the same manner as the BG electrodes are tuned to positive or negative voltages. Lastly, these exact gate voltage scans were also conducted on other parts of the device, such as at a boundary between one of the BGs and the extended TMD region, or away from the split-gate electrodes altogether (appendix section \ref{appendix:dev4}). In all these measurements no spectral signatures, as observable in Fig.\,\ref{fig:device4_0D} {\bfB} and {\bfD}, were visible, leading us to conclude that the faint resonances indeed originate from the gap region in between the split-gate electrodes.

%%%%%%%%%%%%%%%%%%%%%%%%%%%%%%%%%%%%%%%%%%%%%%%%%%%%%%%%%%%%%%%%%%%%%%%%
\section{Next steps}
%%%%%%%%%%%%%%%%%%%%%%%%%%%%%%%%%%%%%%%%%%%%%%%%%%%%%%%%%%%%%%%%%%%%%%%%

While the preliminary measurements conducted on the split-gate geometry show great promise towards realization of 0D quantum-confined excitons, making a more definite claim will require further experimental investigation. Broadband reflectance spectroscopy provided a good initial starting point for identifying the voltage configurations suitable for achieving 0D quantum confinement. However, the weak oscillator strength of the corresponding spectral features does not permit a quantitative analysis of the resonance line shape in this measurement configuration. Alternatively, a PL experiment was attempted, but did not prove to be beneficial. In the device investigated the spectral range just below the exciton resonance appears to be dominated by defect emission, which potentially masks the spectral features associated with 0D excitons (appendix section \ref{appendix:dev4}).

To better isolate these features we have started efforts aimed at performing electromodulation spectroscopy \cite{Yu2010}, a methodology which has already been successfully employed in various settings to better isolate faint spectral signatures embedded in a noisy background \cite{Alen2003,Barre2022}. Given the sensitivity of $\xqc$ states to electric fields, alternating voltages with distinct frequencies applied to the left and right split-gate electrodes will lead to a corresponding variation in their resonance energy, which is proportional to the amplitude of the gate voltage modulation. By using monochromatic laser light to monitor this modulation in reflectance, and utilizing lock-in detection to demodulate at the sum- or difference-frequency, we can effectively exclude signals, including noise, at all other frequencies. In this manner, a significantly enhanced signal-to-noise ratio can be achieved. This experimental framework will also prove advantageous for detecting the potentially weak electroluminescence signal originating from the gap region. Ultimately, to conclusively demonstrate that excitons are localized in 0D confinement potentials and that photon blockade has been achieved, photon statistics need to be measured using a Hanbury Brown-Twiss setup. This would unequivocally establish the existence of nonlinearities at the single-photon level.

% {\color{red}
% \begin{itemize}
%     \item remove quasi-0d, make only 0D
% \end{itemize}
% }

% This would unequivocally prove the quantum nature of the emitted light.
% Ultimately, definitively claiming that excitons are localized in 0D confinement potentials and photon blockade has been achieved will require measuring photon statistics using a HBT setup to truly demonstrate the quantum nature of the emitted light.

% In parallel, device voltage configurations should be identified that allow measuring electroluminescence, while ensuring that the doping 
% weak oscillator strength does not permit a quantitative analysis of the line shape: oscillator strength and line width
% 

% \begin{itemize}
    % \item Data from cross-gate sample (QC-02, QC-01)
    % \item High-performance semiconductor quantum-dot single-photon sources - section on radiative cascade
    % \item Interfacing single photons and single quantum dots with photonic nanostructures - section on semiconductor quantum dots + p392
    % \item Turschmann - Multi-emitter systems, page 11
    % \item PhD thesis Palacios Barraquero
    % \item exciton-exciton interactions - Lightshift paper
    % \item Maximize oscillator strength - Motivation for quasi-0D (RMP Lodahl p.355)
    % \item can control mode overlap with photonic mode by tunabilty of excitonic wavefucntion
    % \item Imamoglu Carusotto paper
    % \item g2(0) relation given by U - Conventional and Unconventional Photon Statistics
% \end{itemize}
  % \include{Sources/7_Discussion}
  % \include{Sources/Summary}
  % \include{Sources/Appendix}

\bookmarksetup{startatroot}
\backmatter

  \appendix
  %%%%%%%%%%%%%%%%%%%%%%%%%%%%%%%%%%%%%%%%%%%%%%%%%%%%%%%%%%%%%%%%%%%%%%%%
% \setcounter{chapter}{1}% Equivalent to "letter O"
% \renewcommand{\thechapter}{\Alph{chapter}}%
\renewcommand{\thesection}{\Alph{section}}%
\chapter{Appendix}
%%%%%%%%%%%%%%%%%%%%%%%%%%%%%%%%%%%%%%%%%%%%%%%%%%%%%%%%%%%%%%%%%%%%%%%%

\renewcommand{\thefigure}{\Alph{section}.\arabic{figure}}

%%%%%%%%%%%%%%%%%%%%%%%%%%%%%%%%%%%%%%%%%%%%%%%%%%%%%%%%%%%%%%%%%%%%%%%%
\section{Device fabrication procedure}
\label{appendix:fabrication}
%%%%%%%%%%%%%%%%%%%%%%%%%%%%%%%%%%%%%%%%%%%%%%%%%%%%%%%%%%%%%%%%%%%%%%%%

All MoSe$_2$, h-BN and graphene flakes used to assemble the devices presented in this thesis are obtained through mechanical exfoliation of bulk crystals (HQ Graphene MoSe$_2$ and NIMS/2D Semiconductors h-BN) onto $285$\,nm SiO$_2$/Si substrates using wafer dicing tape (Ultron). The flake thicknesses are identified using optical contrast measurements \cite{Ni2007,Li2013,Golla2013} and/or atomic force microscopy. For device 1, embedded via–contacts \cite{Telford2018,Jung2019,Liu2022} are formed by etching holes in the h-BN spacer by means of electron beam lithography and reactive ion etching (Oxford Plasmalab 80Plus). The etching conditions are the following: CHF$_3$:O$_2$ 40:4 sccm gas mixture, $60$\,W RF power and $40$\,mTorr gas pressure. To ensure a smooth bottom surface for the subsequent
metal evaporation in the holes, the etched h-BN layer is transferred to a new SiO$_2$/Si substrate using a standard dry polymer transfer method (also used for stacking, see following paragraph). Next, the pre-defined holes are filled with metal by performing a second lithography step, followed by electron-beam evaporation. The multilayer stack is then created by picking up the top h-BN layer with embedded Via–contacts and laminating it onto a monolayer MoSe$_2$, and a bottom h-BN flake using the aforementioned dry transfer process.

For the dry transfer of flakes \cite{Zomer2014,Pizzocchero2016}, we use a glass slide onto which a hemispherical polydimethylsiloxane (PDMS) stamp is attached. This stamp is covered with a thin
layer of polycarbonate (PC) allowing the sequential pickup of the flakes. All stacking steps are thereby performed in an inert Ar atmosphere inside a glovebox and at a temperature of $120$\,$^{\circ}$C. The finished stack is deposited onto
the target substrate by increasing the temperature up to $150$\,$^{\circ}$C, which allows the PC to delaminate from the PDMS and to be released onto the substrate. By further heating to $170$\,$^{\circ}$C, the PC is torn at the edges. The PC is dissolved by immersing the substrate in chloroform.

%%%%%%%%%%%%%%%%%%%%%%%%%%%%%%%%%%%%%%%%%%%%%%%%%%%%%%%%%%%%%%%%%%%%%%%%
\section{Implementing electrostatic simulations in COMSOL}
\label{appendix:comsol}
\setcounter{figure}{0}
%%%%%%%%%%%%%%%%%%%%%%%%%%%%%%%%%%%%%%%%%%%%%%%%%%%%%%%%%%%%%%%%%%%%%%%%

In this section, we describe additional details about the implementation of our electrostatic simulations (section \ref{sec:1DX_sims}) within the \emph{Electrostatics} package of COMSOL. In the Thomas--Fermi approximation, the electron and hole charge density in the TMD monolayer, $\sigma_\mathrm{n}$ and $\sigma_\mathrm{p}$, respectively,  depend on the local electrostatic potential $V(x)$ (Eqns.\,\ref{eqn:sigma_n} and \ref{eqn:sigma_p}). These functional relations can be defined in COMSOL under \verb|Global Definitions| using an \verb|Analytic Function|. For the electron density we define a function \verb|sigma_n| with the following expression:
\begin{verbatim}
    -Cq_n*(pot-dCB/e_const)*flc2hs((pot-dCB/e_const)/(1 [V]),10e-6)
\end{verbatim}
in which \verb|Cq_n| is a parameter representing the electron quantum capacitance of the TMD monolayer ($=e^2\mathcal{D}(E)$, see definition in section \ref{sec:1DX_sims}), \verb|dCB| is the conduction band offset relative to the Fermi level, \verb|e_const| is the elementary charge and \verb|pot| is the function argument. \verb|flc2hs| is a smoothed Heaviside function, pre-implemented in COMSOL. We use \verb|flc2hs| to describe the onset of electron doping as soon as the Fermi level exceeds the conduction band edge $E_\mathrm{C}$. By increasing the smoothness parameter of the \verb|flc2hs| function finite-temperature effects can be modelled. Similarly, we define the hole charge density \verb|sigma_p| as
\begin{verbatim}
    -Cq_p*(pot+dVB/e_const)*flc2hs(-(pot+dVB/e_const)/(1 [V]),10e-6)
\end{verbatim}
in which \verb|Cq_p| denotes the hole quantum capacitance of the TMD monolayer, \verb|dVB| is the valence band offset relative to the Fermi level and \verb|pot| is the function argument. In this manner, hole doping occurs when the valence band edge $E_\mathrm{V}$ exceeds the Fermi level. The function for the total density \verb|sigma_tot| (Eqn.\,\ref{eqn:sigma}) can then be defined as
\begin{verbatim}
    sigma_n(pot)+sigma_p(pot)
\end{verbatim}
In all the definitions above we ensure that function argument units are in [V] and the function output units are [C/m$^2$].

Next, the device architecture can be implemented as desired in the \verb|Geometry| section. Furthermore, since we model the TMD monolayer truly in the 2D limit, i.e.\,assuming it to have zero thickness, the only properties that need to be defined in the \verb|Materials| section, are the \verb|Relative permittivity| of the encapsulating dielectric. The \verb|Electrostatics| boundary conditions for the gate electrodes are defined using a fixed value \verb|Electric Potential|. The monolayer semiconductor is modelled with a \verb|Surface Charge Density| denoted by \verb|sigma_tot(V)|. Here, \verb|V| is the built-in COMSOL expression for the local \verb|Electric Potential|. To increase computational efficiency, we always employ an adaptive mesh, with element sizes becoming smaller in proximity to the 2D semiconductor. Lastly, we increase the \verb|Maximum number of iterations| to $100$ for the numerical solver to ensure convergence. This setting can be found in \verb|Solver Configurations| $\blacktriangleright$ \verb|Solution| $\blacktriangleright$ \verb|Stationary Solver|$ \blacktriangleright$ \verb|Fully Coupled|.

%%%%%%%%%%%%%%%%%%%%%%%%%%%%%%%%%%%%%%%%%%%%%%%%%%%%%%%%%%%%%%%%%%%%%%%%
\section{Supplementary measurements on doping characteristics of Device 1}
\label{appendix:qpc}
\setcounter{figure}{0}
%%%%%%%%%%%%%%%%%%%%%%%%%%%%%%%%%%%%%%%%%%%%%%%%%%%%%%%%%%%%%%%%%%%%%%%%

\begin{figure}[htb]
    \centering
	\includegraphics[width=7cm]{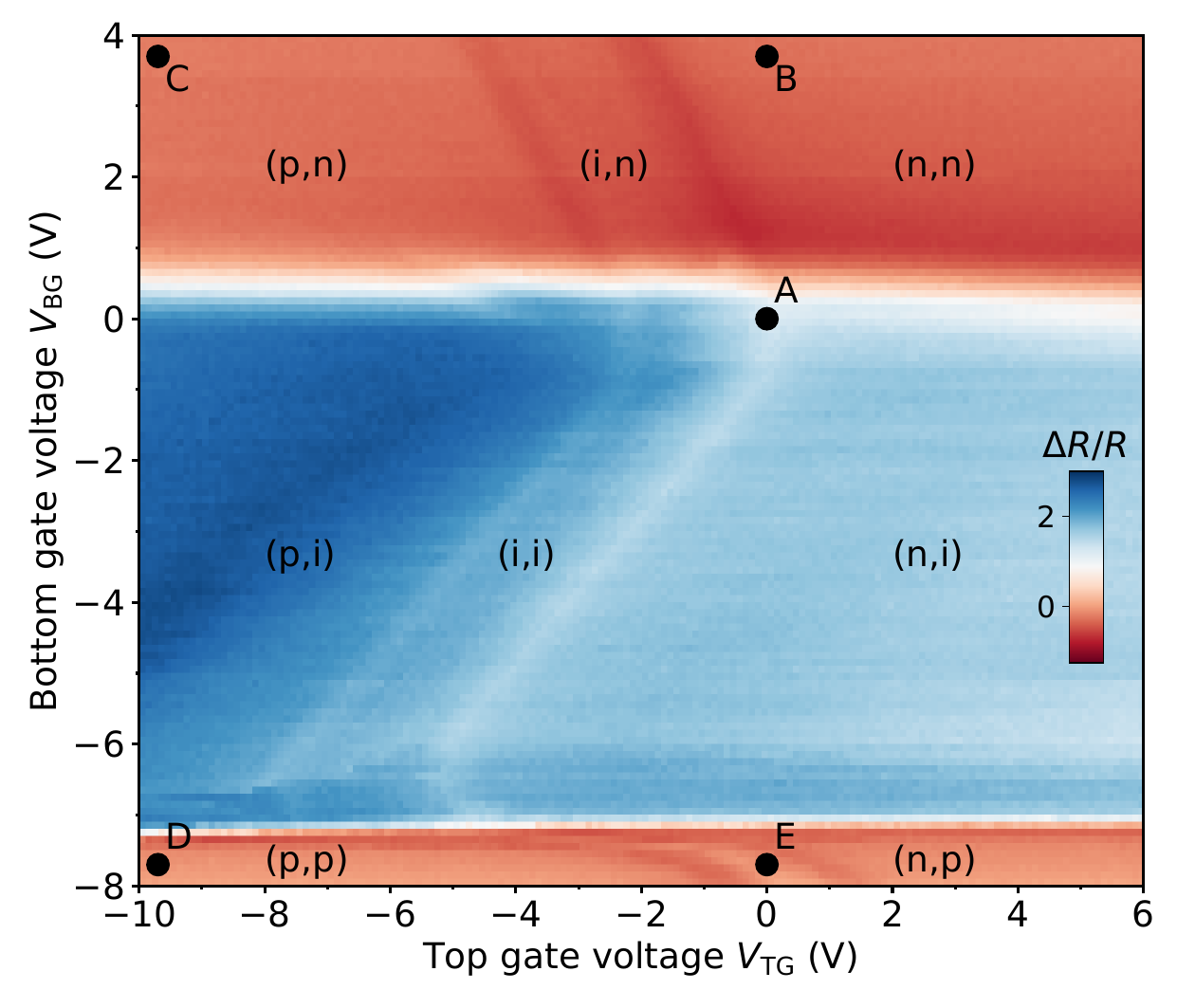}
	\caption{\textbf{Doping characteristics of Device 1.}
    Tracking the charge-density-dependent energy shifts of the repulsive polaron resonance, obtained by measuring differential reflectance $\Delta R/R$, allows us to map the doping configuration in Device 1 (as also detailed in Fig.\,\ref{fig:3_11_Doping}). The doping state in regions I and II is labelled following the convention (II,I). In addition, we indicate five distinct voltage configurations, points A -- E, through which the gate voltages are swept while monitoring the source-drain current through the quantum point contact in device 1.
    }  
	\label{fig:app_doping_device1}
\end{figure}

In this section, we revisit the doping characteristics of Device 1. Previously, by measuring differential reflectance $\Delta R/R$ in the vicinity of the TG edge and tracking the charge-density-dependent energy shifts of the repulsive polaron resonance, we could map the entire doping configuration of Device 1 as a function of $\vtg$ and $\vbg$. In particular, since the TG is optically transparent, the doping state of region II (under the TG) can be directly identified optically. Here, we depict the resulting charging map again for reference (Fig.\,\ref{fig:app_doping_device1}), which clearly indicates that all doping configurations are possible in our device. This is surprising considering that the TMD monolayer is only electrically contacted in region I, and not in region II. Hence, we hypothesized in section \ref{chap:1D:sec:alternative_configs} that charge injection in region II arises through photoinduced doping.

We can verify this claim by making use of the fact, that the TG in Device 1 can also function as a quantum point contact (QPC) and thus can act as a sensitive probe of the electrostatic potential landscape in the device. In particular, it can allow us to infer whether hole injection in region II is possible without any optical illumination when region I is n-doped. This is accomplished by monitoring the TG voltage necessary to pinch off the source-drain current $I_\mathrm{SD}$ through the QPC under different conditions. We emphasize that these measurements are conducted without any optical illumination and with a source-drain bias $V_\mathrm{SD}$ maintained at $2\,$V.

\begin{figure}[htb]
    \centering
	\includegraphics[width=14cm]{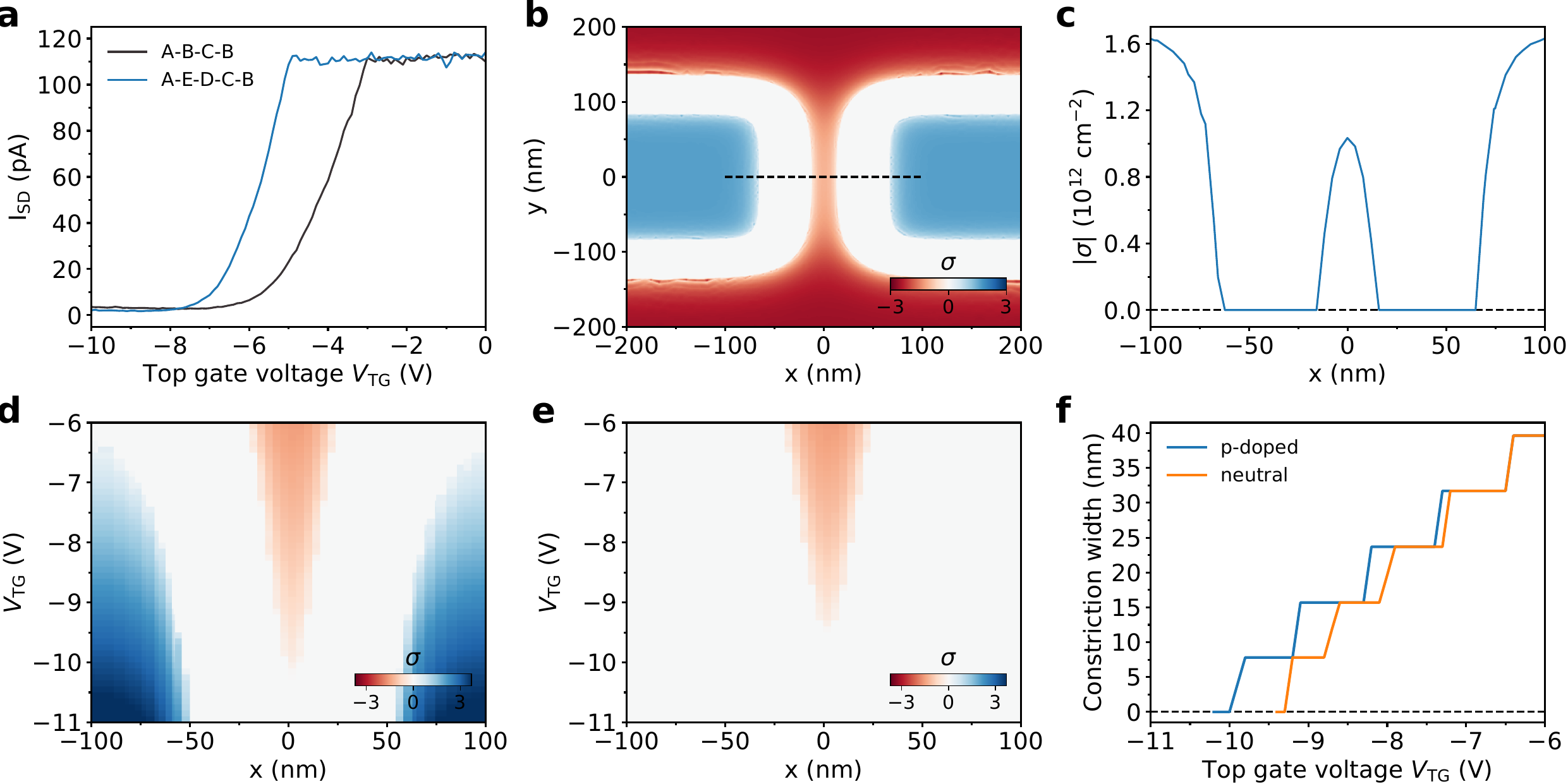}
	\caption{\textbf{Doping characteristics in the vicinity of the quantum point contact.}
    (\bfA) Source-drain current $I_\mathrm{SD}$ through the quantum point contact as a function of $\vtg$ without any optical illumination. The gate voltages are swept through the points A-B-C-B (black) and A-E-D-C-B (blue). $I_\mathrm{SD}$ is measured in the last step when sweeping from point C to point B. A difference in pinch-off voltage for the two cases indicates the presence of holes under the TG for the path A-E-D-C-B. (\bfB) Charge density configuration in Device 1 as determined through electrostatic simulations for $\vbg=4\,$V and $\vtg=-8\,$V. Hole doping is indicated in blue, electron doping in red. The white region depicts the depletion region in between the two doped areas. The magnitude of charge density across the black dashed line is shown in (\bfC). (\bfD) Evolution of the charge density configuration along the black dashed line in (\bfB) as $\vtg$ is varied. (\bfE) Evolution of the charge density configuration in the constriction, as in (\bfD), with hole doping under the TG being suppressed. (\bfF) Width of the constriction, corresponding to the electron-doped area, as $\vtg$ is scanned. A gate voltage of smaller magnitude is required to pinch off the conducting channel completely if the region under the TG remains charge-neutral (orange curve), in comparison to when it is p-doped (blue curve). The step-like features are artifacts of the simulation. At the expense of increased computation time, with a finer mesh such features can be avoided.
    }  
	\label{fig:app_QPC_current}
\end{figure}

In one scenario, the gate voltages are swept though points A-B-C (Fig.\,\ref{fig:app_doping_device1}), after which $I_\mathrm{SD}$ is measured when $\vtg$ is increased from C to B (black curve in Fig.\,\ref{fig:app_QPC_current}). In moving from A to B, the device is globally n-doped. However, as the channel is pinched off by going from B to C and back, hole injection should be suppressed, since the device is not illuminated throughout. In a second scenario, the gate voltages are scanned though points A-E-D-C-B (Fig.\,\ref{fig:app_doping_device1}), and $I_\mathrm{SD}$ is measured in the last step, when increasing $\vtg$ from C to B (blue curve in Fig.\,\ref{fig:app_QPC_current}). Here, by first scanning $\vbg$ from A to E, the device becomes p-doped globally. Next, decreasing $\vtg$ by sweeping to D creates a surplus of holes under the TG. In this manner, when $\vbg$ is swept back to C, it is explicitly ensured the region II is p-doped. If now $\vtg$ is increased from C to B, it can be observed in Fig.\,\ref{fig:app_QPC_current} that the QPC becomes conducting at a different pinch-off voltage $\vtg$, when compared to the first measurement. This disparity between the two measurements is a clear indication that only in the second scenario (path A-E-D-C-B) region II is p-doped, while it remains charge-neutral in the first case (path A-B-C-B).

The fact that when hole doping is suppressed, a gate voltage $\vtg$ smaller in magnitude is necessary to pinch off the conducting channel, can be verified by performing electrostatic simulations of the charge density configuration in the vicinity of the QPC. In Fig.\,\ref{fig:app_QPC_current} {\bfB} we depict the doping configuration at $\vbg=4\,$V and $\vtg=-8\,$V, just prior to pinch-off. In region I the TMD is n-doped (red region), while region II is p-doped (blue region). The magnitude of charge density across the black dashed line is shown in Fig.\,\ref{fig:app_QPC_current} \bfC. This corresponds to the charging configuration realized by taking the path A-E-D-C-B. In particular, we focus our attention on the evolution of charge density along the black dashed line, as $\vtg$ is varied. Figs.\,\ref{fig:app_QPC_current} {\bfD} and {\bfE} depict the two scenarios discussed above, in which p-doping in region II is permitted (path A-E-D-C-B) and suppressed (path A-B-C-B), respectively. Notably, in the latter case, the channel becomes insulating already at $\vtg\approx-9\,$V, while in the former setting $\vtg\approx-10\,$V need to be applied to pinch off the QPC completely. This trend is depicted more clearly in Fig.\,\ref{fig:app_QPC_current} \bfF, which illustrates the constriction width, corresponding to the n-doped area, as $\vtg$ is decreased. Evidently, presence of charge carriers in region II, screens the potential modulation in the device, such that effectively a greater potential gradient between TG and BG is required for pinch-off. Hence, this measurement clearly demonstrates that without optical illumination p-doping in region II cannot be achieved when region I is n-doped. Therefore, this further highlights the importance of photoinduced doping in realizing the charging map shown in Fig.\,\ref{fig:app_doping_device1}.

%%%%%%%%%%%%%%%%%%%%%%%%%%%%%%%%%%%%%%%%%%%%%%%%%%%%%%%%%%%%%%%%%%%%%%%%
\section{Supplementary measurements on Device 4}
\label{appendix:dev4}
\setcounter{figure}{0}
%%%%%%%%%%%%%%%%%%%%%%%%%%%%%%%%%%%%%%%%%%%%%%%%%%%%%%%%%%%%%%%%%%%%%%%%

\begin{figure}[htb]
    \centering
	\includegraphics[width=14cm]{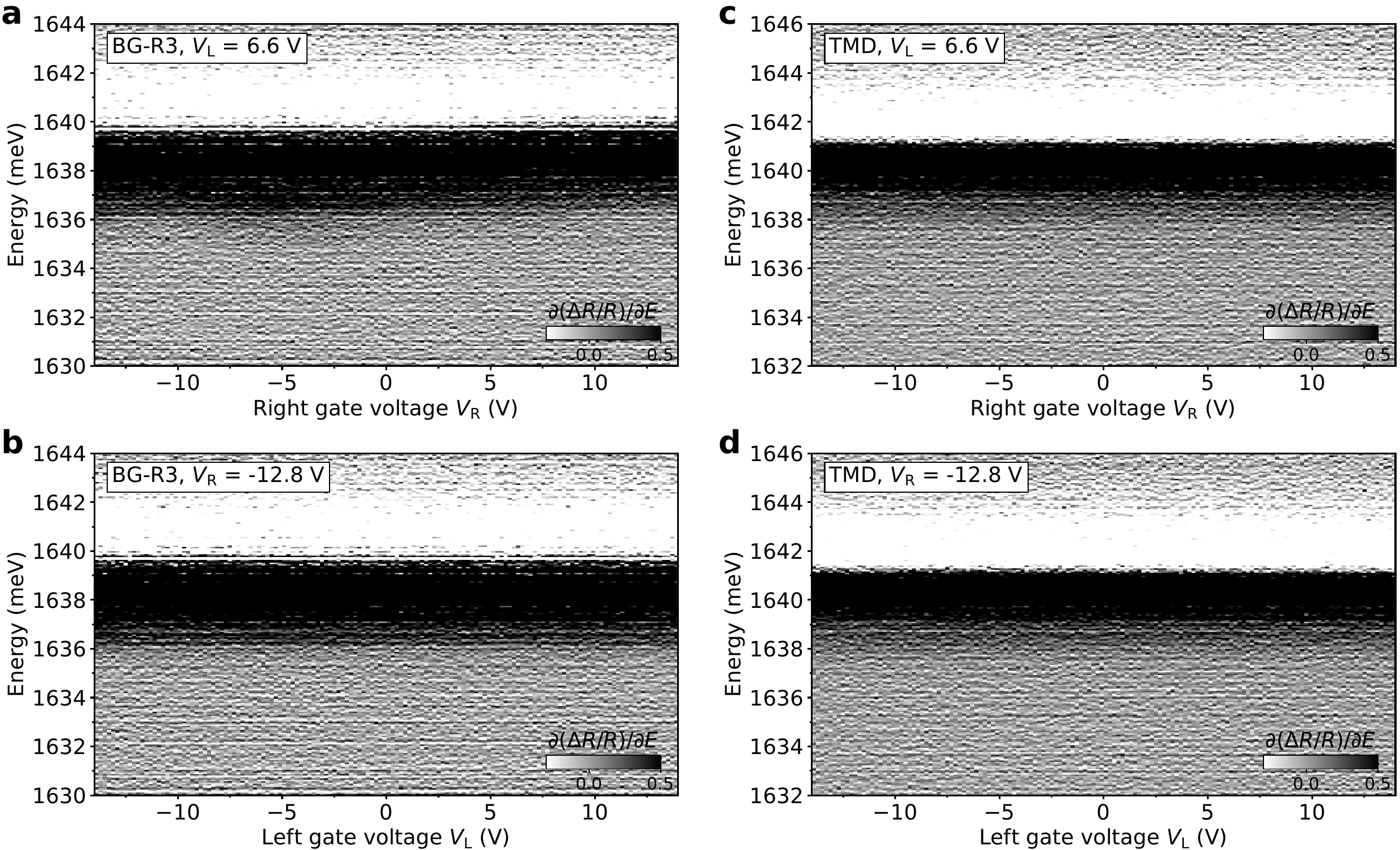}
	\caption{\textbf{Scans of split gate electrode voltages performed away from gap region.}
    (\bfA), (\bfB) Derivative of reflectance contrast $\partial(\Delta R/R)/\partial E$ for fixed $\vl=6.6\,$V and varying $\vr$, and fixed $\vr=-12.8\,$V and scanned $\vl$, respectively. These measurements are conducted at an interface between BG-R3 and the extended TMD region. (\bfC), (\bfD) Derivative of reflectance contrast $\partial(\Delta R/R)/\partial E$ at the same voltage configurations, performed in a region away from the BG electrodes. In all cases $\vtg=-0.4\,$V is maintained. Other than a neutral exciton resonance $\xfree$, no other spectral features can be distinguished in these datasets.
    }  
	\label{fig:device4_verification}
\end{figure}

To further corroborate our claim that the faint resonances observed in Fig.\,\ref{fig:device4_0D} indeed originate solely from the gap region in between the split-gate electrodes, we have conducted similar $\vl$-$\vr$ scans for fixed $\vtg=-0.4\,$V on different regions of Device 4. Fig\,\ref{fig:device4_verification} {\bfA} and {\bfB} show the derivative of reflectance contrast $\partial(\Delta R/R)/\partial E$ for the same voltage configurations as in Fig.\,\ref{fig:device4_0D} {\bfB} and {\bfD}, acquired at a boundary between BG-R3 and the extended TMD region. Evidently, across the entire scanned range of $\vl$ and $\vr$ the voltage asymmetry between TG and BG electrodes is not sufficient to create a redshift greater than the 2D exciton linewidth for 1D excitons. Fig\,\ref{fig:device4_verification} {\bfC} and {\bfD} depict the result of the same measurement, performed in a region away from the split gate electrodes. In all cases, no monotonically redshifting resonances can be discerned. These observations provide further evidence that the faint optical signatures seen in Fig.\,\ref{fig:device4_0D} indeed originate solely from gap region, and thus are associated with 0D excitons.

\begin{figure}[htb]
    \centering
	\includegraphics[width=8cm]{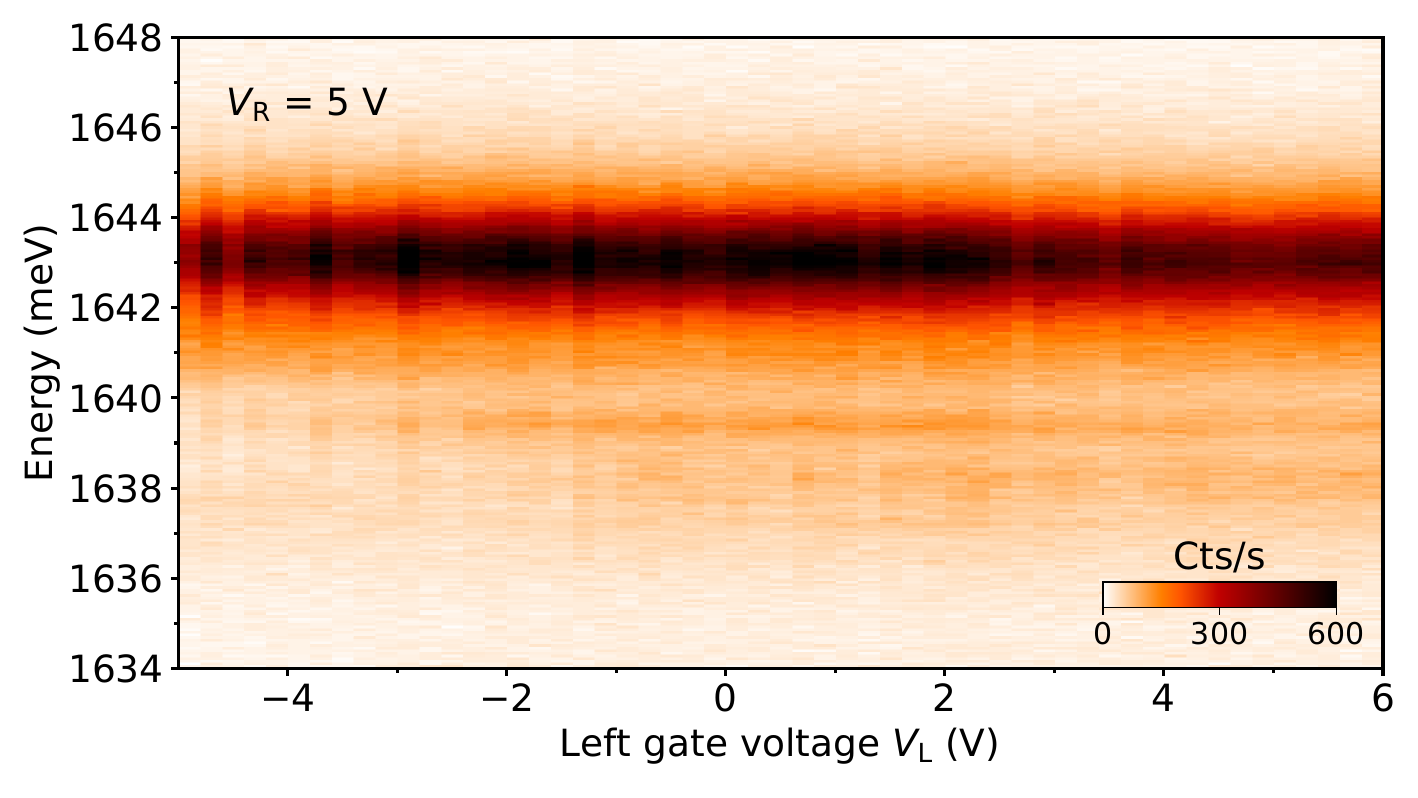}
	\caption{\textbf{Photoluminescence measurement on gap region.}
    The spectral range just below the exciton resonance exhibits defect emission, which potentially masks the spectral features associated with 0D excitons.
    }  
	\label{fig:device4_PL}
\end{figure}

In an attempt to better isolate the 0D exciton resonance from the free exciton resonance $\xfree$ we also conducted a PL experiment (Fig.\,\ref{fig:device4_PL}). However, this measurement did not prove to be beneficial, since the spectral range below the free exciton resonance appears to be dominated by multiple narrow lines which do not shift with voltage. Therefore, we conclude that these features are associated with localized defects in the TMD monolayer.

  \printindex
  \printbibliography

\end{document}